\documentstyle[aps,epsf,epsfig]{revtex}
\begin{document}

\title{Asymptotic properties of turbulent magnetohydrodynamics\\
}
\author{
Arjun Berera$\;$\thanks
{E-mail: ab@ph.ed.ac.uk; 
PPARC Advanced Fellow}
and David Hochberg$\;$\thanks
{E-mail: hochberg@laeff.esa.es}$^{, \dagger \dagger}$
}
\bigskip
\address{ 
$^{*}$Department of Physics and Astronomy, 
University of Edinburgh, Edinburgh EH9 3JZ, Great Britain \\
$^{\dagger}$Laboratorio de Astrof\'\i sica  Espacial y  F\'\i sica
Fundamental, Apartado 50727, 28080 Madrid, Spain\\
$^{\dagger \dagger}$Centro de Astrobiolog\'\i{}a 
(Associate Member of NASA Astrobiology Institute), 
CSIC-INTA, Carretera Ajalvir, Kilometer 4, 28850 Torrej\'on de Ardoz, 
Madrid, Spain }
\bigskip
\date{Version 1.00; 7 March 2001; \LaTeX-ed \today}
\maketitle

\bigskip

{\small

{{\bf Abstract}: The dynamic renormalization group (RG) is used to
study the large-distance and long-time limits of viscous and
resistive incompressible magnetohydrodynamics subject to random
forces and currents. The scale-dependent viscosity and magnetic 
resistivity are
derived and used for carrying out RG-improved perturbation theory. 
This is applied to derive both the asymptotic
scaling and the overall proportionality 
coefficients for both
the velocity and magnetic field correlation functions as well
as the kinetic and magnetic energy density spectral functions.
The Kolmogorov, Iroshnikov-Kraichnan, as well as other energy spectra,
formally can be obtained by suitable choice of injected noise, although
the method limits the validity of these energy spectra only
to the asymptotic regime.   
Injection of a random magnetic helicity is considered, its
RG-improved spectral density derived, and its contribution to the
velocity and magnetic field correlation functions determined.
The RG scaling solutions are used to determine
information at asymptotic scales about energy and helicity cascade
directions and mixing between magnetic and kinetic energy.
Some of the results found here also are shown to be valid
for the Navier-Stokes hydrodynamic equation.
} 

\medskip

}  

\section{Introduction}
\label{intro}

The nonlinear equations of magnetohydrodynamics (MHD) admit
turbulent solutions that can cover a wide range of spatial
scales. One expects to see the turbulent regime of MHD as the
most probable state of a magnetofluid present
in physical systems ranging from the solar corona 
(Van Ballegooijen 1985; 1986) and solar
wind (Burlaga 1991)
to galactic magnetic fields (DeYoung 1980; 1992). MHD turbulence in the early
universe may have even left a detectable present-day signature at  
the cosmological scale (Brandenburg et al. 1996).

To study questions such as the asymptotic scaling behavior
in magnetic field correlations, the presence or absence of
kinetic and magnetic energy cascades, 
the helicity spectrum and its corresponding cascades, as
well as the effect of random perturbations, 
one needs to consider
the evolution of the magnetofluid over a wide range of spatial
and temporal scales. 
In general, turbulence is an essential
feature in coming to terms with these problems, whose rigorous treatment calls 
for an analysis of fully nonlinear MHD.  

{}For the regime that interests us, the large-distance and long time behavior
of the fluid velocity and magnetic fields, the equations of MHD 
lend themselves to an analytical treatment by applying
the methods provided by the 
dynamic renormalization group (RG). 
The aim of the RG is to describe quantitatively how the dynamics of a system
evolves as one changes the length and time scales of the phenomena under study.
The RG is an excellent tool for exploring the physics of complicated physical
systems that are characterized as having 
many interacting degrees of freedom and are subject to
fluctuations covering many different length scales. 
RG methods were
first used in the study of hydrodynamic (non-magnetic) turbulence 
by Forster, Nelson and  Stephen (1977).
They studied the large-distance and long-time
properties of velocity correlations generated by the Navier Stokes equations
for a randomly stirred incompressible fluid with a variety of
Gaussian noise spectra.  More recently these methods were
used by
Yakhot and Orszag (1986) and Dannevik, Yakhot and Orszag 
(1987) in an attempt to
predict several parameters in 
fully developed hydrodynamic turbulence. 
A detailed critique of their use of the RG can be found
in the paper by Eyink (1994).
Some recent reviews of 
the RG in hydrodynamic turbulence may be found in 
Frisch (1995) and McComb (1995).

There have been rather fewer applications of the RG to magnetized fluids, 
probably due to the complexity of the problem and the lack of experimental
data for turbulent magnetic fields in laboratory plasmas. Indeed, 
as indicated above, MHD turbulence
is much more likely to be of interest at astrophysical and cosmological
scales. 
The previous works, that apply RG to MHD turbulence for addressing various
specific problems, have been carried out by
Fournier, Sulem and Pouquet (1982), Longcope and Sudan
(1991)  and more recently
by Camargo and Tasso (1992).  A recent survey of
MHD turbulence can be found in the monograph by 
Biskamp (1993).

In this paper, RG will be applied to examine the 
asymptotic properties of turbulent MHD with a stochastic
force term.  Our analysis treats the full MHD equations,
and thus differs from
Longcope and Sudan (1991), who treated a type of "reduced
MHD".  Also, we treat all nonlinear terms as given in the MHD 
equations on an equal footing, 
in contrast to Fournier et. al. 
(1982), who weighted the inertial nonlinearity
and Lorentz force differently.
In this respect, the starting equations for us are the same as
those of Camargo and Tasso (1992), although our treatment
differs significantly from their work.  In regards to the calculation
technique, we will cast the stochastic MHD equations into
a functional path integral formalism, in contrast
to the direct iterative solution of the equations of motion
carried out in (Camargo and Tasso 1992).  The functional path integral
formalism permits use
of useful field theory methods,
which are much more compact and efficient than direct
iterative expansions.  Although in this paper
we use this formalism only to implement our one-loop RG
analysis,
the formalism can be applied much more generally
to study stochastic MHD.  With respect to results,
we differ from all the previous works
(Fournier et al. 1982; Longcope and Sudan 1991; Camargo and Tasso
1992).
We will examine cascade directions and
energy mixing, present solutions of the correlation functions
in the scaling regime and implement a RG improved perturbation theory.
All of these results are being explored
and presented here for the first time.  {}Furthermore, we also
will implement a RG analysis when magnetic helicity is
present, and to our knowledge no such purely analytic treatment
of this problem has been 
carried out before in the literature.

The paper is organized as follows.
In Section \ref{sect2} we introduce the
stochastic equations of incompressible MHD.
These equations then are 
written in terms of Elsasser variables,
which make manifest the symmetry between the velocity and magnetic field 
for incompressible MHD (Elsasser 1950).
This symmetry is useful in checking the correctness of the renormalization
group equations that will be derived and for understanding the nature
of the correlation functions and energy 
spectra for the velocity and magnetic fields.
In the stochastic equations,
the stochastic random forces and currents are intended 
to model {\it both} random initial
conditions and subsequent disturbances and fluctuations. 
The noise therefore serves
a two-fold purpose and it  
summarizes at a coarse-grained level our ignorance about the
details of the numerous small-scale phenomena. {}For example, as applied
to the problem of galactic or even cosmic fields, the noise would
account for the many diverse
particle physics mechanisms 
(e.g., early phase transitions, vacuum bubble
collisions, fluctuating Higgs gradients, superconducting cosmic
strings, etc.) that have been proposed for 
the generation of seed magnetic fields 
(for a review see eg. Enqvist (1998)).
In this particular context then, the noise simulates in part the effects
of the particle-physics degrees of 
freedom at the micro-scale. 
Although we will not examine models to motivate the
noise forces, we emphasize that the noise need {\it not} be
a stirring force (in the strict sense of turbulence modelling) but also 
can account, in part, for real physical fluctuations 
and perturbations intrinsic
to the system.

In Section \ref{sect3} we turn to the renormalization group analysis
of incompressible MHD.  {}First the physical motivations are discussed
for implementing the RG techniques to
MHD.  Then Subsec. \ref{sect3A} examines the naive scaling
properties of stochastic MHD under independent space and time
rescalings. Here, two fundamental scaling
exponents are introduced, that encode all the 
information pertaining to the asymptotic
behavior of the solutions of MHD.  Subsec. \ref{sect3B}
describes the two step RG procedure of coarse graining
and rescaling.  It then explains the relation between
scale invariant solutions of the MHD equations
and fixed points of the RG analysis.  In particular,
at a fixed point the MHD equations
become scale invariant, and a unique solution exists for the pair of
scaling exponents,
which are calculable by the RG procedure. 
To fix ideas, a simple
example of the RG procedure is given in the context
of linearized MHD.
Of course this example does not
account for the {\it nonlinear} interactions, so crucial for establishing and
maintaining turbulence.
In Subsec \ref{sect3C} the RG analysis is done for the full
MHD equations, in particular including the nonlinear
terms.  As alluded to above,
most of the applications of RG to hydrodynamics and MHD require
one to iterate the equations of motion to a particular order in the
nonlinear coupling.  This can result in rather cumbersome
expressions after a few iterations. 
Here, we prefer to exploit the streamlined functional
integral and allied diagrammatic methods which 
allow for the rapid identification of the
one-particle irreducible (1PI) diagrams whose renormalization then yields
directly the independent RG equations for the parameters appearing in the MHD
equations. Since these techniques lie somewhat out of the mainstream of the
paper and can be applied more generally to the problem,
we relegate them
to a series of technical Appendices, 
where the complete renormalization of the
response function, noise spectral function and interaction vertex is carried
out in full detail.  In Subsec \ref{sect3C}
mainly the differential RG flow equations are presented,
and the calculations leading to these equations
are outlined with references to the appropriate
places in the Appendices for the various details.

In Sec. \ref{sect4}, the RG flow equations are analyzed with a  pair of
dimensionless couplings.  Then in this section,
the fixed points are calculated, their stability
properties are elucidated and the associated critical exponents are 
obtained for each fixed point.
All these results are given for any spatial dimension
$d$ and for a general class of stochastic noise forces,
characterized by a single spectral force exponent $y$,
that ranges from short ($y<0$) to long ($y>0$) range.
In Section \ref{sect5}
we derive and solve differential equations for the scale dependent viscosity
and resistivity, derive the scaling forms obeyed by the velocity and magnetic
field correlation functions, and derive both the 
scaling form and RG improved form of the kinetic 
and magnetic energy spectra.  Here,
an expression is said to be {\it renormalization group improved} if the
bare parameters in that expression are replaced by their scale-dependent
forms, typically calculated to some order in perturbation theory. Thus, an
improved quantity, whether it be a correlation function, 
an energy spectrum, etc.,
is one which combines perturbation theory with the renormalization 
group (Hochberg et al. 1999a).  In Sec. \ref{sect6}, we turn to consider
the injection of random magnetic helicity, which is modeled
with a power law spectrum,
and compute the RG-improved helicity
spectrum. In this Section the correlation functions are 
recomputed, now taking
into account the helicity contribution. Then once again,
these correlation functions are computed in the RG improved form. 
An alternative way to understand
scaling and the approach to scaling 
is given in Section \ref{sect7}, which makes
use of the Callan-Symanzik equation whose solution yields the
improved correlation function automatically.
Sec. \ref{sect8} is concerned  with energy and helicity cascades
in hydrodynamics and MHD. A methodology for extracting
information about cascades from our RG calculations 
is developed and then implemented.  The predictions from
our RG analysis are given for cascades and
cascade directions at the asymptotically largest length scales.
{}Finally, in Section \ref{concl} 
we draw conclusions and summarize our work.
Here some discussion is provided, extending from
Subsect \ref{sect5C}, on the mixing of magnetic and
kinetic energy at asymptotic scales.

As mentioned above, the Appendices contain most of the
technical calculations of our RG analysis
as well as a derivation of the
functional path integral formalism that
is generally applicable to stochastic MHD.
In particular, the Appendices are organized as follows.
In Appendix A we set up the dynamic functional
for incompressible  non-relativistic stochastic MHD and use
this to derive the basic elements needed for a Feynman diagram
representation of the solutions of the MHD equations and define
the response function, noise spectral function and vertex function. 
In Appendix B we carry out the momentum-shell integration needed to
renormalize the response function. {}From this we obtain the 
one-loop corrections to the kinematic and magnetic viscosities. 
The one-loop renormalization of the noise spectral
function is carried out in Appendix C. {}For $y>-2$,
which corresponds to noise spectra from long range
down to reasonably short range, we find that there
is in fact no one-loop 
correction to the noise amplitude in the hydrodynamic
limit. 
The 
one-loop renormalization of the vertex function vanishes
identically for all gaussian forcing functions that are
temporally white, but for arbitrary spatial correlations, 
as demonstrated in Appendix D. Some useful
identities needed for averaging products of unit vectors over
the unit $d$-sphere are collected in Appendix E.

\section{Stochastic MHD Equations}
\label{sect2}

The equations of motion for viscous and resistive, incompressible, 
non-relativistic, magnetohydrodynamics are given by (Biskamp 1993)
\begin{equation}\label{NRfluid}
\frac{\partial {\bf v}}{\partial t} + ({\bf v \cdot}\vec \nabla)
{\bf v} = -\frac{1}{\rho}\vec \nabla \big( p + \frac{B^2}{8\pi} \big) +
\frac{1}{4\pi \rho}({\bf B \cdot}\vec \nabla){\bf B} +
\nu \nabla^2 {\bf v} + \vec \eta_v,
\end{equation} 
and
\begin{equation}\label{mag}
\frac{\partial {\bf B}}{\partial t} =
({\bf B \cdot} \vec \nabla){\bf v} - 
({\bf v \cdot} \vec \nabla){\bf B} + 
\frac{c}{4\pi \sigma}\nabla^2{\bf B} + \vec \eta_B,
\end{equation}
subject to the constraints
\begin{equation}
\vec \nabla \cdot {\bf v} = 0,
\end{equation}
and
\begin{equation}
\vec \nabla \cdot {\bf B} = 0,
\end{equation}
where ${\bf v}$ denotes the fluid velocity, ${\bf B}$ the magnetic field, $p$
the fluid pressure and $\rho$ the density. 
The speed of light is $c$ and $\sigma$ is the 
electrical conductivity of the magnetofluid.
We allow for the presence of 
random forces and random currents, denoted by $\vec \eta_v(\vec x,t)$ and 
and $\vec \eta_B(\vec x,t)$, respectively. 
We assume these forces are Gaussian with zero mean, 
$\langle \vec \eta_v \rangle =  \langle \vec \eta_B \rangle = 0$.
The random vector forces and currents 
are specified by their two-point correlation functions, which for
convenience we specify directly in Fourier 
space and in terms
of their individual vector components as
\begin{eqnarray}\label{vnoise}
\langle (\vec \eta_v)_n(\vec k,\omega)
(\vec \eta_v)_m(\vec k',\omega')\rangle &=& 2D_v \,g(k) (2\pi)^{d+1}
{\bf P}_{nm}(\vec k)\delta(\omega + \omega')
\delta^d(\vec k + \vec k'),\\
\label{Mnoise}
\langle (\vec \eta_B)_n(\vec k,\omega)
(\vec \eta_B)_m(\vec k',\omega')\rangle &=& 2D_B \, g(k) (2\pi)^{d+1}
{\bf P}_{nm}(\vec k)\delta(\omega + \omega')
\delta^d(\vec k + \vec k'),\\
\label{vBnoise}
\langle (\vec \eta_v)_n(\vec k,\omega)
(\vec \eta_B)_m(\vec k',\omega')\rangle &=& 0.
\end{eqnarray}
Here, $d$ is the number of space dimensions,
the indices are $n,m = 1,2, \cdots, d$, and  
the angular brackets denote the average taken over the fluctuations. 
The presence of both kinetic and magnetic helicities can be taken into account
and modeled by means of suitably defined noise terms. We defer the treatment
of random helicity to a later section.
The transverse projection operator ${\bf P}_{nm}(\vec k)$ 
is needed to ensure the
noise is solenoidal, 
i.e., $\vec k \cdot \vec \eta_v = \vec k \cdot \vec \eta_B = 0$, 
and hence compatable with 
incompressibility  $(\vec \nabla \cdot {\bf v} = 0)$ 
and the 
absence of magnetic monopoles $(\vec \nabla \cdot {\bf B} = 0)$. 
The Fourier transform of the random forces is given by
\begin{equation}\label{Fourier}
\vec \eta_{v,B}(\vec x,t) = 
\int\frac{d^d k}{(2\pi)^d} \int \frac{d\omega}{2\pi}\,
\vec \eta_{v,B}(\vec k,\omega) \, \exp(i\vec k \cdot \vec x - i\omega t),
\end{equation}
and this is the convention used throughout this paper.
The magnitudes of the random forces and currents are given by $D_v \geq 0$ 
and $D_B \geq 0$,
respectively, and so provide a measure of the size of the fluctuations.  
The power law function
$g(k) = k^{-y}$ with exponent
$y$ is intended to describe the noise spectrum in the inertial range. 
Eq. (\ref{vBnoise}) states that the random currents 
and forces are uncorrelated.
The noise exponent $y$ will be treated as a model parameter.
Note depending on the sign of $y$, the noise fluctuations are either more
correlated at short distances $(y < 0)$ or at large distances $(y > 0)$.

We now cast the above coupled vector stochastic 
partial differential equations into a 
manifestly {\it symmetric} form first by
defining ${\bf B} \rightarrow {\bf B}/\sqrt{4\pi \rho}$, $p \rightarrow p/\rho$
and $p_* = (p + \frac{1}{2}B^2)$.  
Then, following Elsasser (1950), we 
add and subtract the two equations of motion
Eqs. (\ref{NRfluid}) and (\ref{mag}) to obtain
\begin{eqnarray}\label{Elsasser1}
\frac{\partial \vec P}{\partial t} + 
(\vec Q \cdot \vec \nabla) \vec P &=&
-\vec \nabla p_* + \gamma_{+} \nabla^2 \vec P
+ \gamma_{-} \nabla^2 \vec Q + \vec \eta_P, \\
\label{Elsasser2}
\frac{\partial \vec Q}{\partial t} 
+ (\vec P \cdot \vec \nabla) \vec Q &=&
-\vec \nabla p_* + \gamma_{+}\nabla^2 \vec Q
+ \gamma_{-}\nabla^2 \vec P + \vec \eta_Q,
\end{eqnarray}
where 
\begin{eqnarray}\label{Elbasis}
\vec P = \vec v + \vec B, &\qquad& \vec Q = \vec v - \vec B, \\
\label{gammanu}
\gamma_{+} = \frac{1}{2}(\nu + \nu_B) , &\qquad& \gamma_{-} = \frac{1}{2}(\nu - \nu_B), \\
\label{PQnoise1}
\vec \eta_P = \vec \eta_v + \vec \eta_B, &\qquad& \vec \eta_Q = \vec \eta_v - \vec \eta_B.
\end{eqnarray}
In terms of these Elsasser variables, the noise statistics 
is completely determined from
Eqs. (\ref{vnoise}) - (\ref{vBnoise}) together 
with the definitions Eq. (\ref{PQnoise1}) as
\begin{eqnarray}\label{PQnoise2} 
\langle (\vec \eta_P)_n(\vec k,\omega)
(\vec \eta_P)_m(\vec k',\omega')\rangle &=& 2[D_v + D_B] \,g(k) (2\pi)^{d+1}
{\bf P}_{nm}(\vec k)\delta(\omega + \omega')
\delta^d(\vec k + \vec k'),\nonumber \\
{ }&=& \langle (\vec \eta_Q)_n(\vec k,\omega)
(\vec \eta_Q)_m(\vec k',\omega')\rangle, \\
\label{PQnoise4}
\langle (\vec \eta_P)_n(\vec k,\omega)
(\vec \eta_Q)_m(\vec k',\omega')\rangle &=& 2[D_v - D_B] \,g(k) (2\pi)^{d+1}
{\bf P}_{nm}(\vec k)\delta(\omega + \omega')
\delta^d(\vec k + \vec k').
\end{eqnarray}
We henceforth write $\nu_B = \frac{c}{4\pi \sigma}$, 
to indicate the magnetic resistivity.
This pair of dynamical equations 
Eqs. (\ref{Elsasser1}) and (\ref{Elsasser2})
is symmetric under the interchange 
$\vec P \leftrightarrow \vec Q$ and $\vec \eta_P \leftrightarrow \eta_Q$.
{}Furthermore, in the limit of vanishing magnetic field, $\vec P = \vec Q$
and vanishing magnetic noise, $\vec \eta_P = \vec \eta_Q$, the pair
reduces to two identical
copies of the Navier Stokes equation with a random noise source. These
symmetries and limits are useful in checking results derived from these
equations and can be used
as a diagnostic in comparing our calculations with other RG results obtained
from MHD and from incompressible hydrodynamics.

{}For incompressible MHD, we can take the 
divergence of both equations 
Eqs. (\ref{Elsasser1}) and (\ref{Elsasser2}) to 
eliminate the pressure term $\nabla p_*$ since $\vec \nabla \cdot {\bf v} 
= \vec \nabla \cdot {\bf B} = 0$. This yields 
\begin{equation}\label{EOS}
p_* = -\frac{1}{\nabla^2}\partial_n \partial_k (Q_k P_n),
\end{equation}
which can be regarded as the equation of state for this system. We note that
in solving for $p_*$, there are no boundary terms that contribute to
Eq. (\ref{EOS}).
As in Eqs. (\ref{NRfluid}) and (\ref{mag}), the 
stochastic random forces and currents
can be assumed to be
solenoidal, since any longitudinal components can be absorbed into the pressure
$p_*$.
This simplifies the equations 
even further.  To do this,  we introduce 
the transverse projection operator, which is the same one appearing above in the
noise correlation functions 
\begin{equation}\label{projector}
{\bf P}_{jn} = \Big( \delta_{jn} - \partial_j \frac{1}{\nabla^2} \partial_n \Big),
\end{equation}
for $j,n = 1,2, \cdots, d$.  Of course, this projector is a nonlocal 
operator in coordinate space (as it involves the
Green function $\nabla^{-2}$), but
its Fourier transform is local and easier to handle in actual computations.  
Transforming Eq. (\ref{projector}) using Eq. (\ref{Fourier}), gives
\begin{equation}\label{Fprojector}
{\bf P}_{jn}(\vec k) = \Big( \delta_{jn} - \frac{k_j \,k_n}{k^2} \Big).
\end{equation}
Now $p_*$ can be eliminated from 
Eqs. (\ref{Elsasser1}) and (\ref{Elsasser2}) to give,
in coordinate space,
\begin{eqnarray}\label{Elsasser3}
\frac{\partial P_j}{\partial t} + \lambda_0 {\bf P}_{jn}\partial_l
(Q_l P_n) - \gamma_+ \nabla^2 P_j - \gamma_{-}\nabla^2 Q_j = (\eta_{P})_j ,
\\
\label{Elsasser4}
\frac{\partial Q_j}{\partial t}  + \lambda_0 {\bf P}_{jn}\partial_l
(P_l Q_n) - \gamma_{+}\nabla^2 Q_j - \gamma_{-}\nabla^2 P_j = (\eta_{Q})_j.
\end{eqnarray}
This is the final form of the stochastic MHD equations that serves as the 
starting point for
our renormalization group analysis. The presence of the projection operator
makes it preferable to express these 
equations in component form, as we have done here. 
We introduce a parameter $\lambda_0 = 1$ 
in front of the 
nonlinear terms, which is useful for organizing perturbative
expansions of the solutions of these equations. This is merely a
bookkeeping device, since our true (and dimensionless) perturbation
expansion parameter will be defined and calculated below.
{}Finally, note that
correlations (and other functions) in the fluid velocity and magnetic 
fields can be computed
in terms of these Elsasser fields, since the two sets
of variables  
are linearly related as
\begin{eqnarray}\label{physicalv}
{\vec v}(\vec x,t) &=& \frac{1}{2}\left[ {\vec P}(\vec x,t) +
{\vec Q}(\vec x,t) \right], \\
\label{physicalB}
{\vec B}(\vec x,t) &=& \frac{1}{2}\left[ {\vec P}(\vec x,t) -
{\vec Q}(\vec x,t) \right].
\end{eqnarray} 

\section{Renormalization Group (RG) Analysis of MHD}
\label{sect3}

The renormalization group (RG) is a systematic method for studying a system
at different length scales. One of the key motivations for
implementing this method, and the main application in this
paper, is to verify a scaling hypothesis.  In vague terms
a scaling hypothesis is related to the statement that 
in the asymptotic regime, the correlations in a system all 
can be expressed in terms of a single suitable length scale.
{}For the renormalization group method, vindication of a scaling 
hypothesis is related to showing that the underlying dynamical
equations are self-similar at different scales, up to an overall
rescaling of the space-time coordinates and the dynamical
variables of the system.  Sometimes self-similar
solutions are referred to as fixed points, since the parameters
of the system do not change under the RG transformation.
{}For hydrodynamics and
magnetohydrodynamics, the scaling hypothesis that has
been considered 
(Forster et al. 1977; Yakhot and Orszag 1986; Dannevik et al.
1987; Fournier et al. 1982; Longcope and Sudan 1991;
Camargo and Tasso 1992)
and will be considered here, is
associated with the asymptotic regime of large length and
time separations in the correlation functions.

In this Section, the above statements will be explained in detail
and then applied.  Subsection \ref{sect3A} first will explain the scale
transformations of the MHD equations
that are important for the renormalization group
procedure.  Subsection \ref{sect3B} then explains the renormalization group
in the context of MHD.  {}Finally, Subsection \ref{sect3C} presents
the results of applying the renormalization group to MHD.
This last subsection requires the technically
most involved aspect of the calculation in this paper.
{}For ease of presentation, all the technical steps have
been given in the Appendices and Subsec. \ref{sect3C} simply
quotes the results and gives some guidance to
the Appendices.

\subsection{Scale Transformation}
\label{sect3A}

In preparation for a dynamical renormalization group analysis, and as an
important and necessary preliminary step,  
the stochastic MHD equations Eqs. (\ref{Elsasser3}) and (\ref{Elsasser4}) 
are submitted to
independent global rescalings of 
space and time as
\begin{eqnarray}\label{scaling}
{\vec x} &=& s{\vec x}', 
\qquad \Rightarrow \qquad \vec \nabla = \frac{1}{s}
\vec \nabla',
\nonumber \\
{t}  &=&  s^z\, {t'} \qquad \Rightarrow 
\qquad \frac{\partial}{\partial t} = s^{-z}\frac{\partial}{\partial t'},
\end{eqnarray}
and the (Elsasser) fields as
\begin{equation}\label{fieldscaling}
{\vec P}({\vec x}, t) = s^{\chi}{\vec P'}({\vec x}',t'), \qquad
{\vec Q}({\vec x}, t) =  s^{\chi}{\vec Q'}({\vec x}',t'),
\end{equation}
where $s>1$.  To fix ideas, one can think of this rescaling as
examining the original unprimed system in a new primed
coordinate system, with the two related by the
above rescalings.
  
The above transformation involves two exponents,
the dynamic exponent $z$ and the
``roughness'' exponent $\chi$.
The choice of the above scale transformation is made
because in the renormalization group analysis, the number of
independent space and time rescalings 
one must consider depends on the degree of symmetry and anisotropy
of the system under study. 
As we are considering homogeneous and isotropic 
magnetohydrodynamic turbulence, this
scaling introduces two a-priori independent exponents. 
Alternatively, for example in reduced 
(2+1)-dimensional MHD where one has a dominant background magnetic field, say
along the $z$-direction, 
one scales the 
$(x,y)$-plane and the perpendicular 
$z$ directions independently (Longcope and Sudan 1991)
and this introduces 
{\it three} independent scaling exponents: $z,\chi,\xi$.

An important step in examining self-similar behavior in the
renormalization group approach is to examine the original
unprimed equations in the primed coordinate system
and compare the form of the equations.  The precise
relevance of this step will be explained in the next
subsection.   However, at a procedural level,
in performing this step, one term in both set of
equations always can be made similar by dividing through
by an appropriate factor.  {}For us, we will
choose this term to be the time derivative terms
$\partial P/\partial t$, $\partial Q/\partial t$.
Thus Eqs. (\ref{Elsasser3}) and (\ref{Elsasser4}) when
expressed in the primed coordinates are
\begin{eqnarray}\label{scaled3}
\partial_{t'} P'_j({\vec x}',t') 
+ s^{z + \chi -1}\lambda_0\, {\bf P'}_{jn}\partial'_l
(Q'_l P'_n) &=& s^{z-2} \,\gamma_{+}\nabla^{'2} P'_j({\vec x}',t') - 
s^{z-2}\, \gamma_{-}\nabla^{'2} Q'_j({\vec x}',t') \nonumber \\
&+& s^{z- \chi + \frac{1}{2}(y -d -z)}\,\eta_{P_j}({\vec x}',t'),
\end{eqnarray}
and similarly for the $\vec Q$-equation. 
There are three points about Eq. (\ref{scaled3}) to note.
{}First, in arriving at Eq. (\ref{scaled3}) a factor of $s^{-z+\chi}$
was divided through the equation so that the time derivative terms
in this and
Eqs. (\ref{Elsasser3}) and (\ref{Elsasser4}) are the same.
Second, the transformation behavior of the noise functions
is fully specified by the space and time rescalings
in Eq. (\ref{scaling}) and the form 
of the noise correlations in 
Eqs. (\ref{vnoise}) - (\ref{vBnoise}).
Third, the MHD equations generically have 
an implicit short distance cut-off
$\Delta x > \Delta x_0$ or in momentum space 
$|{\bf k}| < \Lambda \sim 1/\Delta x_0$. {}For distances less
than this cut-off the hydrodynamic approximation is
invalid in properly representing the physics.  
The continuum description of the fluid breaks down.
The scale transformations
above also act on this short distance cut-off so that
in the primed frame this cut-off becomes
$\Delta x'_0 = s \Delta x_0$ or,
equivalently $\Lambda' = \Lambda/s$. 
Since $s>1$, we see that the cut-off (or lattice spacing)
has been scaled up (in real space) towards the infrared. 

We compare this scaled equation of motion Eq. (\ref{scaled3})
with its unscaled version Eq. (\ref{Elsasser3}). If the scaled
equation Eq. (\ref{scaled3}) is to describe the same dynamics, albeit 
at the larger scale $(s > 1)$, then this
implies that the coefficients appearing in it must change accordingly. 
By ``same dynamics'' we mean of course that the mathematical {\it form} of the
dynamical equation does not change as one changes the scale of the problem.
This implies the following scaling behavior of the
parameters appearing in the MHD equations:
\begin{eqnarray}\label{list}
\lambda_0' &=& s^{z + \chi -1} \lambda_0, \nonumber \\
\gamma'_{+} &=& s^{z-2}\, \gamma_{+},\nonumber \\
\gamma'_{-} &=& s^{z-2}\, \gamma_{-}, \nonumber \\
A' &=& s^{z - 2\chi - d + y} A, \nonumber \\
B' &=& s^{z - 2\chi - d + y} B.
\end{eqnarray}
We have defined 
the noise-amplitude combinations $A = D_v + D_B$ and $B = D_v - D_B$ (Thus,
the zero magnetic noise limit $D_B = 0$ is had by taking $A=B$). 

It is important to appreciate that Eq. (\ref{scaled3}) with the
definitions Eq. (\ref{list}) and 
Eqs. (\ref{Elsasser3}) and (\ref{Elsasser4}) are simply two
expressions of the same equation.  In particular no
requirement of scale invariance is made.  We simply have
required that
any changes they suffer due to 
this scaling transformation be absorbable into {\it redefinitions} of the 
(bare) parameters
appearing in the equations. 

There is one final notational comment in regards to the
scale transformation.  Due to the central
role that this transformation plays in the
renormalization group analysis, it is standard to assign
all quantities in the dynamical equation a scaling dimension.
Typically the momentum coordinates,
which scale as $s^{-1}$, is referred to as having scaling
dimension 1, so that length has scaling dimension $-1$.
Similarly time, the fields, and all the parameters in 
Eq. (\ref{list}) have scaling dimension given by minus the exponent to
which $s$ is raised.
\subsection{Renormalization Group Procedure}
\label{sect3B}

The RG transformation is now rather standard fare and is discussed
in a number of excellent texts and monographs 
(Ma 1976; Amit 1978; Zinn-Justin 1996; Cardy 1996).
Here we briefly review
the RG procedure.
The RG is a symmetry transformation that carries a system
from one length scale to another through a two step
procedure, coarse graining and rescaling.  As with any
symmetry transformation, such as rotation, translation etc.,
it is not necessary that the system under examination be
invariant to this transformation.  However, should this be the
case, then is as also the case with any symmetry transformation, important
information about the system can be deduced without detailed
knowledge of its entire dynamics.  In particular, invariance under
the RG transformation will imply scale invariance in the system.
The existence of such an invariance implies
the behavior of correlation functions at 
different scales can be understood 
relative to their behavior at one given scale.

Let us review the two step RG procedure.  Consider the exact
MHD Eqs. (\ref{Elsasser1}) and (\ref{Elsasser2})
or equivalently Eqs. (\ref{Elsasser3}) and (\ref{Elsasser4}).
As stated earlier, there is an implicit assumption
that these equations are only valid down to some distance scale
$x > \Delta x_0$ or in momentum space $k < \Lambda \sim 1/\Delta x_0$.
This is the smallest length scale at which the MHD dynamics,
Eqs. (\ref{Elsasser1}) and (\ref{Elsasser2}),  
are meant to be valid.  Given this starting point, the RG is implemented
in the following two steps:

\medskip

\noindent
{\bf Coarse-graining}:  The first step is to solve the dynamics
at the very shortest length scales.
In Fourier space we integrate the solutions 
of the stochastic equations of motion over a thin
momentum shell 
$\Lambda/s \leq |\vec k| < \Lambda$ where $s = 1 + \delta$ and
$0 < \delta << 1$. Physically, this step serves to integrate out the fast
or short wavelength modes 
and thins out the degrees of freedom. This elimination process leads to
``running'' in the parameters 
(which depend on $s= e^{\ell}$) 
appearing in the MHD equations. The short distance cut-off or lattice spacing
gives rise to a corresponding momentum cut-off $\Lambda/s$.
The resulting MHD equations after this step will be given by
\normalsize
\begin{eqnarray}\label{cge}
\partial_{t}P^<_j &+&
\lambda^<_0(\ell) {\bf P}_{jn}\partial_l
(Q^<_l P^<_n) - \gamma^<_+(\ell) \nabla^2 P^<_j \nonumber \\
& - & \gamma^<_{-}(\ell)\nabla^2 Q^<_j - (\eta^<_{P})_j 
+ {\rm less \hspace{0.1cm} relevant} = 0,
\end{eqnarray}
and likewise for $Q_j$.
Clearly in performing this step additional terms will appear
in the dynamic equations.  These terms have not been 
explicitly written above, but have been referred to
suggestively as "less relevant", since we will see later
that for the leading asymptotic regime they will not be important.
Nevertheless, at this stage if we retain all the terms in the
coarse grained equations Eq. (\ref{cge}), including the
"less relevant" terms, the dynamics from these equations is exactly
the same as from the original MHD
Eqs. (\ref{Elsasser1}) and (\ref{Elsasser2}),  
for computing correlations
for length scales larger than $\Delta x > 2\pi s/\Lambda$.
In particular,
correlations computed from
Eqs. (\ref{Elsasser1}) and (\ref{Elsasser2}) or 
Eqs. (\ref{cge}) will be the same,
i.e.
\begin{equation}
\langle P_i({\vec x}_1,t_1) P_j({\vec x}_2, t_2) \rangle
= \langle P_i^<({\vec x}_1,t_1) P_j^<({\vec x}_2, t_2) \rangle,
\hspace{1cm} {\rm etc.}
\end{equation}

\medskip

\noindent
{\bf Rescaling}:
The second step is to represent the above coarse grained Eqs. (\ref{cge})
in the primed coordinates based on the rescalings 
in Eqs. (\ref{scaling}) and (\ref{fieldscaling})
with $s=e^{\ell}$.  
In the primed coordinates, the MHD equations become
\normalsize
\begin{eqnarray}\label{rse}
\partial_{t'}P'_j &+&
\lambda'_0(\ell) {\bf P}_{jn}\partial'_l
(Q'_l P'_n) - \gamma'_+(\ell) \nabla'^2 P'_j \nonumber \\ &-& 
\gamma'_{-}(\ell)\nabla'^2 Q'_j - s^{(y-z-d)/2}(\eta_{P})_j(x',t')
+ {\rm less \hspace{0.1cm} relevant \hspace{0.1cm} rescaled}
=0,
\end{eqnarray}
and likewise for $Q$. 
In these equations
\begin{eqnarray}\label{list2}
\lambda_0'(\ell) &=& s^{z + \chi -1} \lambda_0^<, \nonumber \\
\gamma'_{+}(\ell) &=& s^{z-2} \gamma_{+}^<,\nonumber \\
\gamma'_{-}(\ell) &=& s^{z-2} \gamma_{-}^<, \nonumber \\
A'(\ell) &=& s^{z - 2\chi - d + y} A^<, \nonumber \\
B'(\ell) &=& s^{z - 2\chi - d + y} B^<.
\end{eqnarray}
Note, if we solve Eqs. (\ref{rse})
and examine correlations at a distance $d=(x_1'-x_2')$,
this is equivalent to solving for the same
correlation in the original 
Eqs. (\ref{Elsasser1}) and (\ref{Elsasser2}),  
at distance
$e^{\ell}d=(x_1-x_2)$.

At this point we can consider Eqs. (\ref{rse}) in place of the
original MHD 
Eqs. (\ref{Elsasser1}) and (\ref{Elsasser2}),  
and repeat these two steps iteratively.
A very interesting thing can happen upon sufficient iteration,
which is a key practical importance of the RG approach.
The set of parameters in iterative versions of Eqs. (\ref{rse})
could be the same after a certain number of successive RG
transformations. 
These points in the parameter space are called {\it fixed points}.
At the fixed point,
correlations computed from either of two successive
equations clearly are identically
the same, since the successive equations
are exactly the same. This fact has nontrivial consequences, since 
due to the scaling relations between the two
sets of equations, it also means we can learn about the
behavior of correlations at different scales of the
original MHD equations.
In particular consider the field 
$P({\vec x},t)$, $P'({\vec x}',t')$ 
(likewise $Q$) from
two successive iterations of Eqs. (\ref{rse}).
If these two successive versions of Eqs. (\ref{rse}) are
exactly the same then it is an identity that (dropping the 
indices)
\begin{equation}
\langle P({\vec x}_1,t_1) P({\vec x}_2, t_2) \rangle
= \langle P({\vec x}'_1,t'_1) P({\vec x}'_2, t'_2) \rangle ,
\end{equation}
and likewise for $Q$.
However,  based on the rescaling 
Eqs. (\ref{scaling}) and (\ref{fieldscaling}) we also have
\begin{equation}
\langle P(x_1,t_1) P(x_2, t_2) \rangle
= e^{2\chi \ell} \langle P(e^{-\ell}x_1,e^{-z\ell}t_1) 
P(e^{-\ell}x_2, e^{-z \ell}t_2) \rangle,
\end{equation}
which therefore gives a relation between correlations
at two different length/time scales.

To help focus these ideas, let us consider a simple example of a fixed
point solution in RG for the case of MHD without the nonlinear
terms.  In this case, the MHD dynamics from
Eqs. (\ref{Elsasser1}) and (\ref{Elsasser2}) are 
\begin{eqnarray}\label{f1}
\partial_{t}P_j
- \gamma_+ \nabla^2 P_j 
- \gamma_{-}\nabla^2 Q_j - (\eta_{P})_j =0,
\end{eqnarray}
\begin{eqnarray}\label{f2}
\partial_{t}Q_j
- \gamma_+ \nabla^2 Q_j 
- \gamma_{-}\nabla^2 P_j - (\eta_{Q})_j =0.
\end{eqnarray}

Implementing the RG procedure, the first step is coarse graining.
Since this is a free (and hence, linear) theory, this step is trivial.
We simply remove the highest modes between
$\Lambda/s < |k| < \Lambda$ from the $P$ and $Q$ fields.
The second step then is rescaling which
based on Eqs. (\ref{scaling}) leads to  
\begin{eqnarray}\label{f1rs}
\partial_{t'}P'_j (x',t')
- s^{z-2} \gamma_+ \nabla'^2 P'_j 
- s^{z-2} \gamma_{-}\nabla'^2 Q'_j - 
s^{z-\chi+\frac{1}{2}(y-d-z)} \eta_{P_j}(x',t') =0,
\end{eqnarray}
and similarily for $Q$.
Observe if
\begin{eqnarray}
\label{fp}
z&=& 2 \nonumber\\
\chi &=& \frac{y-d+z}{2},
\end{eqnarray}
then both Eqs. (\ref{f1}) and (\ref{f1rs}) are the same, and so
this is a fixed point of the RG.

Using this fact to compute correlations at two different scales
in the momentum space correlation functions yields
\begin{eqnarray}
\label{rgeq}
\langle P(k_1,\omega_1) P(k_2,\omega_2) \rangle
&=& e^{2(\chi+d+z)\ell}  \langle P'(e^{\ell} k_1,e^{z \ell}\omega_1)
P'(e^{\ell} k_2,e^{z\ell}\omega_2) \rangle \nonumber\\
&=& e^{2(\chi+d+z)\ell} \langle P(e^{\ell} k_1,e^{z\ell}\omega_1)
P(e^{\ell} k_2,e^{z\ell}\omega_2) \rangle,
\end{eqnarray}
and similarily for $Q$.  Since this is a free theory, we can check
this result of the RG procedure from the exact correlation functions.
{}For this note that the above fixed point Eq. (\ref{fp})
is the same for both linearized MHD Eqs. (\ref{f1}) and (\ref{f2})
or linearized hydrodynamics, which arises from these equations
by identifying $P=Q$.  Also, the exact solutions of both 
linearized MHD and linearized hydrodynamics have the same
basic form, but the latter are much less elaborate.
Thus, it is sufficient for our demonstrative purpose here to quote
the simpler case of linearized hydrodynamics, for which
the exact two-point correlation function is
\begin{equation}
\label{corr}
\langle P(k_1,\omega_1) P(k_2,\omega_2) \rangle
\sim const \times \frac{k_1^{-y} \delta^d({\vec k}_1+{\vec k}_2) 
\delta(\omega_1+\omega_2)}
{\omega_1^2 + (\nu k_1^2)^2},
\end{equation}
where $\nu = \gamma_++\gamma_-$ from Eq. (\ref{gammanu}).
The scaling transformations Eqs. (\ref{scaling}) and (\ref{fieldscaling})) 
applied to this imply
\normalsize
\begin{eqnarray}
\label{corr2}
e^{2(\chi+d+z)\ell} \langle P(e^{\ell} k_1,e^{z\ell}\omega_1)
P(e^{\ell} k_2,e^{z\ell}\omega_2) \rangle
&\sim& e^{2(\chi+d+z-4-y-d-z) }const \times \nonumber \\ 
& & \frac{k_1^{-y} 
\delta^d({\vec k}_1+{\vec k}_2) \delta(\omega_1+\omega_2)}
{\omega_1^2 + (\nu k_1^2)^2}.
\end{eqnarray}
Comparing the scaling behavior from the RG analysis
Eq. (\ref{rgeq}) and the exact
solution, Eq (\ref{corr2}), we see they agree.

In summary, returning to the opening statements of this section, finding
fixed points in the RG procedure corresponds to identifying 
scale invariance in the MHD equations with respect to the
scale transformations Eqs. (\ref{scaling}) and (\ref{fieldscaling}).
As one practical matter, since the iterative 
procedure in Eqs. (\ref{cge}) and (\ref{rse})
is focussed on the behavior of the parameters, it is
standard and convenient to express the RG
procedure through a set of differential equations
for the evolution of the parameters, 
eg. $d\lambda(\ell)/d{\ell}, d\gamma_+(\ell)/d{\ell}$, etc.
The fixed points then correspond to the solutions
where all parameters have zero evolution,
that is,  $d\lambda(\ell)/d{\ell} = d\gamma_+(\ell)/d{\ell} = \cdots = 0$.

\subsection{Differential Renormalization Group Equations for MHD}
\label{sect3C}

This subsection presents the RG evolution equations for MHD.
There are several steps to this calculation, all of which
have been carried out in full detail in the Appendices.  Here these steps
only will be outlined with reference to the appropriate Appendices
for the complete details, and then the results will be given.
For solving the equations of stochastic MHD, in Appendix A
they have been Fourier transformed and
cast into a dynamic functional formalism.  This formalism allows a 
systematic loop-expansion, which in other words is a perturbation 
expansion in the amplitude of the random fluctuations.
The basic building blocks for constructing the perturbative expansion
are expressed diagrammatically through the Feynman rules in Fig. 1.
Based on this approach,
the information contained in the 
explicit solutions of $\vec P, \vec Q$ is equivalently contained,
but much more compactly,
in the response function, noise spectrum and 
vertex function, whose definitions and diagrammatic representations
(expressed to one-loop order)
are given in Figs. (2a)-(2c). Thus, we work directly with these objects as they
will yield, to a given order in perturbation theory, the corrections to the
MHD parameters $\lambda_0, \gamma_{+}, \gamma_{-}, A$ and $B$. 
As stated in the last subsection, these
corrections result from moving to ever larger length scales while taking into
account the fluctuations from the smaller scales.  
We have calculated the one loop corrections to the response function,
the noise spectral function and the nonlinear vertex function in the 
Appendices B, C and D respectively, 
and from them, have obtained the corresponding one-loop
corrections to the parameters $\lambda_0, \gamma_{+},\gamma_{-},
A$ and $B$ in the hydrodynamic limit (i.e., the limit of large distance and
long time). We point out that in arriving at the RG flow equations below, we
have taken the limit $\omega \rightarrow 0$ at the outset of the calculations
as described in the Appendices. 
This is equivalent to taking the limit $t \rightarrow \infty$, and so
serves to ``project'' out the asymptotic time limit.  
{}Finally the second step of the RG procedure is
to rescale the coarse grained quantities based on 
Eqs. (\ref{scaling}) and (\ref{fieldscaling})

Applying these steps 
to the one-loop expressions Eqs.
(\ref{vertex4}), (\ref{Anoise}), (\ref{Bnoise}),
(\ref{elements}), and (\ref{elements2}),
computed in the Appendices, we
obtain the following set of one-loop renormalization group equations for
nonrelativistic incompressible MHD:

\begin{eqnarray}\label{RGE1}
\frac{d \lambda_0}{d \ell} &=& \big(z + \chi -1 \big) \lambda_0, \nonumber \\
\frac{d A}{d \ell} &=& \big(z - 2\chi - d + y \big) A, \nonumber \\
\frac{d B}{d \ell} &=& \big(z - 2\chi - d + y \big) B, \nonumber \\ 
\frac{d (\gamma_{+} + \gamma_{-})}{d \ell} &=& (\gamma_{+} + \gamma_{-}) \left\{
(z- 2) + \frac{{\cal S}}{2} \lambda_0^2 A_d \Lambda^{d-y-4} \frac{(A + B)}
{(\gamma_{+} + \gamma_{-})^3} \right\}, \nonumber \\
\frac{d (\gamma_{+} - \gamma_{-})}{d \ell} &=& (\gamma_{+} - \gamma_{-}) \left\{
(z- 2) + \frac{{\cal S}}{2} \lambda_0^2 A_d \Lambda^{d-y-4} \frac{(A - B)}
{(\gamma_{+} - \gamma_{-})^3} \right\},
\end{eqnarray}
where 
\begin{equation}\label{Ad}
A_d = \frac{(d^2 - y - 4)S_d}{d(d+2)(2\pi)^d},
\end{equation}
and $S_d = \frac{2\pi^{d/2}}{\Gamma(d/2)}.$
Here, ${\cal S} = 4$ is the
numerical symmetry factor associated to the diagram representing the
one-loop correction to the response function (see, Fig. (2a)), which has been
calculated in Appendix B. 
The short-distance or ultraviolet cut-off
(or, lattice spacing) is $\Lambda$, below which the hydrodynamic 
or continuum description of the fluid breaks
down. We can set $\Lambda = 1$ without loss of generality.
In fact, none of the RG results depend explicitly on this cut-off. 
In Section \ref{sect7}
we provide an alternative way to understand scaling which makes implicit use
of the short-distance divergences that result from formally taking the limit
$\Lambda \rightarrow \infty$ in the one-loop expressions.
It is clear that Eq. (\ref{RGE1}) 
describe the scale dependence of the parameters
of MHD. Their solution therefore yields the effective parameters
$\lambda_0(\ell),A(\ell),B(\ell),\gamma_{+}(\ell)$ and $\gamma_{-}(\ell)$.
Substitution of these scale-dependent parameters into the MHD equations
Eqs. (\ref{Elsasser3}) and (\ref{Elsasser4}) 
yields the effective dynamical equations
corresponding to this scale as parametrized by $\ell \geq 0$.

In terms of the {\it dimensionless} couplings defined by
\begin{equation}\label{dimensionless}
g_{\pm} = \frac{\lambda_0^2}{2} A_d \Lambda^{d-y-4} \frac{(A \pm B)}
{(\gamma_{+} \pm \gamma_{-})^3},
\end{equation}
the RG flow can be summarized by means of the two equations
\begin{eqnarray}\label{RGE2}
\frac{d g_{+}}{d \ell} &=& g_{+}\Big([4-d+y] - 3{\cal S} g_{+} \Big),\\
\label{RGE3}
\frac{d g_{-}}{d \ell} &=& g_{-}\Big([4-d+y] - 3{\cal S} g_{-} \Big),
\end{eqnarray}
which shows that flow takes place in a two-dimensional coupling or
parameter space with axes $(g_{+},g_{-})$. Here, ${\cal S} = 4$ is the
same numerical symmetry factor as mentioned above.

By shutting off the one-loop corrections (by putting $\lambda_0 = 0$)
in these equations Eq. (\ref{RGE1}), one immediately recovers the naive 
scaling in Eq. (\ref{list}). This demonstrates the importance of the nonlinear
interaction for scaling behavior. 
We see that the scale dependence of the viscosity
and resistivity is modified from its naive behavior. 
However, the noise amplitudes and the formal expansion parameter do not
get modified at one-loop; see Appendices C and D for
further details. 

\section{Fixed point solutions and exponents}
\label{sect4}

The solutions of the differential RG equations 
Eq. (\ref{RGE1}) (or equivalently Eqs. (\ref{RGE2}) and (\ref{RGE3})) 
tell us how the parameters
appearing in the MHD equations Eqs.
(\ref{Elsasser3}) and (\ref{Elsasser4}) change 
with length and time scale.
At a 
fixed point, as the name implies, the parameters no 
longer change under a further RG transformation, and one
is instead in the scaling regime, where the dynamics is scale invariant. To find
the renormalization group fixed points, we set the differential 
RG equations to zero, which yields 
algebraic equations. 

The RG equations Eqs. (\ref{RGE2}) and (\ref{RGE3}) 
have the two fixed point solutions:
\begin{eqnarray}\label{triv}
(i)\,\, (g^*_{+},g^*_{-}) &=& (0,0) \qquad {\rm and},\\ 
\label{nontriv}
(ii)\,\, (g^*_{+},g^*_{-}) &=& \Big(\frac{4-d+y}{12}, \frac{4-d+y}{12}\Big).
\end{eqnarray}
We wish to emphasize the fact that $g_{\pm}$ provides the bona-fide 
expansion parameter: it is dimensionless and is less than one whenever
$(4-d+y) < 12$. This latter condition is satisfied for all the noise
spectra and spatial dimensions considered in this paper. 
The stability properties of each fixed point is determined by performing
a linear analysis about each point. That is, we expand the couplings
$g_{\pm} = g^*_{\pm} + \delta g_{\pm}$ where $\delta g_{\pm}$ is a small
fluctuation, and then substitute this expansion into 
Eqs. (\ref{RGE2}) and (\ref{RGE3}) to obtain the
RG equations for the linearized perturbations
about the fixed points Eqs. (\ref{triv}) and (\ref{nontriv}).
This yields
\begin{eqnarray}
(i)\,\, \frac{\ d \delta g_{+}}{d \ell} &=& (4-d+y) \delta g_{+} + O(\delta g_{+}^2),\\
(ii)\,\,\frac{d \delta g_{+}}{d \ell} &=& -(4-d+y) \delta g_{+} + O(\delta g_{+}^2).
\end{eqnarray}
Identical equations
hold for $\delta g_{-}$ by simply replacing 
$\delta g_{+} \rightarrow \delta g_{-}$ in
these equations. 
These results tell us that when 
$(4-d+y) > 0$, the trivial fixed
point Eq. (\ref{triv}) is infrared unstable 
and repulsive (since the perturbation grows
for $\ell \rightarrow \infty$) while the non-trivial fixed point
Eq. (\ref{nontriv}) is infrared stable and attractive. 
What this means physically is the following. We imagine probing
the system at a given initial length and time scale. There is a
corresponding set of parameters $\lambda_0, A,B, \gamma_{+}, \gamma_{-}$ 
associated to this scale. Geometrically, this can be represented by a 
point in (a five-dimensional) 
parameter space. Then, tracing the effective dynamics of the system at ever
larger length and time scales generates a directed path in this space which
starts off from this initial point. This flow therefore is always driven away
from 
the repulsive fixed point 
and towards the attractive fixed point, which therefore acts as
a ``sink''. 
Therefore, the asymptotic dynamics is uniquely determined by the attractive
fixed point. 
If $(4-d+y) < 0$, then the stability
properties 
of these two fixed points are interchanged. 
In this case, all RG flows, irrespective of where they start off in parameter
space, inevitably end up in the neighborhood of the 
trivial fixed point $\lambda_0 = 0$. This
implies the asymptotic dynamics is linear (and trivial). 
We see the stability properties of fixed
points depend on the spatial dimension $d$ and on the noise spectrum through
the noise exponent $y$. {}For the Kolmogorov spectrum $y = d$ and for 
the Iroshnikov-Kraichnan spectrum $y = d - 1/4$, 
the trivial fixed point Eq. (\ref{triv}) is always repulsive 
while the non-trivial fixed point Eq. (\ref{nontriv}) is
always attractive in the infrared. 
Similar conclusions hold for a white noise spectrum, $y = 0$, provided the
dimension $d < 4$. 
{}Finally, the fixed point critical exponents are 
obtained by substituting the values of the fixed points into the 
original set of RG equations Eq. (\ref{RGE1}), which yields the solutions
\begin{eqnarray}
\label{fpsol}
(z,\chi) &=& \Big( 2, \frac{1}{2}(2+y-d) \Big) \qquad, {\rm for\, fixed\, point\, (i)},\\
(z,\chi) &=& \Big( \frac{1}{3}(2+d-y), \frac{1}{3}(1+y-d) \Big)
\qquad , {\rm for\, fixed\, point\, (ii)}.  
\end{eqnarray}
The exponents for the trivial fixed point are simply the generalization
of those of the Edwards-Wilkinson model of surface growth subject to 
spatially correlated noise (Barab\'asi and Stanley 1995).
Note furthermore that the exponents satisfy $\chi + z = 1$ at
the nontrivial fixed point (i.e., when $\lambda_0 \neq 0$). This in fact had
better be the case as dictated by the renormalization group equation
for $\lambda_0$ in Eq. (\ref{RGE1}) which admits a nontrivial fixed
point solution iff $z + \chi -1 = 0$. So, this serves as an important
consistency check on our calculation of the exponents. {}Finally, note that
for either fixed point, the dynamic exponent is non-zero, $z \neq 0$.

\section{Scaling and the approach to asymptotia}
\label{sect5}

With the fixed point solutions obtained in the
previous sections we now turn to the asymptotic
scaling behavior (in the infrared)
of the velocity and magnetic field correlation functions
and energy density spectra.

\subsection{Renormalized scale dependent viscosity and resistivity}
\label{sect5A}

The approach to the (nontrivial) fixed point can be quantified and is
very useful for carrying out RG-improvement. 
The effective viscosity $\nu(k)$ and magnetic resistivity $\nu_B(k)$
at the scale $k$ result from all turbulent motions with wave numbers
above $k$ (i.e., at shorter length scales). To derive the renormalized
viscosity and resistivity, we integrate Eqs.
(\ref{elements}) and (\ref{elements2}) over a finite band of momenta
from the cut-off $\Lambda$ to $\Lambda/s$, but without rescaling.  

{}From taking first the sum and then the difference 
of these expressions, we derive the following infinitesimal recursion
relations 
\begin{eqnarray}
\label{dnu}
\frac{d \nu}{d \ell} &=& \frac{2A_d (A + B)}{\nu^2} \Big(\frac{\Lambda}{
e^{\ell}}\Big)^{(d-y-4)},\\
\label{dnuB}
\frac{d \nu_B}{d \ell} &=& \frac{2A_d (A - B)}{\nu_B^2} \Big(\frac{\Lambda}{
e^{\ell}}\Big)^{(d-y-4)},
\end{eqnarray}
where we have set $\lambda_0 = 1$ and have used $s = e^{\ell}$. We assume that
the noise amplitudes, $A,B$  have achieved their asymptotic 
(i.e., near the fixed point) values.  

Integrating Eq. (\ref{dnu}) yields the relation
\begin{equation}
\nu^3(\ell) = \nu^3(0) + \frac{6A_d(A+B)}{(4+y-d)}\, \Big(
(e^{-\ell}\Lambda)^{d-y-4} - {\Lambda}^{d-y-4} \Big),
\end{equation}
where $\nu(0)$ is the initial value of the viscosity. When a 
large range of wavenumbers is eliminated this takes 
a scaling form independent of initial properties and the ultraviolet cut-off
$\Lambda$,
\begin{equation}\label{nukay}
\nu(k) = \Big(\frac{6A_d(A+B)}{4+y-d}\Big)^{\frac{1}{3}}\, k^{-(4+y-d)/3},
\end{equation}
where $k = e^{-\ell}\Lambda$ 
and $A+B = D_v > 0$ is the magnitude of the random force fluctuations.   

By a similar argument, we find that the $k \rightarrow 0$ scaling form
of the magnetic resistivity is given by
\begin{equation}\label{nubkay}
{\nu}_B(k) = \Big(\frac{6A_d(A-B)}{4+y-d}\Big)^{\frac{1}{3}}\, k^{-(4+y-d)/3},
\end{equation}
where $A - B = D_B > 0$ is the magnitude of the random current fluctuations.

Thus we see that the viscosity and magnetic resistivity tend to increase
as $k \rightarrow 0$ (that is, for large length scales) provided
$(4 + y - d) > 0$, which is the condition that the RG trajectories
flow towards the nontrivial fixed point. This trend 
of increasing effective viscosity has also been
reported as a prediction of the RG
in pure Navier-Stokes (Dannevik et al. 1987) and in incompressible
MHD (Longcope and Sudan 1991; Camargo and Tasso 1992), where 
in addition there is
an increasing effective magnetic resistivity. 
These scale dependent
quantities will be used repeatedly later on when we carry out the
renormalization group improvement of the correlation functions and
spectral densities.

\subsection{Scaling form of the correlation functions}
\label{sect5B}

The scaling properties of the $\vec P$-field correlations, the 
$\vec Q$-field  correlations as well as the cross correlations
provide direct information on the large-distance and long-time
behavior of the fluid velocity and magnetic field via the simple
linear relations written in Eqs.
(\ref{physicalv}) and (\ref{physicalB}). These correlations form
part of a larger block-matrix structure as 
given in Eqs. (\ref{correlationz}),
(\ref{correlation}) and
(\ref{bareresponse}), but we can easily work out the scaling
properties of the individual elements of this block array and further note
from direct inspection of those expressions 
that $\langle P_j P_n \rangle = \langle Q_j Q_n \rangle$
and $\langle P_j Q_n \rangle = \langle Q_j P_n \rangle$. So, there are
really only two independent types of field correlation functions to consider,
whether these are expressed in terms of the Elsasser variables or
in terms the velocity and magnetic fields. In this Section, we derive the exact 
asymptotic scaling form of the correlation functions up to (as of yet) 
an unknown scaling function. This illustrates the power and limitations of
pure scaling arguments. After the inclusion of magnetic helicity, we return
to the correlation functions and will determine their scaling functions
explicitly by means of RG-improved perturbation theory.

It will prove extremely useful to have the scaling form of these correlations 
available in both 
the space and time domain as well as in the Fourier 
(momentum and frequency) domain. 
In the scaling regime,
the $\vec P$-field scales as  
\begin{equation}
P_j(\vec x,t) = s^{\chi}\, P'_j(\vec x',t')
= s^{\chi} P'_j(s^{-1} \vec x,s^{-z}t) = 
s^{\chi}\, P_j(s^{-1}\vec x,s^{-z} t),
\end{equation}
and similarly for $\vec Q$. 
The middle expressions are direct consequences of the
scale transformations Eqs. (\ref{scaling}) and (\ref{fieldscaling}) 
whereas in the scaling
regime, the MHD equations are self-similar and then
the rightmost equality also holds.  
Thus in the scaling regime the $P-P$ correlation function scales as
\begin{eqnarray}\label{fullscaling}
\langle P_j(\vec x, t)P_n(0,0)\rangle &=& s^{2\chi}
\langle P_j(s^{-1} \vec x,s^{-z} t)P_n(0,0)\rangle,\nonumber \\
C_{P_j P_n}(\vec x, t) &=& 
s^{2\chi}\, C_{P_j P_n}(s^{-1} \vec x, s^{-z} t).
\end{eqnarray}
Here by $C_{P_j P_n}$, we mean the correlation function between the
$jth$ and the $nth$ components of the $P$-field. 
Note that because of space and time translational invariance, 
the correlation for the
fields depend only on the spatial separation 
$\vec x$ and temporal separation $t$. 
The above equality relates the correlation 
function at two different scales, whose
separation is measured by $s > 1$.
We emphasize that Eq. (\ref{fullscaling}) is exact,
when the system is in the scaling regime, that is to
say, in the vicinity of a fixed
point. 

We now will see what further information can be extracted from the 
expression Eq. (\ref{fullscaling}). 
There is 
an implicit tensor structure 
in these functions, due to the vector indices, 
which we make explicit below in the Fourier domain.
Consider spatial correlations at equal times (i.e., $t = 0$) then from
Eq. (\ref{fullscaling}) we have that
\begin{equation}\label{spacescaling}
C_{P_j P_n}(\vec x,0) = |\vec x|^{2\chi}\, \hat C_{P_j P_n}(1,0),
\end{equation}
whereas temporal correlations at the same spatial point 
(i.e., $\vec x = 0$) scale as
\begin{equation}\label{timescaling}
C_{P_j P_n}(0,t) = t^{\frac{2\chi}{z}}\, \hat C_{P_j P_n}(0,1),
\end{equation} 
where we have chosen $s=|\vec x|$ in the former and $s= t^{1/z}$ in the 
latter. 
Here, $\hat C_{P_j P_n}(1,0), \hat C_{P_j P_n}(0,1)$ are 
just constant tensor coefficients. 
The general scaling form of the correlation function valid for
arbitrary spatial and temporal arguments can be written
as 
\begin{equation}\label{generalscaling}
C_{P_j P_n}(\vec x, t) = |\vec x|^{2\chi}\, \hat 
C_{P_j P_n}\Big( \frac{t}{|\vec x|^z} \Big),
\end{equation}
where the scaling function $\hat C$ obeys the following limits:
\begin{eqnarray}\label{u1}
u \rightarrow 0 \qquad \Rightarrow \qquad 
\hat C_{P_j P_n}(u) &\rightarrow& \hat C_{P_j P_n}(1,0), \\
\label{u2}
u \rightarrow \infty \qquad \Rightarrow \qquad 
\hat C_{P_j P_n}(u) &\rightarrow& \hat C_{P_j P_n}(0,1)\, u^{\frac{2\chi}{z}}.
\end{eqnarray}
Note that Eqs. (\ref{spacescaling}) 
and (\ref{timescaling}) are just special cases of Eq. (\ref{generalscaling}).

Identical results hold for the $Q-Q$ correlation function by simply 
replacing the index labels 
$P \rightarrow Q$ from Eq. (\ref{fullscaling}) to Eq. (\ref{u2}). 
{}For the cross-correlations $P - Q$, just replace one of the $P$ symbols
in Eq. (\ref{fullscaling}) to Eq. (\ref{u2}) by a $Q$. 
This is as far as one can go employing scaling arguments. 
However, this is enough
for determining how the power-law correlations depend on the exponents
$z$ and $\chi$, which are obtained from solving for the RG fixed points.

{}Fourier transforming Eq. (\ref{fullscaling}) 
yields equivalent information, but 
expressed instead in the
momentum and frequency domain
\begin{equation}\label{pipi}
C_{P_j P_n}(\vec k,\omega) = k^{-d - 2\chi - z}\, {\bf P}_{jn}(\vec k)
\, \hat C_{P P}\Big( \frac{\omega}{k^z} \Big),
\end{equation}
with of course, the scaling functions $\hat C_{P P} =  \hat C_{Q Q}$. 
The scaling of the cross-correlation
function involves an a-priori independent scaling function, and thus we have
\begin{equation}\label{pipsi}
C_{P_j Q_n}(\vec k,\omega) = k^{-d - 2\chi - z}\, {\bf P}_{jn}(\vec k)
\, \hat C_{P Q}\Big( \frac{\omega}{k^z} \Big).
\end{equation}
Note that Eqs. (\ref{pipi}) and (\ref{pipsi}) show that the tensor structure
of the correlation functions is carried by the projection operator, as was to
be expected. 

It should be clear from these considerations 
that the asymptotic behavior of the correlation functions
and the quantities derived from them, are all completely determined in terms of
the two exponents $z$ and $\chi$ calculated above.  
Thus, using Eqs. (\ref{physicalv}) and (\ref{physicalB}) 
to transform back to the physical 
velocity and magnetic fields, we immediately find
in the space and time domain
\begin{eqnarray}
\label{vv}
\langle v_j(\vec x,t) v_n(0,0) \rangle &=&  |\vec x|^{2\chi}\,
\left\{ \hat C_{P_j P_n}\Big(\frac{t}{|\vec x|^z}\Big) +
\hat C_{P_j Q_n}\Big(\frac{t}{|\vec x|^z}\Big) \right\},\\
\label{bb}
\langle B_j(\vec x,t) B_n(0,0) \rangle &=&  |\vec x|^{2\chi}\,
\left\{ \hat C_{P_j P_n}\Big(\frac{t}{|\vec x|^z}\Big) -
\hat C_{P_j Q_n}\Big(\frac{t}{|\vec x|^z}\Big) \right\},\\
\label{vb}
\langle v_j(\vec x,t) B_n(0,0) \rangle &=& 0 .
\end{eqnarray}
Note there is no net averaged cross-helicity. This is consistent with the
fact that no helicity has been injected into the system.  

{}From Eqs. (\ref{vv}) - (\ref{vb}),
the spatial dependence of both fluid velocity and magnetic field correlations
in the vicinity of the non-trivial IR stable fixed point scale as
\begin{equation}
\langle v_j(\vec x,0)v_n(0,0)\rangle \sim \langle B_j(\vec x,0)B_n(0,0)\rangle \sim
r^{\frac{2}{3}(1+y-d)},
\end{equation}
where $r = |\vec x|$ and the 
roughness exponent $\chi = \frac{1}{3}(1+y-d)$ at the infrared
stable fixed point Eq. (\ref{nontriv}).
Note that for the Kolmogorov spectrum, $y=d$, and we recover the two-thirds law
($r^{\frac{2}{3}}$) 
for the velocity correlation function in fully developed {\it hydrodynamic}
turbulence (Frisch 1995)
(and we predict that the spatial dependence of the magnetic field
correlation function is a power-law obeying a two-thirds law for this noise spectrum).
{}For the Iroshnikov-Kraichnan spectrum, the spatial correlation scales
instead as $r^{\frac{1}{2}}$. 
{}For the temporal correlations, from Eqs. (\ref{vv}) - (\ref{vb})
they scale as
\begin{equation}
\langle v_j(0,t)v_n(0,0)\rangle \sim \langle B_j(0,t)B_n(0,0)\rangle \sim
t^{2\chi/z} = t^{\frac{2(1+y-d)}{(2+d-y)}}, 
\end{equation}
where the dynamic exponent $z=\frac{1}{3}(2+d-y)$.
{}For the Kolmogorov spectrum, the temporal correlations
scale as $\sim t$, whereas for the I-K spectrum, they scale
as $\sim t^{6/7}$. By marked contrast, a white noise spectrum $y=0$ yields
spatial correlations that scale as $r^{-\frac{4}{3}}$ and temporal correlations
that scale as $t^{-\frac{4}{5}}$ in $d=3$ dimensions.

To help understand the above results about the correlation functions
in the scaling regime, the case of $d=3$, $y=-2$ hydrodynamics will be computed
explicitly.  The above scaling relations of the correlation
functions all apply to hydrodynamics by setting 
the magnetic field to zero, ${\vec B}=0$.   
{}For this case $4-d+y < 0$, thus the trivial fixed
point is stable and so it is simple to
obtain the exact correlation functions by inverse Fourier transforming
Eq. (\ref{corr}) to yield
\begin{eqnarray}
\label{c3m2}
C_{PP}^{(y=-2)}({\vec x},t) & = & const. \times
\left[ \exp(-\frac{|{\vec x}|^2}{4\nu t}) \right]/
(\nu t)^{3/2} \nonumber \\
& = & const. \times \frac{1}{|\vec x|^3} \left[
(\frac{|\vec x|^2}{\nu t})^{3/2}
\exp(-\frac{|{\vec x}|^2}{4\nu t}) \right] .
\end{eqnarray}
Let us verify this solution is consistent with the
general scaling forms given above.
Since at this trivial fixed point for $d=3$, $y=-2$
from Eq. (\ref{fpsol}) $z=2$, $\chi = -3/2$,
it follows that the above solution is consistent with
the general scaling forms 
Eqs. (\ref{fullscaling}) and (\ref{generalscaling}).
{}Furthermore, from the general scaling form Eq. (\ref{timescaling})
one expects $C_{PP}(0,t) \sim t^{-3/2}$
which is consistent with the above solution.
{}For $t=0$, the inverse Fourier transform is ill-defined,
so comparison is not relevant.  This is not inconsistent
with the general scaling form Eq. (\ref{spacescaling}) since
these general considerations do not guarantee
that the function $C_{PP}(1,0)$, for example in
Eq. (\ref{spacescaling}), necessarily is well defined.

\subsection{Energy spectrum}
\label{sect5C}

A quantity of fundamental importance in turbulence research is the
energy spectrum which contains, in part, information regarding the
flow of energy across different length scales.  The energy
spectrum also is key in
defining and calculating the spectral energy density, which is the
energy
density per mode or wavenumber.   
The transfer of energy from smaller to larger 
(or larger to smaller) length scales 
as a system evolves in time
is known as an inverse (direct) 
energy cascade. We can investigate whether and under
what conditions, randomly forced stochastic MHD 
in the absence of magnetic helicity, 
subject to the noise spectrum 
of Eqs. (\ref{vnoise}) and (\ref{Mnoise}), 
will exhibit energy cascades, and more generally, what is the
asymptotic behavior of the energy spectrum. 
To do so, we will begin 
by focusing attention on the quadratic energy densities for the Elsasser
fields and make direct use of the asymptotic fixed point 
scaling properties for the correlation
functions deduced in the previous subsection.  
After this is done, we then turn to a complete calculation of the
spectral energy density functions for both the 
physical fluid velocity and 
magnetic field by means of improved perturbation theory using the
results obtained in Subsection \ref{sect5A} 
for the renormalized viscosity and 
magnetic resistivity.
This will lead
to identical scaling relations 
as obtained from the ``pure'' or naive scaling arguments, but with 
the added advantage of
yielding the explicit form of the individual scaling functions, valid to the
one-loop order in perturbation theory at which we are working. Moreover,
the example will serve to demonstrate how RG-improvement works in
practice. 

Consider 
therefore the total energy $E_{PP}$ in the Elsasser 
$\vec P$ field averaged over the random fluctuations
(this will serve as a useful guide when we come to treat
all the fields together below)
\begin{eqnarray}\label{allsteps}
E_{PP} &=&
\frac{1}{2} \int d^d{\vec x} \, \langle  \vec P(\vec x ,t){\bf \cdot}
 \vec P(\vec x,t) \rangle,\nonumber \\
&=& \frac{1}{2} {\rm tr} \int d^d{\vec x} \, 
\langle  P_j(\vec x ,t)
 P_k(\vec x,t) \rangle,\nonumber \\
&=& \frac{1}{2} {\rm tr} \int d^d{\vec x} \, \int \frac{d^d {\vec k}}{(2\pi)^d}
\, \int \frac{d^d {\vec k'}}{(2\pi)^d} \, \exp(i(\vec k + \vec k')\cdot \vec x)\,
\langle  P_j(\vec k ,t)
 P_k(\vec k',t) \rangle,\nonumber \\
&=& \frac{1}{2} {\rm tr} \int \frac{d^d {\vec k}}{(2\pi)^d}
\langle  P_j(\vec k ,t)
 P_k(-\vec k,t) \rangle,\nonumber \\
&=& \frac{1}{2} {\rm tr} \int \frac{d^d {\vec k}}{(2\pi)^d}
\int \frac{d\omega}{2\pi}\, \int \frac{d\omega'}{2\pi}\,
\exp(-i(\omega + \omega')t)\, \langle  P_j(\vec k ,\omega)
 P_k(-\vec k,\omega') \rangle,\nonumber \\
&=& \frac{V}{2} {\rm tr} \int \frac{d^d {\vec k}}{(2\pi)^d}
\int \frac{d\omega}{2\pi}\, C_{P_j P_k}(\vec k,\omega),
\end{eqnarray}
where we have used $\langle  P_j(\vec k ,\omega)
 P_k(\vec k',\omega') \rangle = (2\pi)^{d+1} 
\delta^d(\vec k + \vec k')\delta(\omega + \omega') C_{P_j P_k}(\vec k,\omega)$,
and the $d$-dimensional spatial volume factor is 
$V = (2\pi)^d \, \delta^d(\vec 0)$.
{}From the identity in the last line of Eq. (\ref{allsteps}), we can
immediately identify the energy density $(E_{PP}/V)$, which
is given by the indicated double integral of the correlation function. 
The {\it spectral} energy density, i.e., the
energy density per unit wavenumber, is therefore given by 
\begin{equation}\label{spectralE}
E_{PP}(k) = [\frac{1}{2}S_d/(2\pi)^{d+1}]\, k^{d-1} 
\int_{-\infty}^{\infty} d\omega \,\,
{\rm tr} \, C_{P_jP_k}(k,\omega),
\end{equation}
since clearly 
\begin{equation}
E_{PP} = V \int_0^{\infty} dk \, E_{PP}(k).
\end{equation}
By making use of Eq. (\ref{pipi}) which holds in the scaling
regime, we can discover how the spectral energy density scales
as a function of the injected random noise spectrum. Inserting the
expression Eq. (\ref{pipi}) into Eq. (\ref{spectralE}) yields
\begin{equation}\label{spectralE2}
E_{PP}(k) = \frac{(d-1)S_d}{2(2\pi)^{d+1}}\, k^{\alpha}\, 
\int_{-\infty}^{\infty} du \,
\hat C_{PP}(u).
\end{equation}
This follows after defining 
the exponent 
\begin{equation}
\label{enexp}
\alpha = -1-2\chi = -\frac{5}{3}+ \frac{2}{3}(d-y), 
\end{equation}
and 
making use of the
exponent identity $z + \chi = 1$, which holds at the non-trivial 
infrared stable fixed point Eq. (\ref{nontriv}). 
The 
steps leading to the derivation of the spectral energy
density function $E_{Q Q}(k)$
for the $Q$-field are identical to those outlined above, and in fact, we have
$E_{P P}(k) = E_{Q Q}(k)$, as a
consequence of the equality between the scaling functions
$\hat C_{P P} = \hat C_{Q Q}$.  
We remark at this point that the choice of noise exponent $y=d$ 
formally leads to the
Kolmogorov energy density spectrum, since 
this yields $\alpha = -\frac{5}{3}$, and the spectral energy density then
scales as 
\begin{equation}\label{latetime}
E_{PP}(k) \sim k^{-\frac{5}{3}}. 
\end{equation}
By way of contrast, at the unstable and trivial fixed point, the
spectral density would scale instead as
\begin{equation}\label{trivlatetime}
E_{PP}(k) \sim k^{-3}
\end{equation}
for $y=d$. The difference in the scaling is due to the presence 
Eq. (\ref{latetime}) or
absence Eq. (\ref{trivlatetime}) 
of the nonlinear terms in the equations of motion which
act to mix distinct modes.

We note of course that the overall magnitude of the spectral density
Eq. (\ref{spectralE2}) 
is determined in part by the scaling function $\hat C_{PP}$.
It has been common practice in many, if not all,  RG analyses of turbulence to
deduce only the power law scaling exponent of the physical quantities of
interest while leaving the overall proportionality factors undetermined.
This could lead to the impression that the RG is incapable of
determining these overall factors. 
In fact, the renormalization group not
only yields a determination of the scaling exponents but 
also provides quantitative information regarding the overall constant
factors.      
This is achieved 
to a given order in perturbation theory by means of RG-improved
perturbation theory (Hochberg et al. 1999a). 
We will make use of this technique
to complete our discussion of the spectral energy density.
We now turn to this task, which will be carried out in two
steps. {}First we calculate the energy corresponding to the free-field
limit, where the correlation functions are known explicitly, and then
we use RG to improve the resulting energy by use
of our perturbation theory. This results in the corrected
form of the spectral energy, to the same order of perturbation theory
at which we are working.  
We begin again at the level of the Elsasser fields and 
then at the end of the calculation make
use of the linear transformation back 
to the physical fluid velocity and magnetic
fields.  The net result will be a one-loop
expression for both the fluid velocity and magnetic field spectral
energy densities. The calculation of $E_{PP}(k)$ carried out in detail
above will serve as a useful template for this purpose.

We define an energy function involving both the 
Elsasser fields $P$ and $Q$ at once 
(which is actually a $2 \times 2$ matrix, as
we indicate with the dyad) 
\begin{eqnarray}\label{totalE}
\stackrel{\leftrightarrow} E 
&=& \frac{V}{2} {\rm tr} \int \frac{d^d \vec k}{(2\pi)^d} 
\int_{-\infty}^{\infty} \frac{d\omega}{2\pi} \, 
[\stackrel{\leftrightarrow}{\bf C}(\vec k,\omega)]_{mn},\nonumber \\
&=& V \int_0^{\infty}dk
\stackrel{\leftrightarrow} E(k),
\end{eqnarray}
where the energy spectral density function (also a $2 \times 2$ matrix) is given by
\begin{equation}\label{bigspectralE}
\stackrel{\leftrightarrow} E(k) = 
[\frac{1}{2}S_d/(2\pi)^{d+1}] k^{d-1} 
\int_{-\infty}^{\infty} d\omega \,
{\rm tr} \,[\stackrel{\leftrightarrow}{\bf C}(\vec k,\omega)]_{mn}.
\end{equation}
This expression is exact. We can calculate it to one-loop order by
means of RG-improved perturbation theory. This proceeds by the two
above mentioned steps.  That is to say,
first we evaluate this spectral density using the free-field correlation
function (denoted with the zero subscript)
\begin{equation}
[\stackrel{\leftrightarrow}{\bf C}_0(\vec k,\omega)]_{mn} =
{\bf P}_{mn}(\vec k)\, {\cal G}_0(\vec k,\omega)D_0(k)
{\cal G}_0(-\vec k,-\omega),
\end{equation}
and then improve the resulting expression by replacing the bare
parameters by their one-loop asymptotic scaling forms (in the present
case, only the viscosity $\nu$ and 
magnetic resistivity $\nu_B$ receive nontrivial
corrections at one-loop order, so these are the only MHD parameters
that get improved to this order).
The free correlation function appearing above is defined and calculated
in Eqs. (\ref{correlation}) and 
(\ref{bareresponse}), where further details can
be found.
The frequency integration is straightforwardly performed by means of the 
residue theorem. {}From Eqs. (\ref{correlation}) 
and (\ref{bareresponse}), we immediately see
that there are two simple poles in the lower complex frequency plane and two
simple poles in the upper half plane.  
Closing the contour in the lower half plane 
and taking the radius of the semicircular contour to infinity
yields the 
desired integral Eq. (\ref{bigspectralE}), and we obtain 
(at zero-loop order, that is, for the free non-interacting theory)
\begin{eqnarray}\label{result}
\stackrel{\leftrightarrow} E_0(k) 
&=& \big(\frac{d-1}{8} \big)\frac{S_d}{(2\pi)^d} k^{d-3} D_0(k)\,
\left\{ \frac{1}{\nu} 
\left[ \matrix{1&1\cr   
         1&1} \right]
+ \frac{1}{\nu_B} 
\left[ \matrix{1&-1\cr   
         -1&1} \right] \right\}\nonumber \\
&=& 
\big(\frac{d-1}{4} \big)\frac{S_d}{(2\pi)^d} k^{-3 + (d-y)}
\left\{ \frac{A+B}{\nu} 
\left[ \matrix{1&1\cr   
         1&1} \right]
+ \frac{A-B}{\nu_B} 
\left[ \matrix{1&-1\cr   
         -1&1} \right] \right\}.
\end{eqnarray}
The first line holds for {\it arbitrary} spatially correlated noise $D_0(k)$.
We now renormalization group {\it improve} this expression
by replacing the viscosity and magnetic resistivity by their 
(one-loop) asymptotic scale dependent forms $\nu(k)$ and
$\nu_B(k)$ calculated above in Eqs. (\ref{nukay}) and (\ref{nubkay}) to obtain
the one-loop spectral density
\begin{eqnarray}\label{improved}
\stackrel{\leftrightarrow} E(k) &=& 
(\frac{d-1}{4})\frac{S_d}{(2\pi)^d} 
\Big(\frac{6 A_d}{4+y-d}\Big)^{-\frac{1}{3}}\,
k^{-\frac{5}{3} +\frac{2}{3}(d-y)}\,
\nonumber \\
&\times& \left\{ (A+B)^{\frac{2}{3}} \,
\left[ \matrix{1&1\cr   
         1&1} \right]
+ (A-B)^{\frac{2}{3}} \, 
\left[ \matrix{1&-1\cr   
         -1&1} \right] \right\}.
\end{eqnarray}
This has been evaluated using 
the specific noise spectrum as written in Eq. (\ref{correlation}).
The dimension dependent constant $A_d$ is given in
Eq. (\ref{Ad}).
{}From this expression we can immediately read off the 
individual spectral energy functions for each Elsasser variable
since each entry of this array Eq. (\ref{improved}) 
involves a quadratic pairing of the
Elsasser fields, that is
\begin{equation}
\stackrel{\leftrightarrow} E(k)
=
\left( \matrix { E_{P P}(k) & E_{P Q}(k) \cr
 E_{Q P}(k) & E_{Q Q}(k) }
\right).
\end{equation}

By means of the transformations given in 
Eqs. (\ref{physicalv}) and (\ref{physicalB})
we readily obtain the spectral energy
densities for the physical fluid velocity and magnetic fields,
\begin{eqnarray}\label{trans1}
E_V(k) &=& \left\{
E_{P P}(k) + E_{P Q}(k) \right\}, \\
\label{trans2}
E_B(k) &=& \left\{
E_{P P}(k) -  E_{P Q}(k) \right\}, 
\end{eqnarray}
respectively.
With these expressions in hand, we can now obtain in detail the relative
influence of kinetic and magnetic noise.
{}From Eqs. (\ref{improved}),(\ref{trans1}), and (\ref{trans2}), we 
deduce the individual spectral energy densities, namely
\begin{equation}\label{vspectral}
E_V(k) = (\frac{d-1}{2})\frac{S_d}{(2\pi)^d}
k^{-\frac{5}{3} + \frac{2}{3}(d-y)}
\Big(\frac{6A_d}{4+y-d}\Big)^{-\frac{1}{3}}
(A+B)^{\frac{2}{3}},
\end{equation}
for the kinetic energy spectrum and 
\begin{equation}\label{Bspectral}
E_B(k) = (\frac{d-1}{2})\frac{S_d}{(2\pi)^d}
k^{-\frac{5}{3} + \frac{2}{3}(d-y)}
\Big(\frac{6A_d}{4+y-d}\Big)^{-\frac{1}{3}}
(A-B)^{\frac{2}{3}}, 
\end{equation}
for the magnetic energy density spectrum. The ratio of the magnetic to kinetic
energy spectra is given by
\begin{equation}
\frac{E_B(k)}{E_V(k)} = 
\Big(\frac{D_B}{D_v}\Big)^{\frac{2}{3}},
\end{equation}
demonstrating that this ratio depends only on the ratio
of the magnitudes of magnetic and kinetic noises. 

As promised, we
have succeeded in calculating the scaling functions for
both the fluid velocity 
and magnetic field spectral functions.  To one-loop, these results
may be read off from
Eqs. (\ref{vspectral}) and (\ref{Bspectral}). Moreover, the difference
in the two spectral functions is due solely from the 
dependence on the noise amplitudes. Thus for example, in the limit
of vanishing magnetic noise, we have $A = B$. In this limit,
the magnetic resistivity Eq. (\ref{nubkay}) 
does not renormalize and moreover, the
magnetic spectral energy density Eq. (\ref{Bspectral})
vanishes identically.
Conversely, in the limit of zero kinetic noise, the
fluid viscosity Eq. (\ref{nukay}) would not renormalize and the
kinetic energy density Eq. (\ref{vspectral}) would vanish identically. 
Another observation is that when both kinetic and magnetic noises
are present, both spectral functions scale in the same way.

The scaling behavior of the energy spectrum has a special
interest due to the significance of
the Kolmogorov spectrum as well as others, such as
the Iroshnikov-Kraichnan spectrum.  In hydrodynamics,
for a freely decaying turbulent fluid, which in
particular means without an external driving force,
Kolmogorov gave general arguments that the energy
spectrum would behave as $k^{-5/3}$.
{}For forced hydrodynamics and/or MHD, it is clear from
Eq. (\ref{improved}) that the energy spectrum can have a range of behavior,
with the $5/3$ form not being unique.  In particular,
for a specific choice, namely $y=d$, the
Kolmogorov spectrum emerges.  However, since the validity of
the present RG analysis only is in the
asymptotic region, this spectral behavior only occurs in the
$k \rightarrow 0$ limit.  This differs from the region relevant
for freely decaying turbulence, which extends from some intermediate
value of $k$ up to the higher $k$ regime.  Nevertheless, an interesting
hypothesis can be motivated 
based on Kolmogorov's
universality arguments for this spectrum.
If the energy spectrum in the forced case has the
same behavior as found for freely decaying turbulence, then
the asymptotic limit of other
hydrodynamic quantities,
such as the Prandtl number,  Reynolds number, and the viscosity
also should be the same in both cases.  Following this
hypothesis, it would
allow the stochastic hydrodynamic equations for
$y=d$ to be used as a tool for studying various properties about
freely decaying turbulence.  Similarly,
other types of noise (e.g., Iroshnikov-Kraichnan) 
could be studied,  by
choosing the noise exponent $y$ accordingly. 
In this paper, this direction of reasoning will not be
developed or applied, but it is worth noting it here.

\section{Magnetic Helicity}
\label{sect6}

Up to now we have been considering the effects of both
random forces and random currents in turbulent MHD and the observable
consequences these lead to at large spatial separations 
and for long times. One can
also assess the influence of magnetic helicity on the
asymptotic behavior of the magnetized turbulent fluid. 
This is especially interesting since helicity can arise naturally
in rotating turbulent systems such as are encountered in astrophysics
(e.g., in accretion disks. see Papaloizou and Lin 1995) 
and galactic dynamics. 
In MHD,
the magnetic helicity is an independent quadratic invariant
conserved in the limit of nondissipative turbulence. In three dimensions ($d=3$), it
is given by
\begin{equation}\label{maghelicity1}
H = \frac{1}{2} \int d^3{\vec x} \,\, {\bf A \cdot \bf B},
\end{equation}
where $\bf A$ is the vector potential. In 2d MHD, the 
corresponding quadratic invariant is given instead 
by the square of the magnetic
potential (Biskamp 1993).
{}For {\it hydro}dynamics, the analogous quantity of interest is the kinetic
helicity, but in nondissipative MHD, this is not a conserved quantity.  

In much the same way as was done for the 
energy spectral functions derived above, one
can derive a spectral density for magnetic helicity and use
this to study the 
helicity spectrum at very large spatial and temporal scales. 
We will do so for $d=3$ dimensions.
{}First, we note that Eq. (\ref{maghelicity1}) is a gauge-invariant
quantity.  Thus, by means of the relation between magnetic field
and vector potential ${\bf B} = {\bf \nabla \times A}$,
in the Coulomb gauge we can write
\begin{equation}\label{vecpot}
{\bf A}(\vec x) = \frac{1}{4\pi}
\int d^3{\vec y} \, \frac{{\bf \nabla \times B}(\vec y)}{|\vec x - \vec y|}.
\end{equation}
With this, the helicity can be expressed entirely in terms of the magnetic
field as
\begin{equation}\label{maghelicity2}
H = \frac{1}{8\pi} \int d^3{\vec x} \, \int d^3{\vec y}\,
\frac{({\bf \nabla \times B}(\vec y))}{|\vec x - \vec y|}\cdot
{\bf B}(\vec x).
\end{equation}
By Fourier transforming the magnetic field 
and the vector potential Eq. (\ref{vecpot})
to wavenumber space, it allows the helicity to be expressed as 
\begin{equation}\label{maghelicity3}
H = \frac{1}{8\pi} \int \frac{d^3 \vec k}{(2\pi)^3}\,
k^{-2} [i\vec k \times \vec B(\vec k,t)] {\bf \cdot} \vec B(-\vec k,t).
\end{equation}
We now define a
helicity density (helicity per unit spatial volume) and an
associated  helicity spectral density. 
{}For this, first Fourier transform 
the integrand in Eq. (\ref{maghelicity3}) 
to frequency variables and take the stochastic average of $H$ 
over the random fluctuations. This allows
the averaged magnetic
helicity to be expressed as
\begin{equation}
H = V \, \int_0^{\infty} dk \,\, {\cal H}(k),
\end{equation}
where $V = (2\pi)^3 \delta^{3}(\vec 0)$ 
is once again a spatial volume factor and 
the helicity spectral density is given by
\begin{equation}\label{Hspectral}
{\cal H}(k) = \frac{i}{8\pi}
\int \frac{d\Omega_3}{(2\pi)^3} 
\int_{-\infty}^{\infty} \frac{d\omega}{2\pi} \,
\epsilon_{jnm} k_n 
\{ C_{P_mP_j}(\vec k, \omega) - C_{P_mQ_j}(\vec k, \omega)\}.
\end{equation}
Here, $\epsilon_{ijm}$ is the fully antisymmetric tensor density
in three dimensions.
The indicated integrations are taken over 
the unit sphere and frequency, respectively.
In arriving at this final expression, we have averaged over all the sources
of random fluctuations. We also point out that ${\cal H}$ is a real
quantity; this can be checked directly in Eq. (\ref{maghelicity3}) 
by using $\vec B(\vec k,t)^* = \vec B(-\vec k,t)$
together with the symmetry properties of $\epsilon_{ijm}$.
In arriving at this final expression, we have written the magnetic
field correlation function in terms of 
Elsasser field correlations $(C_{PP},C_{PQ})$.

It is clear that the magnetic helicity spectral density 
(as  well as the total helicity) vanishes identically in
a system subject strictly to 
non-helical forces and currents. This is 
because the resulting 
magnetic field correlation function is proportional to a 
symmetric tensor, 
$<B_m(\vec k,t) B_i(-\vec k,t)>\, \propto {\bf P}_{mi}(\vec k)$.
The same symmetry characteristics are of course shared by the
correlation functions for the Elsasser variables.  
It is furthermore clear 
from Eq. (\ref{Hspectral}) that to have a {\it net} averaged
helicity, the correlation function must contain an antisymmetric
contribution, and that this helical contribution must in fact be proportional
to the epsilon tensor.  In other words, to include helicity
one must ``build'' an antisymmetric contribution
to the correlation function using 
only the the metric,
or Kronecker delta $\delta_{ij}$, 
products of the wavevector $\vec k_m$, 
and epsilon itself $\epsilon_{lnk}$.
This we will do dynamically by injecting an initial spectrum of random
magnetic helicity into the MHD equations
by means of a straightforward extension of the noise term. We will
then track this random helicity dynamically by means of the renormalization
group applied to helical MHD and compute the asymptotic 
limit of the helicity spectrum.

The presence of random helicity in three dimensions can be incorporated 
in the system by extending the
noise spectrum Eq. (\ref{covariance}) as
\begin{equation}\label{covariance2}
[\stackrel{\leftrightarrow}{\bf \Gamma}(\vec k,\omega)]_{mn} = D_0(k)\,
{\bf P}_{mn}(\vec k) -i F_0(k)\epsilon_{mnl} \, k_l,
\end{equation}
where the nonhelical contribution arising from
the random forces and currents is that given in Eq. (\ref{covariance}) 
and the second term involving the epsilon tensor is due entirely to helicity.
In higher ($d > 3$) spatial dimensions, the ``extra'' 
indices on the corresponding epsilon
tensor would be contracted by additional factors of the momentum vector. 
The noise function (matrix) $D_0(k)$ is specified in Eq. (\ref{correlation}). 
The noise function (matrix) associated with the helical contribution
can be modeled by taking the following
initial power law spectrum for random helicity,
\begin{equation}\label{F_0}
F_0(k) = 2 h_B \, k^{-w} \left[ \matrix {1  & -1 \cr
               -1 & 1  }\right], 
\end{equation}
where $h_B > 0$ is the amplitude of the helical noise and the number $w$ is an
exponent (analogous to the exponent $y$ characterizing nonhelical noise).
Both are free parameters. 
We have written the helical noise directly
in the Elsasser field basis Eq. (\ref{Elbasis}). In terms of the physical
fields, this source Eq. (\ref{F_0}) represents adding a helical noise
term to the right hand side of Eq. (\ref{Mnoise}), leaving
both Eqs. (\ref{vnoise}) and (\ref{vBnoise}) unaltered. One can add 
a random kinetic helicity to the right hand side of Eq. (\ref{vnoise}) if
one wishes, by a simple modification of Eq. (\ref{F_0}).  
To our knowledge, in contrast to the kinds of scaling arguments 
leading to the derivation
of the Kolmogorov spectrum, there are no arguments that
can shed light on what kind of helicity spectrum one might expect
to see for example in a freely decaying
turbulent helical fluid.  As such, there are no
specifically interesting values of $w$ of which
to take note.

In a series of 
Appendices, we have set up and carried out the renormalization of the
response, noise and vertex functions associated with the MHD equations
in the absence of helicity, $h_B \equiv 0$. 
Now that we have modified the noise spectrum
in Eq. (\ref{covariance2}), we must repeat the renormalization 
group calculations taking
into account the new helical contribution. If we refer to the elementary
Feynman diagrams in Fig. (1a)-(1c), we immediately see that the only
elementary diagram to be modified is the correlation function, Fig (1b).
On the other hand, the bare response
and vertex functions are unaffected by the presence of Eq. (\ref{F_0}).   
When computing the loop diagrams, rather than 
Eq. (\ref{correlation}), one must use
\begin{equation}\label{corrhel}
[\stackrel{\leftrightarrow}{\bf C}_0(\vec k,\omega)]_{mn} =
{\cal C}_0(\vec k,\omega) {\bf P}_{mn}(\vec k) -
i{\cal C}^F_0(\vec k,\omega) \epsilon_{mnl}k_l,
\end{equation}
where ${\cal C}^F_0$ is obtained from ${\cal C}_0$ 
in Eq. (\ref{correlation}) by replacing
$D_0(k)$ by $F_0(k)$.
In terms of the diagrammatic language, 
the response function diagram, Fig. (2a), 
and vertex function diagram,
Fig. (2c), are unaffected, 
while the ``blob'' in the diagram Fig. (2b) for the correlation
function is modified by the addition (insertion) of the helical noise.
All the calculations in the Appendices are to be carried out with this
modified diagram and setting $d=3$, as we are considering helical
noise in three dimensions. The details and mechanics of the calculations
are very similar to those presented in the Appendices.  The
net result of this procedure simply is that 
at one loop order the random
helicity in Eq. (\ref{covariance2}) does not  
renormalize the response function, the noise spectrum
nor the 
vertex function in the hydrodynamic limit, i.e., for $k \rightarrow 0$ and for
$t \rightarrow \infty$.  
In the case of the response and vertex functions, the angular
integrations are responsible for yielding null results in the
hydrodynamic limit. The non-helical noise spectrum does not renormalize
for $w > -1$, which is a mild condition we impose on this exponent. 
This means that the
original set of renormalization group equations Eq. (\ref{RGE1})
(as well as their solutions, fixed points and stability properties)
remain unaffected (at least at one-loop) by the presence of random
magnetic helicity. 

Independently from this, observe that 
the helicity density 
never contributes to the system's total energy Eq. (\ref{totalE})
(since helicity is traceless). Conversely, the energy density never contributes
to the helicity (the energy is linked to a purely symmetric tensor).
Nevertheless, the helicity has an impact on the
dynamics of the system, 
generating its own
contribution to the velocity and magnetic field correlations, as is
clear from Eq. (\ref{corrhel}). We come back to this important
point after calculating the helicity spectral density by means of
RG-improved perturbation theory at one-loop.  

\subsection{Helicity spectral density}
\label{sect6A}

In analogy with the energy spectrum, we can calculate the
helicity spectrum for both, non-interacting and interacting
dynamics. 
To this end, we define the following helicity (matrix) function (in three dimensions)
\begin{eqnarray}\label{totalH}
\stackrel{\leftrightarrow} H &=& \frac{V}{8\pi}i 
\int \frac{d^3 \vec k}{(2\pi)^3} 
\int_{-\infty}^{\infty} \frac{d\omega}{2\pi} \, 
k^{-2} \, \epsilon_{jmn}\,k_j\,
[\stackrel{\leftrightarrow}{\bf C}(\vec k,\omega)]_{mn},\nonumber \\
&=& V \int_0^{\infty}dk
\stackrel{\leftrightarrow} {\cal H}(k)
\end{eqnarray}
where the helicity spectral density function (also a $2 \times 2$ matrix) is given by
\begin{equation}\label{spectralH}
\stackrel{\leftrightarrow} {\cal H}(k) = 
\frac{i}{8\pi} \int \frac{d\Omega_3}{(2\pi)^3}
\int_{-\infty}^{\infty} \frac{d\omega}{2\pi}\, \epsilon_{jnm} k_n [
\stackrel{\leftrightarrow} {\bf C}(\vec k,\omega)]_{mj}.
\end{equation}
We need to know the correlation function (for the Elsasser fields)
$[\stackrel{\leftrightarrow} {\bf C}(\vec k,\omega)]_{mj}$ in order
to determine the helicity spectrum. We will obtain this spectrum by
means of (1-loop) RG-improved perturbation theory. So, we first evaluate
the spectrum Eq. (\ref{spectralH}) in the non-interacting field limit and
then improve the resultant expression in order to obtain the corrected
spectrum, which is valid at one-loop. {}For the first step, we simply replace
$[\stackrel{\leftrightarrow} {\bf C}(\vec k,\omega)]_{mj}$ by its free-field
limit as given in Eq. (\ref{corrhel}). 
Note in comparing Eq. (\ref{corrhel}) with Eqs. (\ref{totalH}) and 
(\ref{spectralH}), we see that only the antisymmetric
part of the full correlation function contributes to the helicity and to the
helicity spectrum, which should come as no surprise.  
Carrying out this replacement yields for the zeroth-order helicity spectrum
\begin{equation}\label{zerothH}
\stackrel{\leftrightarrow} {\cal H}_0 (k) = \frac{S_3}{4\pi(2\pi)^3}\,
k^2 \, \int_{-\infty}^{\infty}
\frac{d\omega}{2\pi} \, {\cal C}_0^F(\vec k,\omega),
\end{equation}
where the factor $S_3$ results from the integration over
angles (in $d=3$), and we must integrate 
\begin{equation}
{\cal C}_0^F(\vec k,\omega) = {\cal G}_0(\vec k,\omega)
F_0(k) {\cal G}_0(-\vec k,-\omega)
\end{equation}
over the frequency.  
This involves the identical
frequency integration performed for the calculation
of the energy spectrum,
in arriving at Eq. (\ref{result}).  Thus,
direct use can be made of that step, without
any further calculation, to obtain
\begin{eqnarray}
\stackrel{\leftrightarrow} {\cal H}_0(k) &=& 
\frac{S_3}{(2\pi)^3} F_0(k)\,
\left\{ \frac{1 }{\nu}
\left[ \matrix{1&1\cr   
         1&1} \right]
+ \frac{1}{\nu_B} 
\left[ \matrix{1&-1\cr   
         -1&1} \right] \right\} \nonumber \\
&=& \frac{S_3 h_B}{4\pi(2\pi)^3} \, \frac{k^{-w}}{\nu_B} 
\left[ \matrix{1&-1\cr 
         -1&1} \right] ,
\end{eqnarray}
for the zeroth-order value of the spectrum
(which is also the spectrum corresponding to the
trivial fixed point).
The matrix products are worked out making use
of Eq. (\ref{F_0}).
The RG-improved helicity spectrum is therefore given by
\begin{eqnarray}
\stackrel{\leftrightarrow} {\cal H}(k) &=& 
\left( \matrix{ {\cal H}_{PP}(k) &  {\cal H}_{PQ}(k)\cr
            {\cal H}_{QP}(k) & {\cal H}_{QQ}(k)} \right) \nonumber \\
&=& \frac{h_B}{4\pi} \frac{S_3}{(2\pi)^3}
\Big(\frac{6A_3(A-B)}{1+y}\Big)^{-\frac{1}{3}}\,
k^{-w + \frac{1}{3}(1+y)}\,
\left[ \matrix{1&-1\cr   
        -1&1} \right] ,
\end{eqnarray}
which indicates this helicity spectral matrix refers to the 
Elsasser $P,Q$ field basis. In terms of
the physical fields, the magnetic helicity spectral density 
is given by 
\begin{eqnarray}\label{helispec}
{\cal H}_M(k) &=& {\cal H}_{PP}(k) - {\cal H}_{PQ}(k) \nonumber \\
&=&  \frac{h_B}{2\pi} \frac{S_3}{(2\pi)^3}
\Big(\frac{6A_3\, D_B}{1+y}\Big)^{-\frac{1}{3}}\,
k^{-w + \frac{1}{3}(1+y)}.
\end{eqnarray}
We note that the other linear combination yields the {\it kinetic}
helicity spectrum, but this is identically zero, 
${\cal H}_V = {\cal H}_{PP} + {\cal H}_{PQ} = 0.$ This is consistent
with the fact that no kinetic helicity has been injected into the
dynamics. We see the difference in scaling of the helical
spectrum as determined from the fixed points: apart from
numerical prefactors, the magnetic helicity
scales as
\begin{equation}
\label{frhelsp}
{\cal H}_M(k) \sim k^{-w},
\end{equation}
for the free, non-interacting theory (trivial fixed point)
while it scales as
\begin{equation}
\label{nthelsp}
{\cal H}_M(k) \sim k^{-w + \frac{1}{3}(1+y)},
\end{equation}
at the IR stable nontrivial fixed point
(interacting theory).

\subsection{Renormalization Group Improved Correlation Functions}
\label{sect6B}

Just as the general scaling analysis used for obtaining the
general form of the energy density spectra
is rigorously complemented by improved perturbation theory, the
scaling behavior derived above for the field correlation functions can be
complemented by renormalization group improvement. Thus, the
scaling {\it form} of the correlations 
displayed in Eqs. (\ref{pipi}) and (\ref{pipsi}) is correct, 
as it stands.  However, this only makes explicit 
the scaling exponent, while lumping together
with the scaling function
the as yet unknown proportionality constants.  To obtain
the latter, a bit more work is required. Once again, renormalization group
improved perturbation theory leads directly to the answer.

Having introduced the magnetic helicity, we opt for improving the
full correlation function Eq. (\ref{corrhel}), which contains
both kinetic and helicity effects, 
since the same basic steps
are needed for improving each contribution separately. 
Improvement of a particular quantity consists, as we have already seen, in
replacing the bare viscosity and magnetic resistivity by their (one-loop)
running expressions Eqs. (\ref{nukay}) and (\ref{nubkay}).

{}From the structure in Eq. (\ref{corrhel}), we see that this is tantamount to
improving the product of response functions Eq. (\ref{bareresponse}) that
appear in both the ${\cal C}_0$ and ${\cal C}_0^F$ contributions, as can
be checked in Eq. (\ref{correlation}). Since both noise matrices commute
with the product of response function matrices, the problem reduces to
improving the following matrix function
\begin{eqnarray}\label{prodresponse}
{\cal G}_0(\vec k,\omega){\cal G}_0(-\vec k,-\omega) &=& f(\nu,\nu_B; k,\omega)\,
\left[ \matrix{1&0\cr   
         0&1} \right], \\
f(\nu,\nu_B; k,\omega) &=& \frac{\gamma_{+} k^{-4}}{(\gamma_{+} + \gamma_{-})^2
(\gamma_{+} - \gamma_{-})^2}\, C\Big(
\frac{\omega}{k^2} \Big),
\end{eqnarray}
where the dimensionless (in the free-field limit) 
scaling function $C$ is explicitly given by
\begin{equation}\label{bigC}
C\Big(
\frac{\omega}{k^2} \Big) =
\frac{\frac{\omega^2}{(\gamma_{+}k^2)^2} + \frac{\gamma_{-}^2}{\gamma_{+}^2}}{
\Big(1 + \frac{\omega^2}{[k^2(\gamma_{+}+\gamma_{-})]^2}\big)
\big(1 + \frac{\omega^2}{[k^2(\gamma_{+}-\gamma_{-})]^2} \Big)}.
\end{equation}
We make no special distinction between functions of $\gamma_{+},\gamma_{-}$ and
functions of $\nu,\nu_B$, since these viscosity and resistivity
parameters are linearly related 
via Eq. (\ref{gammanu}), and are freely interchangeable.

{}From Eq. (\ref{prodresponse}) and the commutativity of the block matrices, we
can express Eq. (\ref{corrhel}) as
\begin{eqnarray}
[\stackrel{\leftrightarrow}{\bf C}_0(\vec k,\omega)]_{mn} &=&
\left[ \matrix{\langle PP \rangle & \langle PQ \rangle \cr   
         \langle QP \rangle & \langle QQ \rangle } \right]_{mn}, \nonumber \\
&=& f(\nu,\nu_B; k,\omega) 
\left\{ 2k^{-y} \left[ \matrix{A&B\cr   
         B&A} \right]
{\bf P}_{mn}(\vec k) +
2h_B k^{-w} \left[ \matrix{1&-1\cr   
         -1&1} \right]
\epsilon_{mnl}k_l \right\}.
\end{eqnarray}
Of course, the term proportional to $h_B$ holds only in $d=3$, while the remainder
of this expression can be 
evaluated in any spatial dimension. 
This yields the individual correlations in terms of the Elsasser fields, but we can skip
immediately to the quadratic correlations involving the fluid velocity and magnetic fields 
by means of the linear relations 
given in Eq. (\ref{Elbasis}). {}From this, it implies
\begin{eqnarray}\label{individualv}
\langle v_i v_j \rangle &=& f(\nu,\nu_B; k,\omega)\, 
(A+B) k^{-y} \left\{ {\bf P}_{ij}(\vec k) + \frac{h_B}{A + B}\,
k^{y-w}\, \epsilon_{ijm}k_m \right\}, \\
\label{individualB}
\langle B_i B_j \rangle &=& f(\nu,\nu_B; k,\omega)\, 
(A-B) k^{-y} \left\{ {\bf P}_{ij}(\vec k) + \frac{h_B}{A - B}\,
k^{y-w}\, \epsilon_{ijm}k_m \right\}, \\
\label{individualvB}
\langle v_i B_j \rangle &=& f(\nu,\nu_B; k,\omega)\, 
h_B\, k^{-w}\, \epsilon_{ijm} k_m.
\end{eqnarray}

The improved correlations are obtained immediately upon making the
replacements mentioned above. We see that this operation affects
only the overall factor $f(\nu,\nu_B; k,\omega)$, which is common to each
individual correlation function.  As such, this is the only function that
needs to be improved, with the correlations in 
Eqs. (\ref{individualv}), (\ref{individualB}),
and (\ref{individualvB}) improved
by replacing
\begin{equation}
f(\nu,\nu_B; k,\omega) \rightarrow 
f(\nu(k),\nu_B(k); k,\omega).
\end{equation}
{}From Eqs. (\ref{prodresponse}), (\ref{nukay}) and (\ref{nubkay}), we find that
\begin{eqnarray}\label{fimp}
f(\nu(k),\nu_B(k); k,\omega) &=& \frac{\big(\nu(k) + \nu_B(k) \big)^2 k^{-4}}{
4 \nu(k)^2 \nu_B(k)^2}\, C\Big( \frac{\omega}{k^z} \Big),\nonumber \\
&=& \frac{(a + b)^2}{4(a^2\, b^2)}\, k^{-4 + \frac{2}{3}(4 + y -d)}\,
C\Big( \frac{\omega}{k^z} \Big), 
\end{eqnarray}
where the constants 
\begin{equation}
a = \Big( \frac{6 A_d (A+B)}{4+y-d} \Big)^{\frac{1}{3}}, \qquad
b = \Big( \frac{6 A_d (A-B)}{4+y-d} \Big)^{\frac{1}{3}}.
\end{equation}
We point out that the scaling function $C$ is also modified by the
running of the viscosity and resistivity as should be clear from
Eq. (\ref{bigC}), and this is accounted for by the exponent $z$.  

Substituting Eq. (\ref{fimp}) into 
Eqs. (\ref{individualv}), (\ref{individualB}), and (\ref{individualvB})
gives the complete structure of the correlation functions correct at one-loop
order. Note that the net exponent 
$-4 + \frac{2}{3}(4 + y - d) - y \equiv -d -2\chi -z$, so 
that Eqs. (\ref{individualv}) and (\ref{individualB}) 
reproduce the same scaling
(for $h_B = 0$, i.e. zero helicity) as we found in Eqs. (\ref{pipi}) and
(\ref{pipsi}). However, we also explicitly obtain the associated
scaling functions. We see that the presence of random magnetic helicity
modifies these correlations, since the helical contribution scales
as $k^{y-w+1}$ {\it relative} to the non-helical part.
{}Furthermore, the relative
strength of this term goes as the ratio of the amplitudes of the helical 
to the non-helical fluctuations: $h_B/(A+B), h_B/(A-B)$.

\section{The Callan-Symanzik equation}
\label{sect7}

In this section, we derive the scaling behavior of the correlation functions 
in the Fourier domain from
a distinct but entirely complementary point-of-view provided by the methods of field theory.
In this approach, the renormalization of the ultraviolet 
(short-distance) divergences
of the stochastic field theory for MHD formulated in Appendix A leads
to certain partial differential equations, frequently 
referred to as Callan-Symanzik
equations\footnote{Equations of this type are typically referred to as
Callan-Symanzik equations, though strictly speaking, this name refers
to a specific type of renormalization group equation resulting from so-called
mass-subtraction (in quantum field theory). 
A mass-independent subtraction leads to a different
equation, similar to the one actually derived here, and might be more
properly called a t'Hooft-Weinberg equation. A clear distinction
between Gell-Mann-Low, Callan-Symanzik and t'Hooft-Weinberg renormalization
group equations is drawn by (Gross 1981).} 
for the one-particle
irreducible (1PI) Green functions.  The exact solutions 
of these equations provide
considerable information that goes well beyond perturbation theory. 
These solutions tell us how the 1PI functions scale near
the fixed points
of the theory and permit one to quantify the approach to the fixed
points. Moreover, the concept and mechanics of RG-improvement is given
a rigorous basis by means of the Callan-Symanzik equation. 
We therefore devote some space in the present section to 
an application of these ideas to stochastic MHD field theory. 
In what follows, we consider non-helical MHD. The extension to the helical
case is straightforward but will not be carried out here.

To begin, we first tabulate the 
canonical dimensions (in units of momentum) of all the parameters
and fields 
appearing in the
generating functional for MHD. These are easily obtained by noting that the
action ${\cal A}$ in the dynamic functional, 
\begin{equation}
Z = \int [{\cal D}\vec \sigma][{\cal D}\bf \Phi]\exp ({\cal A}) ,
\end{equation}
must be dimensionless: $[{\cal A}] = 0.$ 
We recall that we have employed a short-hand notation for the pair
of Elsasser fields: 
${\bf \Phi} = (\vec P, \vec Q)$, and the conjugate fields
$\vec \sigma$ are defined in Appendix A.
We can read off the explicit structure
of the action $\cal A$ directly from Eqs. (\ref{full}), (\ref{firstpass}),
(\ref{inverseresponse}) and (\ref{covariance}).
{}From this zero-dimensionality condition we therefore find 
that (where $[X] = d_X$ denotes the dimension of $X$ in units of momentum)
\begin{equation}\label{paramdims}
[\gamma_{\pm}] = d_{\gamma}, \qquad [\omega] = d_{\omega} = 2 + d_{\gamma}, 
\end{equation}
\begin{equation}
[A] = [B] = d_A = d_B, \qquad
[\stackrel{\leftrightarrow}{\Gamma}] = d_{\Gamma} = d_A - y,
\end{equation}
\begin{equation}
[\lambda] = d_{\lambda} = \frac{(4-d+y)}{2} +\frac{3}{2}d_{\gamma} -
\frac{1}{2}d_A, \qquad [g_{\pm}] = 0.
\end{equation}
The Elsasser and conjugate fields have dimensions
\begin{equation}
[{\bf \Phi}] = d_{\Phi} = -3 -\frac{d}{2} -\frac{3}{2}d_{\gamma}
+ \frac{1}{2}d_{\Gamma}, \qquad 
[\vec \sigma] = d_{\sigma} = -1 -\frac{d}{2} -\frac{1}{2}d_{\gamma} -
\frac{1}{2}d_{\Gamma},
\end{equation} 
respectively. Note that the canonical dimension
of the fields depends on the noise exponent $y$ since 
$d_{\Gamma} = d_A - y$.
{}Finally, the two 1PI functions of particular interest to us are the
response function and the noise spectral function. 
The response function
has dimension
\begin{equation}\label{11}
[\stackrel{\leftrightarrow}{\Gamma_{11}}] = d_{11} = 
[(\stackrel{\leftrightarrow}{\bf G}^{-1}_0)_{ij}] =
(2 + d_{\gamma}),
\end{equation}
while the non-helical part of the noise spectral function has dimension
\begin{equation}\label{02}
[\stackrel{\leftrightarrow}{\Gamma_{02}}] = d_{02}
= [(\stackrel{\leftrightarrow}{\bf \Gamma})_{ij}] =
d_A - y.
\end{equation}
We remark that it is 
customary to denote the 1PI Green functions by a capital $\Gamma$, which is not to be
confused with the block array of noise correlation functions. These 1PI functions
are built up out of certain numbers of conjugate fields and physical fields: the
indices on the first two terms 
on left hand side of Eq. (\ref{11}) indicate that one (1) 
conjugate and one (1) physical
field are involved, while in Eq. (\ref{02}), no conjugate field (0) and two (2) physical
fields are involved. We suppress the latin indices here to avoid clutter.   
Note all canonical dimensions can be expressed in terms of the dimension of the 
viscosity (and resistivity)
$d_{\gamma}$, the noise amplitude $d_A=d_B$, and
exponent $y$.

The one-loop corrections to the MHD 
parameters calculated in the Appendices 
have been cut-off 
with a momentum or wavenumber regulator $\Lambda$.
Recall, this actually is a physical cut-off 
in momentum.  It is associated with the
smallest length scale below which the continuum description of the
fluid breaks down and must be replaced by
an atomic/molecular 
description.
If this short-distance 
cutoff is taken to zero ($\Lambda^{-1} \rightarrow 0$), 
i.e. take $\Lambda \rightarrow
\infty$, it would result in
short distance or ultraviolet 
divergences showing up in the one-loop response and correlation
functions that are calculated in the Appendices. 
To systematically remove these ultraviolet divergences,
renormalization 
constants would have to be introduced in the parameters and fields that
appear in the above action $\cal A$. 
This may be implemented by 
introducing renormalized parameters (denoted by $Z_a, Z_{\Phi}, Z_{\sigma}$)
and renormalized fields (denoted by an $R$) as 
(Gross 1981; Amit 1978)
\begin{eqnarray}
\xi_a &=& Z^{-1}_a \, \xi^R_a, \\
\label{zedphi}
{\bf \Phi} &=& Z_{\Phi}^{1/2} \, {\bf \Phi}^R, \\
\label{zedsigma}
{\vec \sigma} &=& Z_{\sigma}^{1/2} \, {\vec \sigma}^R .
\end{eqnarray} 
Here the symbol $\xi_a$ is a shorthand notation denoting the collection
of all parameters that appear in the
dynamical equations, $\xi_a=(A,B,\gamma_{+},\gamma_{-},\lambda)$, with the 
label running from
$a =1,..,5$. We will be able to draw some general conclusions about
the asymptotic properties of the correlation functions without having to
explicitly calculate these renormalization parameters.

In the following, we first consider an arbitrary 1PI function, and then specialize
to our response and noise functions to obtain information regarding the
correlation function. 
Keep in mind that for the MHD theory considered here, 
these functions
all have the structure of two-by-two block arrays 
multiplied by a $d$-by-$d$ tensor
factor. We suppress the tensor indices in what follows.  
The renormalization of the 1PI functions containing $N$ factors
of the physical field and $\tilde N$ factors of the conjugate field is given by
\begin{equation}\label{renorm1}
\stackrel{\leftrightarrow}{\Gamma}^R_{N,\tilde N}(\vec k, \omega; \{\xi^R_a\},g;\mu) =
Z^{N/2}_{\Phi}\,Z^{\tilde N/2}_{\sigma} 
\stackrel{\leftrightarrow}{\Gamma}_{N,\tilde N}(\vec k, \omega; \{\xi_a\},g_0;\Lambda).
\end{equation}
In arriving at this expression, we have expressed the bare fields
in terms of the 
renormalized fields by means of Eqs. (\ref{zedphi}) 
and (\ref{zedsigma}).
The label $R$ denotes a renormalized quantity. 
Most importantly, the renormalization requires introducing a finite but
arbitrary
momentum scale $\mu$, which remains in the finite, renormalized expressions. 
As it stands, this equation merely expresses the fact that the function in
question is renormalizable.  In other words, 
the UV divergences can be consistently
subtracted out. This subtraction is done at some arbitrarily chosen
scale $\mu$, and this fact must be recorded in 
the resulting finite renormalized
function. 
Using the fact that the bare, unrenormalized 1PI function does {\it not} depend on this
arbitrary momentum scale $\mu$ (it depends only on the 
ultraviolet cut-off $\Lambda$), 
we can easily derive a differential equation for the
renormalized 1PI function as follows:
\begin{equation}
\Big( \mu \frac{d}{d\mu}\Big)_0 
\stackrel{\leftrightarrow}{\Gamma}_{N,\tilde N}(\vec k, \omega; \{\xi_a\},g_0;\Lambda)
= 0 .
\end{equation}
This implies (see footnote 1)
\begin{equation}\label{tHW}
\Big(\mu \frac{\partial}{\partial \mu} + \beta(g) \frac{\partial}{\partial g}
+ \sum_a \Delta_a \xi_a^R \frac{\partial }{\partial \xi^R_a} 
-\frac{N}{2} \gamma -\frac{\tilde N}{2} {\tilde \gamma} \Big)
\stackrel{\leftrightarrow}{\Gamma}^R_{N,\tilde N}(\vec k, \omega; \{\xi^R_a\},g;\mu) = 0,
\end{equation}
where the coefficient functions appearing in this differential equation are defined by
\begin{eqnarray}\label{Delta}
\Delta_a &=& \big(\mu \frac{\partial \ln Z_a}{\partial \mu}\big)_0, \\
\label{beta}
\beta(g) &=& \big(\mu \frac{\partial g}{\partial \mu}\big)_0, \\
\label{gamma}
\gamma  &=& \big(\mu \frac{\partial \ln Z_{\Phi}}{\partial \mu}\big)_0, \,\,\,
{\tilde \gamma}  = \big(\mu \frac{\partial 
\ln Z_{\sigma}}{\partial \mu}\big)_0.
\end{eqnarray}
The zero-subscript (0) indicates the derivatives 
are taken while holding fixed the bare parameters
This equation Eq. (\ref{tHW}) indicates that a change in the arbitrary
but finite 
subtraction scale $\mu$ 
is compensated for by corresponding changes in the
physical parameters.  
As Eq. (\ref{tHW}) is a linear first-order partial differential
equation, it can be readily solved by 
means of characteristics (Amit 1978). The 
exact solution is given by
\begin{equation}\label{solution}
\stackrel{\leftrightarrow}{\Gamma}^R_{N,\tilde N}
(\vec k, \omega; \{\xi^R_a\},g;\mu) = 
\exp\Big(- \int_1^b \frac{du}{u} [\frac{N}{2}\gamma(u) + \frac{\tilde N}{2}
{\tilde \gamma}(u)] \Big)\,
\stackrel{\leftrightarrow}{\Gamma}^R_{N,\tilde N}
(\vec k, \omega; \{\xi^R_a(b)\},g(b); b\mu),
\end{equation}
where the equations for the characteristics are 
\begin{eqnarray}
\label{beta2}
\beta(g(b)) &=& b\frac{\partial g(b)}{\partial b}, \\
\label{params}
\Delta_a(g(b)) &=& b\frac{\partial \ln \xi_a^R(b)}{\partial b}.
\end{eqnarray}
These equations are subject to the boundary conditions at $b=1$,  
$g(1) = g$ and $\xi^R_a(1) = \xi^R_a$.
The solution 
can be readily verified by differentiating 
Eq. (\ref{solution}) 
on both sides with respect to $b$ and using the chain rule.  
The set of characteristics define a ruled (hyper) 
surface on which $\stackrel{\leftrightarrow}{\Gamma}^R_{N\tilde N}$
is defined 
and $b$ parametrizes the characteristic curves on this surface. 
It is related to the RG flow parameter via $b = e^{-\ell}$.  
This dimensionless parameter $b$ can be related to 
ratios of momentum (or length) scales, as we will
see below.  
The physical content of the solution Eq. (\ref{solution}) 
is as follows.  A change in momentum scale
(as measured by $b$) is accompanied by a change in the coupling $g(b)$ and
in the other parameters $\xi^R_a(b)$.  
{}Furthermore, the vertex function picks up
an additional overall exponential 
factor depending on the two anomalous dimensions $\gamma$ and
$\tilde \gamma$ that are defined and calculated in Eq. (\ref{gamma}). 
In particular, we see that the anomalous dimensions depend
on the renormalization of the physical and conjugate fields
(wavefunction renormalization). These dimensions control in turn,
the approach to the fixed point, as we will see below.
In a nutshell, this equation relates the 1PI function at two different
scales. We now use this fact to deduce how the correlation functions
depend on scale. 

Define dimensionless scaling functions ($d_{N\tilde N}$ is the canonical
dimension of the 1PI function) 
$\stackrel{\leftrightarrow}{F}_{N \tilde N}$ as
\begin{equation}
\stackrel{\leftrightarrow}{\Gamma}^R_{N\tilde N}(\vec k,\omega;\{\xi^R_a\},g;\mu) =
\mu^{d_{N\tilde N}}\, 
\stackrel{\leftrightarrow}{F}_{N\tilde N}\Big(\frac{k}{\mu}, 
\frac{\omega}{\mu^{2+d_{\gamma}}},
\{ \frac{\xi^R_a}{\mu^{d_{\xi_a}}} \}; g
\Big),
\end{equation}
so that the dimension of 
$\stackrel{\leftrightarrow}{\Gamma}^R_{N\tilde N}$ 
is carried entirely by an appropriate power of
$\mu$. 
The individual arguments of $\stackrel{\leftrightarrow}{F}_{N\tilde N}$
are rendered dimensionless since each is expressed as a 
dimensionless quotient, by making use of the dimensions listed
above in Eq. (\ref{paramdims}). Here, $d_{\xi_a}$ denotes one of
$d_A,d_B,d_{\lambda},d_{\gamma_+},d_{\gamma_{-}}$. 
Then from the solution Eq. (\ref{solution})
\begin{eqnarray}\label{scaledsolution}
\stackrel{\leftrightarrow}{\Gamma}^R_{N,\tilde N}(\vec k, \omega; \{\xi^R_a\},g;\mu) 
&=& \big(b\mu \big)^{d_{N\tilde N}}\,
\exp\Big(-\int^b_1 \frac{dx}{x}[\frac{N}{2}\gamma(x) + \frac{\tilde N}{2}
\tilde \gamma(x)] \Big)\, \nonumber \\
&\times& \stackrel{\leftrightarrow}{F}_{N\tilde N}\Big(\frac{k}{b\mu}, 
\frac{\omega}{(b\mu)^{2+d_{\gamma}}},
\{ \frac{\xi^R_a(b)}{(b\mu)^{d_{\xi_a}}} \}; g(b)
\Big). 
\end{eqnarray}
At this point, we have succeeded in writing each renormalized
1PI Green function as a product involving a dimensionful
power of the momentum scale $\mu$ times an exponential factor
times a dimensionless scaling function.
Now, the correlation function for the 
Elsasser fields $P, Q$, which recall is {\it not}
1PI,  
is  given by 
(see Eq. (\ref{correlation})) the block-matrix product
$\stackrel{\leftrightarrow}{C} = (\stackrel{\leftrightarrow}{\Gamma}_{11})^{-1} 
\stackrel{\leftrightarrow}{\Gamma}_{02} 
(\stackrel{\leftrightarrow}{\Gamma}_{11})^{-1}$.
Thus, from Eq. (\ref{scaledsolution})
\begin{equation}\label{general}
\stackrel{\leftrightarrow}{C}(\vec k,\omega;\{ \xi^R_a \},g;\mu) 
= \big( b\mu \big)^{d_A - y -2d_{\gamma} - 4}\,
\exp\Big(+\int_1^b \frac{dx}{x} \, \gamma(x) \Big)
{\hat C} 
\Big(\frac{k}{b\mu}, \frac{\omega}{(b\mu)^{2+d_{\gamma}}},
\{ \frac{\xi^R_a(b)}{(b\mu)^{d_{\xi_a}}} \}; g(b)
\Big), 
\end{equation}
where ${\hat C} = (\stackrel{\leftrightarrow}{F}_{11})^{-1} 
\stackrel{\leftrightarrow}{F}_{02} 
(\stackrel{\leftrightarrow}{F}_{11})^{-1}$ 
and from Eqs. (\ref{11}) and (\ref{02})
$d_{02}-2d_{11} = d_A - y - 4 - 2d_{\gamma}$. 
Recall that $\hat C$ is a $2\times 2$ block array with a pair
of $d$-dimensional indices. 
Note that in Eq. (\ref{general}) 
the dependence on the conjugate field anomalous
dimension $\tilde \gamma$ has dropped out.

It is worth commenting that provided the 
characteristic equations Eqs. (\ref{beta2}) and (\ref{params})
are solved, Eq. (\ref{general}) yields the exact solution for the correlation
functions in terms of the parameter $b$ which is a measure of the momentum
scale (or inverse length scale) at which one observes the system. 
The exact solution is known if
explicit knowledge is available of the anomalous dimension
$\gamma(x)$ and the solutions of the above characteristic equations.
However, typically these are known only up to some order in perturbation
theory, which usually is to a given number of loops. Thus, the
solution of the Callan-Symanzik equation is only known to the
same number of loops. The important utility of the solution lies in the
fact that it tells us how quantities behave away from
the fixed point and it places the results obtained from
naive scaling arguments on a solid footing.
In general,
the beta function Eq. (\ref{beta}) will not vanish for 
arbitrary $b$ (i.e., far away from fixed points) nor will the coupling 
$g(b)$ be constant.  As such, the
expression Eq. (\ref{general}) 
tells how the fields are correlated in general, that is, everywhere in
parameter space, and not just in the vicinity of the fixed points.    
Of course, 
if the system happens to be at a 
fixed point, $g \rightarrow g^* = {\rm const.}$,
$\gamma \rightarrow \gamma^* = {\rm const.}$, etc.
Choosing $b = k/\mu$ and restoring the block labels 
and tensor indices gives for the $PP$ 
entry of the block array
\begin{equation}\label{improvedc}
C_{P_j P_n}(\vec k,\omega) = k^{d_A-y-2d_{\gamma} - 4}\, 
\Big(\frac{k}{\mu}\Big)^{\gamma^*}\,
{\bf P}_{jn}(\vec k)\,
{\hat C}_{P P} \Big(\frac{\omega}{k^{2+d_{\gamma}}},
\{ \frac{\xi^R_a(k)}{k^{d_{\xi_a}}} \}; g^*
\Big),
\end{equation}
and similarly
for the $PQ$ and $QQ$ entries.
At a fixed point, i.e. $\Delta_a^* = \Delta_a(g^*)$,
\begin{equation}\label{runxi}
\xi_a^R (b) = \xi_a^R (1)\, b^{\Delta_a^*},
\qquad \longrightarrow 
\xi_a^R (k) = \xi_a^R (1)\, \big(\frac{k}{\mu} \big)^{\Delta_a^*},
\end{equation}
which follows from evaluating Eq. (\ref{params}) at a fixed point. 
In particular we see that the correlation function, calculated
up to $n$-loops, is given by the free correlation function but
expressed in terms of the scale dependent (running) couplings
and parameters (calculated up to $n$-loop order). This is 
nothing but the prescription leading to RG-improved 
quantities, but here it is a natural and automatic property of the
solutions of the Callan-Symanzik equation.

Comparing (\ref{improvedc})
to Eqs. (\ref{pipi}) and (\ref{pipsi}) yields the exponent
relations
\begin{equation}\label{compare}
z = 2 + d_{\gamma}, \qquad \chi = 1 - \frac{d}{2} + \frac{1}{2}\big(
d_{\gamma} - d_A + y - \gamma^*).
\end{equation}
These relations allow the various ``engineering'' 
dimensions to be related with the
fixed point exponents calculated earlier. Moreover, we see how the concept
of RG-improvement used throughout this 
paper appears as a rigorous feature of the solutions
of the Callan-Symanzik equation. The equality in Eq. (\ref{improvedc}) 
shows that the correlation function in the vicinity of the 
fixed point
is a power law multiplied by the scaling function. {}Furthermore, the
scaling function
is evaluated with the
running or scale-dependent values of the parameters,
that are computed
in the neighborhood of that fixed point, as exactly 
indicated by Eq. (\ref{runxi}).
Note that if there is any nontrivial wavefunction 
renormalization (indicated by a $\gamma* \neq 0$), it
is automatically subsumed in the exponent $\chi$ as shown in
Eq. (\ref{compare}). 
Since we have
calculated both $z$ and $\chi$ by independent means, there
is no need for a separate calculation of $\gamma^*$ if we
are interested only in what is happening at the fixed point. 

\subsection{Approach to the fixed point}
\label{sect7A}

Observe that the exponent (also called the anomalous dimension) $\gamma(g)$
controls the approach to the scaling regime, which is reflected in the pure power law
behavior of the correlation function 
at the fixed point. In general, it is unlikely the
dynamical system will be sitting exactly at the fixed point.
Therefore one should
study the form of the correlation function 
for values of the coupling that are
in the basin of attraction of the fixed point and 
see how this function depends on
arbitrary initial values of the coupling. 

This can be done systematically by returning to the exponential factor in
Eq. (\ref{general}) and carefully 
expanding this about the (non-trivial) fixed point.  
{}First make a convenient change of variables as
\begin{equation}
\int^b_1 \frac{dx}{x} \gamma(g(x)) = 
\int_0^l dl \,\gamma(g(l)) = \int^{g(l)}_{g(0)}
\frac{\gamma(y)}{\beta(y)} dy,
\end{equation}
where $b = e^l$. Note, this $l$ is not the 
same as the RG flow parameter $\ell$ in Eq. (\ref{RGE1}), but they
are related as $-\ln(b) = \ell$ (Frey and Tauber 1994). 
As already
pointed out, $b$ parametrizes the sense of ``flow'' along the characteristic
curves that arise in the solution of the CS equation and does not
automatically single out either the infrared (decreasing $b$) or 
the ultraviolet (increasing $b$) limit. The
connection to these physical limits is made once we assign infrared and ultraviolet
flow ``directions'' to the characteristics.  
The one-loop beta function can be read off directly from 
Eq. (\ref{RGE2}) and is given by $\beta(g) = 12 g(g^* - g)$.
Solving for the coupling by integrating 
the differential equation Eq. (\ref{RGE2}) gives
\begin{equation}
\frac{g(\ell)}{g^*} = \frac{\frac{g(0)}{g^*-g(0)}\exp(\epsilon \ell)}{
1 + \frac{g(0)}{g^*-g(0)}\exp(\epsilon \ell)}, 
\end{equation}
where $\epsilon = 4-d+y > 0$.
The infrared stable non-trivial fixed point is reached by taking
$b = \frac{k}{\mu} \rightarrow 0$ or $l \rightarrow
-\infty$. In terms of the Wilsonian flow parameter $\ell$, this corresponds to the limit
$\ell \rightarrow \infty$.  
Note this fixed point is reached for 
all initial values $g(0)$ of the coupling.  

Next, Taylor expand the anomalous dimension about the fixed point as
\begin{equation}
\gamma(g) = \gamma^* + \gamma' (g - g^*) + O(g - g^*)^2 .
\end{equation}
With this,  evaluating the integral
\begin{equation}
\int^{g(l)}_{g(0)}
\frac{\gamma(g)}{\beta(g)} dg = \gamma^* l -
\frac{\gamma'}{12} \ln\Big(\frac{g(l)}{g(0)}\Big),
\end{equation} 
it implies the exponential factor 
contributing to the correlation function Eq. (\ref{general}) works out
to be
\begin{eqnarray}
\exp \Big(\int_1^b \frac{dx}{x}\, \gamma(g(x)) \Big) &=& 
b^{\gamma^*}\, \Big(\frac{g(b)}{g(1)}
\Big)^{-\frac{\gamma'}{12}},\nonumber \\
&=& \big(\frac{k}{\mu}\big)^{\gamma^*}\,
\Big( \frac{ \frac{g^*}{g^* - g}\big(\frac{k}{\mu}\big)^{\epsilon}}{
1 + \frac{g}{g^* - g}
\big(\frac{k}{\mu}\big)^{\epsilon}} \Big)^{-\frac{\gamma'}{12}} .
\end{eqnarray}
Recall that the $k$-dependent factor with exponent $\gamma^*$ is already
subsumed into the definition of the fixed point ``roughness'' exponent
$\chi$, as indicated in Eq. (\ref{compare}). However the second term, which
depends on the exponents $\gamma'$ and $\epsilon$,  
clearly is a correction to the
fixed point (power law) form of the correlation function written
in Eq. (\ref{improvedc}). Depending 
on the algebraic sign of $\gamma'$, this extra term, which
controls the approach to the fixed point,  can either
augment or diminish the energy spectrum with respect to its
asymptotic fixed point value Eq. (\ref{latetime}). 
Similar consideration holds 
at the nontrivial fixed point
for the asymptotic
form of the helicity spectrum in Eq. (\ref{nthelsp}).

Some comments are in order about the region away from the
fixed point. {}First, any amplification
or diminution in the spectra densities with respect to their
steady state fixed point values, which is
brought about by this extra term, is a statement about
how the densities evolve as they approach the nontrivial
fixed point.
It is a conceivable possibility that numerical simulations,
due to their spatial and temporal limitations, actually may be
seeing some sufficiently long-lived non-steady state of this type.
This point will
be elaborated upon in Sec. \ref{concl}.
However, we hasten to point out that if the
wavefunction renormalization constant turns out to be equal to
unity, $Z_{\Phi} = 1$, then the anomalous dimension 
Eq. (\ref{gamma}) is identically
zero, $\gamma = 0$. 
When this is the case, then the expontial factor in 
(\ref{general}) reduces to unity, and any corrections
to pure power law scaling will be due entirely to the
scaling function itself. 
{}For incompressible hydrodynamics subject to
random forces, it is known that $Z=1$ 
(DeDominicis and Martin 1979). 
This in 
turn can be
shown to be a consequence of the underlying Galilean invariance of 
the Navier-Stokes equation, which is maintained for all spatially
correlated random forces. Since the equations of nonrelativistic MHD
are also Galilean invariant, we suspect that the  
anomalous dimension Eq. (\ref{gamma}) will also be vanishing. 
This issue will not be discussed further in this paper.

\section{Cascade Directions}
\label{sect8}

One important question of general interest in the study
of hydrodynamic systems is how the nonlinear terms
redistribute energy and, if present, helicity amongst
the various scales.  In particular, if energy and/or helicity
is introduced into the system at a given scale, it then will
be distributed by the subsequent evolution of the hydrodynamic
or MHD equations, either to smaller ({\it direct or normal cascade})
or larger ({\it inverse cascade}) length scales.
In general, this discussion of cascades is restricted
to the inertial range, which corresponds to
length scales sufficiently large that they are
not significantly dissipated by the viscosity term.

It is interesting to reflect on the history of the subject of cascades.
The earliest expectation, based on simulations
and general arguments of the Navier-Stokes equation
in $d=3$, was that a direct cascade (also suggestively called
a normal cascade) would occur
from hydrodynamic evolution.  Subsequently in 
(Kraichnan 1967),
for the Navier-Stokes equation in $d=2$, based on general
observations about the two conserved constants of motion,
enstrophy and kinetic energy per unit mass, it was 
argued that an inverse cascade of energy should occur.
{}For MHD, these arguments of inverse cascade are not applicable
in either $d=2,3$ and so a direct cascade of energy is expected
(Biskamp 1993; Kraichnan 1973).  
On the other hand for MHD,
considerations based on conservation
of magnetic helicity in $d=3$ and its
counterpart, the square of the magnetic potential in
$d=2$ showed that an inverse cascade is expected
of the respective quantities (Frisch et al. 1975).
All the above expectations are based on compelling,
but not rigorous, arguments.  Verification
of these expectations mainly have come from
computer simulations directly of the hydrodynamic/MHD
equations or indirectly through some approximations
to these equations 
(Brandenburg et al. 1996; Lilly 1969; Frisch and Sulem 1984;
Sommeria 1986; Pouquet et al. 1976; Pouquet and Patterson 1978;
Pouquet 1978).
Clearly this approach has its
limitations in the interesting regime of high Reynolds
number, due to finite simulation time and system size restrictions.

Returning from this digression back to our RG analysis,
cascades clearly are associated with time dependent processes.
On the other hand, the RG analysis in this paper examined
steady state solutions of MHD with time translationally
invariant force terms and
only treated the leading asymptotic behavior
at large distance.  From this analysis, information
about the energy spectrum can be obtained for the largest
length scale.  In particular, this means length scales
beyond the largest characteristic scale at which energy
is input from the external force.  As such, the RG analysis
of this paper can determine presence or absence of inverse
cascade, depending whether the infra-red  limit of
the energy spectrum is enhanced or not.  This is a useful
application for RG, since generally some direct cascade always
occurs and the interesting question is whether there also
is some inverse cascade.

In order to extract information
about cascades from this framework, a few points
must be appreciated. 
The force term is continuosly inputing
energy and/or helicity into the system with some prescribed
spectrum.  A cascade for such a case therefore is associated
with the difference in the steady state magnetic
and velocity spectra directly created with 
the force (in the absence of mode mixing) versus
the steady state spectra that emerges from
the subsequent MHD evolution.  {}For example,
if the spectra associated with the force is less enhanced
in the ultraviolet versus the resulting spectra after MHD evolution,
then the MHD dynamics is directing energy flow to smaller length
scales thus exhibiting a direct cascade.  Similarily,
if the input spectra associated with the force is less enhanced
in the infrared versus the resulting spectra from MHD,
then MHD is exhibiting an inverse cascade.  

To proceed with this approach, the implied magnetic and velocity
spectra associated with the force terms must be determined.
The point here is Eqs. (\ref{vnoise}) - (\ref{vBnoise})
give the spectra of the
force, but what we seek is the associated spectra this implies
for the magnetic and velocity fields.  To determine these,
MHD with the force terms, but without the nonlinear terms,
must be examined.  These linearized MHD equations express the
corresponding magnetic and velocity spectra for given input force
spectrum, before there is any mode mixing.  Equivalently stated,
these are the associated input magnetic and velocity spectra
for the given force, prior to any nonlinear MHD evolution.

The procedure then for determining information about cascades
is to compare for a given force term 
the corresponding magnetic and velocity
spectra from the linearized MHD equations
versus those from the full nonlinear MHD equations.
Depending on how the ultraviolet and infrared behavior
of the latter changes compared to the former, cascade
directions can be deduced.

{}Following this criteria, let us determine the cascade
directions predicted at the largest length scales
from our RG analysis.  {}First note that the spectra for 
linearized MHD equivalently stated is the one also for the
trivial fixed point.  Therefore one fact immediately follows
from this criteria.  In the region of stability of the
trivial fixed point, the RG analysis predicts no cascade
whatsoever, at the largest length scales.

Turning next to energy spectra in MHD, 
from Eqs. (\ref{spectralE2}) and (\ref{enexp})
$E(k) \sim k^{\alpha}$ with
$\alpha_{\rm linear} = -3+d-y$ and 
$\alpha_{\rm nontrivial} = -5/3+(2/3)(d-y)$, so that
$\alpha_{\rm linear} \leq (\geq) \alpha_{\rm nontrivial}$
in the region $4-d+y \geq (\leq) 0$.
Thus we see that in the stability region of the
nontrivial fixed point, the input energy
spectra $\sim k^{\alpha_{\rm linear}}$ is more enhanced
in the infrared and less in the ultraviolet
versus the resulting spectra from MHD evolution
$k^{\alpha_{\rm nontrivial}}$.  Thus our RG analysis
of MHD predicts direct energy cascades at the largest length scales.
Since the above case involves the entire stability region
of the nontrivial fixed point, this analysis finds
at the largest length scales,
no prediction of an inverse cascade of energy
in MHD for any $d$.

The RG calculations in this paper easily can be reduced
from MHD to hydrodynamics by identifying the fields
$P=Q$.  The outcome of this is the predictions of the fixed
points for the hydrodynamic case are found to be exactly
the same as given in Sec. IV.  As such,
our analysis also expects direct but no inverse energy cascades
at the largest length scales in hydrodynamics for any $d$.

It is interesting to note that the change from direct cascade
to inverse cascade in the RG analysis occurs right at the
boundary of stability for the nontrivial fixed point.
In particular the unstable region of the nontrivial fixed point
would predict inverse cascade.  In the Conclusion we will
discuss interpretations of this observation.

The same analysis as above also can be done for
helicity. The helicity spectra given in Subsect. \ref{sect6A} are
${\cal H}_M(k) \sim k^\beta$ with
$\beta_{\rm linear} = -w$ from Eq. (\ref{frhelsp}) and 
$\beta_{\rm nontrivial} = -w+(1/3)(1+y)$ from Eq. (\ref{nthelsp}).
The region of direct helicity cascade is
$\beta_{\rm linear} < \beta_{\rm nontrivial}$,
which implies $1+y > 0$. Since this is for $d=3$,
note this last condition also corresponds to
the stability region of the nontrivial 
fixed point, i.e., $4-d+y = 1+y$. 
So once again, RG predicts a direct 
magnetic helicity cascade in the entire
stability region of the nontrivial fixed point
and no cascade in the stability region of the trivial
fixed point.  {}Furthermore, once again the change 
from absence to presence of
inverse cascade for the nontrivial fixed point
is at its boundary of stability.

\section{Conclusion}
\label{concl}

In this paper a Renormalization Group (RG) analysis has been performed
for MHD with a stochastic force term.  Such an analysis enables
the study of highly turbulent MHD in the asymptotic regime
of large length and time scales.  In the format of the RG approach,
the central problem is to determine the effective MHD equations
at large scales that emerge once all short distance
physics is integrated out of the initial fundamental 
MHD equations.  The analysis was carried out for a general class of force
functions Eq. (\ref{vnoise}) - (\ref{vBnoise})
and for arbitrary spatial dimension $d$.
The outcome of this analysis, as detailed in 
Sec. \ref{sect4}, showed
that MHD for this class of force functions, has
two types of asymptotic behavior, depending on
$d$ and the nature of the force.  In one regime, the system
becomes a free theory in the asymptotic or hydrodynamic limit
which
is represented by a trivial fixed point in Sec. \ref{sect4} , 
and in the other regime
the system is a nonlinear interacting theory at
the largest scales, which is represented by the nontrivial fixed
point in Sec. \ref{sect4}.  
{}For both regimes, correlations were computed
in Subsec. \ref{sect5B} 
of the magnetic and velocity fields at large length and time
scales and in Subsec. \ref{sect5C} of the energy spectrum . 
In addition, the viscosity coefficients in the
effective long distance theory were computed in Subsec. \ref{sect5A}.
Helical force also was treated in Sec. \ref{sect6}.
Up to the one-loop
analysis in this paper, the fixed point structure
was unchanged with respect to that in Sec. \ref{sect4}. 
The helicity spectrum at both
fixed points was given in Subsec. \ref{sect6A}. 
It would be interesting
to test the RG predictions against simulations for all these quantities.

Another interesting direction examined was mixing between
velocity and magnetic energy at large scales.  The results
of the one-loop analysis in Sec. \ref{sect5C} indicate that in fact
at the largest scales magnetic and velocity energy
will not mix.  Thus, for example, if only the velocity force was nonzero,
then no magnetic energy would be created at the largest scales and 
visa-versa.  These results of course are only concerned with
the asymptotic large scales.  Clearly the nonlinear interactions
in the MHD equations generally should mix velocity and magnetic
energy and this is seen in simulations 
(Pouquet et al. 1976; Pouquet and Patterson 1978),
which suggest an approximate equipartition of energy.
However, the RG analysis indicates that at increasingly
larger length scales this mixing diminishes.
It would be interesting to see if such trends also are
seen in simulations.

The RG analysis of this paper addressed the steady state solutions
of MHD with time translationally invariant stochastic force terms.
As shown in Sec. \ref{sect8}, 
information about cascade directions also can be deduced
from this framework. To our knowledge, this is the first
time the RG has been used to predict cascade 
directions in turbulence.  
In that section, we examined directions
for energy cascade in both hydrodynamics and MHD and helicity
cascade in MHD.  
In the stability region of the trivial fixed point,
in all cases and in any $d$ the one loop
RG analysis showed no cascade at asymptotic scales.
On the other hand, in the stability region of the
nontrivial fixed point, the RG analysis predicts for the energy
in all cases and for all $d$, direct or normal cascades and
for helicity in MHD also a direct cascade.
{}For energy cascades in $d=2,3$ for MHD and $d=3$ for hydrodynamics,
the results of the RG analysis found here are consistent with general
expectations.  However, for energy cascade in $d=2$
hydrodynamics and helicity cascade in $d=2,3$ MHD, 
some general arguments (Kraichnan 1967; Frisch et al. 1975) 
and simulations (Lilly 1969; Frisch and Sulem 1984; Sommeria 1986;
Pouquet at al. 1976; Pouquet and Patterson 1978; Pouquet 1978)
suggest an inverse cascade.  There are no rigorous
proofs in any of these cases. Nevertheless, the cases where the
RG analysis differs from conventional
expectations deserves further discussion.
Below we identify some possible sources for the discrepancy
and suggest ways they might be resolved. However first, let us consider
the random force terms.
The class of random force functions we consider are
reasonably general in terms of covering a broad range of scales,
and we do not believe the special form of the force is the main
cause for this discrepancy. These have been taken to be power
laws in wavenumber, characterized by a free exponent $y$ that can either be
positive, zero or negative. This is enough to cover most cases of spatially
correlated noise and white uncorrelated noise (if $y=0$).

Returning to the discrepancies, there are several
possible reasons that come to mind for these disagreements between our
RG analysis versus the various simulations and general
arguments. We will
first list them and
and then address each in turn.  The main possibilities are: 
(a) intrinsic limitations of perturbative RG applied to MHD; 
(b) limitations of the one-loop RG analysis which may
be corrected by higher loop calculations;
(c) the RG analysis is correct and the inverse cascades
suggested by simulations are a metastable effect in the
unstable region of the  nontrivial fixed point.     

Concerning point (a), we remark that it has long been known and criticized
that since the {\it formal} (bare and unrenormalized) 
perturbation 
expansion parameter $\lambda_0 = 1$ in the 
Navier-Stokes equation is equal to 
unity, 
the (primitive) perturbation series based upon it are 
of dubious utility (see, e.g. discussion in 
McComb 1995,  and the
references therein). 
Analogous primitive perturbation series constructed for 
MHD would not escape this criticism either.
In response,
we point out that in the {\it loop} expansion carried out here, 
we have defined and expanded our perturbation series in powers
of the amplitude of the noise fluctuations (which could be large or small). 
This is because the noise amplitude serves to count the 
number of loops in the series (represented by a graphical expansion). 
An expansion in
$\lambda_0$ is not the same as an expansion in loops. 
The 
parameter $\lambda_0$  
is only a strict bookkeeping device, or ``flag'', 
allowing us to keep track of the
nonlinear adjective mode-mixing terms.
Note moreover, that
$\lambda_0$ is a dimensionful parameter, and it makes
no physical sense to expand in terms of it. Meaningful perturbation
series must ultimately refer to dimensionless expansion
parameters or couplings. 
The actual perturbative expansion parameter for MHD
is {\it not} $\lambda_0$, but rather, $g_{\pm}$, which as we have
calculated, is less than unity at both the fixed points. 
Thus, it is even possible to have reasonably large fluctuations and
maintain a small expansion parameter, (see Eq. (\ref{dimensionless})).
This coupling $g_{\pm}$
is dimensionless, and provides a bona-fide expansion parameter.
So, we do not believe there to be 
any inherent internal inconsistency in making use of a perturbative
loop calculation of hydrodynamics or MHD, provided we correctly recognize
that the true (and small) expansion parameter is not $\lambda_0$ but
rather Eq. (\ref{dimensionless}). 

Turning to point (b), this
has to do with possible limitations of the one-loop analysis
itself. 
The number of loops
considered is intimately related to the magnitude of the stochastic
fluctuations. If this number is small, then one
reasonably expects a low order loop expansion should
be able to capture the salient physics in the problem. However, if the 
fluctuations are large, one may have to push the
calculation to higher loops. 
Once again however, the noise amplitude is not a dimensionless
parameter, and what counts loops, properly speaking, is the dimensionless
ratio written in Eq. (\ref{dimensionless}). Note that large
effective viscosity and/or magnetic resistivity can counteract
the presence of large stochastic fluctuations. 
The number of terms or graphs in the
loop expansion grows geometrically with the number of loops, so 
ample physical justification would be needed before taking the 
calculational leap. A tempting and quite interesting,
though exceedingly nontrivial, proposal would be
to try and go beyond the loop
expansion by using nonperturbative methods to make
predictions of cascade directions, following 
Wilson's ideas for the exact
renormalization group 
(Wilson and Kogut 1974). 
In any case, neither of these
options have yet been pursued. 

{}Finally, coming to the point raised in (c), it may very well
be the case that 
the RG analysis is correct in all its physical predictions
despite the one-loop order, and there is no actual discrepancy
with the numerical results. The key to the resolution is 
in recognizing that the numerical simulations are always limited
by the finite duration times of the computer runs and the finite size of the
lattice, whereas the RG analysis here refers always to the
asymptotically largest spatial and temporal scales, which
are evidently not accessible to present day computer simulations. 
So, we
are comparing phenomena (cascade directions) 
associated to very different spatial and temporal
regimes. In other words, the inverse cascade might not be
an asymptotic phenomena, but is linked instead to 
intermediate scale physics, and this is evidently what
the simulations are seeing.  One possible explanation
is that simulations are seeing a low frequency mode that
temporarily shifts energy to larger length scales but
eventually shifts the energy to smaller length scales.
The RG analysis in this paper would ignore such modes. 
Let us elaborate further on this suggestion. Both inverse
and direct cascades are possible, but different cascade directions
would come to predominate at different time scales. 
The complete dynamical evolution may occur in two
phases: an earlier inverse cascade phase followed by
a later direct cascade phase. If true, this two-phase behavior would
be an example of a ``cross-over'' phenomenon, familiar from
critical dynamics. Of course, we do not see any
evidence for cross-over from our
one-loop RG calculation, and only a higher-order
(or preferably, non-perturbative) calculation could
begin to shed some light on this possibility
(see, point (b) above).
To get a further hold on the problem, one
could either (1) run the simulations for ever increasing times
and larger lattices, or (2) study the RG away from the vicinity
of the fixed points. In the latter case, we are 
``backing away'' from the leading asymptotic limit 
and the 
terms in the equations of motion 
denoted as ``less relevant'' in Eqs. (\ref{cge})  
and (\ref{rse})    
would now have to be included in the RG analysis.

\section*{Acknowledgments}
We thank Axel Brandenburg and Kari Enqvist for helpful discussions.
We also thank W.D. McComb for a useful discussion after
the paper was completed.
A.B. is funded by the United Kingdom Particle Physics and
Astronomy Research Council (PPARC) and D.H. is
supported in part by Spain's 
National Aerospace Technology Institute (INTA) 
and the Centro de Astrobiolog\'\i a (CAB).
A.B. acknowledges the kind hospitality provided him by   
the CAB in Madrid and D.H. acknowledges
the Department of Physics of the 
University of Edinburgh for the hospitality provided him during
a reciprocal visit.
Travel to and from our respective home institutions between
Madrid and Edinburgh has been made
possible with funds provided in part by CAB (Spain) and PPARC (U.K.).

\appendix
\section{Dynamic generating functional for stochastic MHD}

The purpose of this
Appendix is to set up and apply a powerful and 
extremely useful dynamic functional integral 
technique that leads to an efficient, systematic and organized 
perturbative 
loop-expansion for the solutions of the 
fully nonlinear magnetohydrodynamic equations 
Eqs. (\ref{Elsasser1}) and (\ref{Elsasser2}), when the
latter are subject
to random forces and fluctuations. 
The ability to cast stochastic differential equations in terms of functional
integrals opens the door to the calculation of many physically important quantities
including the so-called effective action and effective 
potential (Hochberg et al. 1999b). 
These
quantities can be used to investigate the onset of pattern formation and pattern
selection and can be used to identify symmetry breaking states of the system. In
this paper, we will use functional methods to deduce the Feynman rules needed
to construct a compact graphical representation of the perturbative
solutions of the coupled nonlinear MHD equations. These rules permit one to identify a
distinguished class of diagrams, the so-called one-particle irreducible (1PI)
graphs.  The renormalization
of these diagrams to a given number of loops will be used
to derive the differential renormalization group 
equations Eq. (\ref{RGE1}), that  govern the
scale dependence of the parameters appearing in the MHD equations. 
Additional details regarding the application
of functional
and path integral methods to stochastic differential equations can be found
in (Martin et al. 1973; DeDominicis and Peliti 1978; Rivers 1987;
Zinn-Justin 1996; Hochberg et al. 1999b)
and references therein. In the following, vector quantities will
generally be denoted in boldface (or sometimes with an arrow) 
and matrix operators indicated by means
of the dyad (double-arrow)
notation.  

Let ${\bf Q}_{soln},{\bf P}_{soln}$ denote a general solution of the coupled
stochastic MHD equations Eqs. (\ref{Elsasser1}) and (\ref{Elsasser2}). 
We pass immediately to a 
functional or path-integral representation for an arbitrary function $F$
of this solution-pair by means of the trivial identity
\begin{equation}\label{soln}
F({\bf Q}_{soln},{\bf P}_{soln}) = \int [{\cal D}{\bf P}]
[{\cal D}{\bf Q}] \, F({\bf Q},{\bf P})\,
\delta[{\bf P}-{\bf P}_{soln}] \, \delta[{\bf Q}-{\bf Q}_{soln}],
\end{equation}
where the path integrals are taken over all
field configurations. The product of delta-functionals selects out  
the particular
field configurations ${\bf P}={\bf P}_{soln}$ and 
${\bf Q}={\bf Q}_{soln}$
corresponding to the solutions of the MHD equations.
They constrain the functional integrals which otherwise freely sum over 
the totality of all 
${\bf Q}$ and ${\bf P}$ field configurations.     

The solutions of the MHD equations depend of course on the 
random noise sources, so that
${\bf P}_{soln} = {\bf P}_{soln}[\vec \eta_{P},\vec \eta_{Q}]$, and
${\bf Q}_{soln} = {\bf Q}_{soln}[\vec \eta_{P},\vec \eta_{Q}]$. 
The solutions themselves can be regarded as functionals of the noise source.
Ultimately, we are only interested in the solutions of MHD averaged over the
random forces and random currents. This statistical average is denoted by angular brackets
and is given by
\begin{equation}\label{average}
\langle 
F({\bf Q},{\bf P}) \rangle = \int [{\cal D}{\vec \eta}]\,
F({\bf Q}_{soln},{\bf P}_{soln}) {\cal P}[{\vec \eta}],
\end{equation}
where the noise distribution functional 
${\cal P}$ is assumed to be 
given (i.e. it is a Gaussian functional for Gaussian noise) and
is normalized to unity
\begin{equation}\label{normalized}
\int [{\cal D}\vec \eta_{P}] [{\cal D}\vec \eta_{Q}] {\cal P}[{\vec \eta_{P}}
,{\vec \eta_{Q}}] =
\int [{\cal D}{\vec \eta}]\, {\cal P}[{\vec \eta}] = 1.
\end{equation}
We abbreviate the $2d$-component noise vector field by 
$\vec \eta = (\vec \eta_{P},\vec \eta_{Q})$, where
$\vec \eta_{P}$ and $\vec \eta_{Q}$ 
each have $d$-components in a $d$-dimensional space.
The goal and objective of the following 
sequence of manipulations consists in choosing a particular form for $F$
and explicitly performing all the functional integrals encountered 
along the way. The resulting functional
then is used to generate all the
statistically averaged products of solutions of the nonlinear
MHD equations to any desired order in the expansion
parameter $\lambda_0$.  
The Feynman rules (see Fig. 1) 
can be read off directly from this final functional.

We need to carry out a change of variables in Eq. (\ref{soln}). To see how this
works, consider first a simple example. 
Let $\vec f(\vec x) = (f_1(x,y),f_2(x,y))$
be a two-component vector-valued function of a vector variable
$\vec x= (x,y)$ and suppose this has a unique root at $\vec x_0$,
$\vec f(\vec x_0) = 0.$ 
Now make use of the functional generalization of the identity
\begin{eqnarray}\label{delta-identity1}
\delta(\vec f(\vec x)) = \delta(f_1(x,y)) \delta(f_2(x,y))
&=& \frac{\delta(\vec x - \vec x_0)}{
\det|\frac{\partial f_{1,2}(x,y)}{\partial(x,y)}|},
\end{eqnarray}
where
\begin{eqnarray}
\frac{\partial f_{1,2}(x,y)}{\partial(x,y)} & = & 
\left( \matrix{ \frac{\partial f_1(x,y)}{\partial x} & \frac{\partial f_1(x,y)}{\partial y}\cr
\frac{\partial f_2(x,y)}{\partial x}  &  \frac{\partial f_2(x,y)}{\partial y} } \right),
\end{eqnarray}
and
\begin{eqnarray}
\delta(\vec x - \vec x_0) &=& \delta(x - x_0) \delta(y - y_0), 
\end{eqnarray}
to express the product of delta functionals in Eq. (\ref{soln}) as
\begin{eqnarray}\label{delta-identity2}
\delta[{\bf P}-{\bf P}_{soln}]\, \delta[{\bf Q}-{\bf Q}_{soln}] &=& {\cal J}
\delta[\partial_{t}P_j + \lambda_0 {\bf P}_{jn}\partial_l
(Q_l P_n) - \gamma_+ \nabla^2 P_j - \gamma_{-}\nabla^2 Q_j - (\eta_{P})_j]
\nonumber \\
&\times&
\delta[\partial_{t} Q_j  + \lambda_0 {\bf P}_{jn}\partial_l
(P_l Q_n) - \gamma_{+}\nabla^2 Q_j - \gamma_{-}\nabla^2 P_j - (\eta_{Q})_j].
\end{eqnarray}
Here, the Jacobian functional determinant ${\cal J}$ is the determinant of the following
$2\times 2$-block array:
\begin{equation}\label{jacobian}
\left( \matrix{
(\partial_{t} -\gamma_{+}\nabla^2)\delta_{jk} 
+ \lambda_0 {\bf P}_{jk}\partial_l (Q_l ) & 
-\gamma_{-}\nabla^2 \delta_{jk} + \lambda_0 {\bf P}_{jn}\partial_k(P_n) \cr
-\gamma_{-}\nabla^2 \delta_{jk} + \lambda_0 {\bf P}_{jn}\partial_k(Q_n) &
(\partial_{t} -\gamma_{+}\nabla^2)\delta_{jk} 
+ \lambda_0 {\bf P}_{jk}\partial_l (P_l ) } \right)
\delta(\vec x- \vec x')\delta(t - t').
\end{equation} 
When the identity Eq. (\ref{delta-identity2}) 
is substituted into Eqs. (\ref{soln}) and (\ref{average}), the
stochastic MHD equations 
Eqs. (\ref{Elsasser3}) and (\ref{Elsasser4})
appear explicitly under the functional integrals as
constraints, and we see that the stochastic dynamics is built in from the outset. 
The next step is to replace each of delta-functionals 
appearing on the left-hand side of Eq. (\ref{delta-identity2}) with 
its (functional) Fourier integral representation. 
That is, for any (vector-valued) space-time field 
${\bf \Phi}(\vec x, t)$,
\begin{equation}\label{Fourier-integral}
\delta[{\bf \Phi}] = \int [{\cal D}{\vec \sigma}] \, 
\exp \Big( i \int d^d{\vec x}\, dt\,\,
{\bf {\vec \sigma} \cdot \Phi} \Big),
\end{equation}
where $\vec \sigma$ is the Fourier conjugate field, 
and has the same number of components
as ${\bf \Phi}$. Also, in the above, the dot product 
indicates a sum over the repeated indices, so that
${\vec \sigma \cdot {\bf \Phi}} = 
\sigma_j {\bf \Phi}_j$, for $j=1,2,\cdots,d$. 
Replacing each delta-functional
by its corresponding Fourier integral introduces two conjugate field variables 
into the stochastic average Eq. (\ref{average}). 
Since there are two Elsasser fields, namely $\bf P$ and $\bf Q$,  
each requiring a separate
conjugate field, we distinguish each individual conjugate
field by a superscript: $\vec \sigma = (\vec \sigma^{P},\vec \sigma^{Q})$.

At this stage what results is a multiple 
functional integral representation for the stochastic average of an arbitrary
function $F$, which depends on the {\it exact} 
solutions of the original pair of MHD equations. That is
\begin{eqnarray}\label{fivefold}
\langle 
F({\bf Q},{\bf P}) \rangle &=& \int [{\cal D}{\vec \eta}]\,
[{\cal D}{\bf P}]\, [{\cal D}{\bf Q}] \,
[{\cal D}{\vec \sigma}^{P}]\, [{\cal D}{\vec \sigma}^{Q}]\,
F({\bf Q},{\bf P}) {\cal J}\, {\cal P}[{\vec \eta}] \nonumber \\
&\times& \exp \Big(  i \int d^d{\vec x}\, dt \,\, \sigma^{P}_j
[\partial_{t}P_j 
+ \lambda_0 {\bf P}_{jn}\partial_l
(Q_l P_n) - \gamma_+ \nabla^2 P_j - \gamma_{-}\nabla^2 Q_j - (\eta_{P})_j]
\Big) \nonumber \\
&\times& \exp \Big(  i \int d^d{\vec x}\, dt\,\, \sigma^{Q}_j
[\partial_{t}Q_j + \lambda_0 {\bf P}_{jn}\partial_l
(P_l Q_n) - \gamma_{+}\nabla^2 Q_j - \gamma_{-}\nabla^2 P_j - (\eta_{Q})_j]
\Big).
\end{eqnarray}
{}Following the standard practice in both quantum and statistical field
theory, now 
choose a convenient form for $F$. 
In particular, let ${\bf J}^{P}, {\bf J}^{Q},
{\bf J}^{\sigma^P}, {\bf J}^{\sigma^Q}$ denote a set of four arbitrary vector 
source functions, one source for each of 
the Elsasser and conjugate fields, respectively. Then let
\begin{equation}\label{F}
F = \exp \big( {\bf P \cdot J{^P}} + {\bf Q \cdot J{^Q}} +
{\vec \sigma^{P} \cdot {\bf J}{^{\sigma^P}}} + 
{\vec \sigma^{Q} \cdot {\bf J}{^{\sigma^Q}}} \big),
\end{equation}
and define the generating functional $Z$ by
\begin{equation}\label{Z}
Z[{\bf J}^{P}, {\bf J}^{Q},
{\bf J}^{\sigma^P}, {\bf J}^{\sigma^Q}] \equiv 
\langle \exp \big( {\bf P \cdot J{^P}} + {\bf Q \cdot J{^Q}} +
{\vec \sigma^{P} \cdot {\bf J}{^{\sigma^P}}} + 
{\vec \sigma^{Q} \cdot {\bf J}{^{\sigma^Q}}} \big)
\rangle.
\end{equation}
We have suppressed the omnipresent integrations over space and time to avoid
clutter (they can be restored at any time). 
The source functions allow one to factor the non-linear (in this case, cubic) 
interaction terms out
from under the functional integral, where they now appear as functional derivative
{\it operators} acting on a simpler and {\it exactly calculable} generating functional. 
Using the
functional generalization of the identity
\begin{equation}
\int du \, g(u) \exp(uJ) = g \Big(\frac{\partial}{\partial J} \Big)
\int du \, \exp(uJ),
\end{equation}
it then follows from Eqs. (\ref{fivefold}), (\ref{F}) and (\ref{Z}), that 
\begin{eqnarray}\label{full}
Z[{\bf J}^{P}, {\bf J}^{Q},
{\bf J}^{\sigma^P}, {\bf J}^{\sigma^Q}] &=&
\exp \Big( i\int d^d{\vec x}\, dt \, \lambda_0 
\frac{\delta}{\delta J_j^{\sigma^P} } {\bf P}_{jn} \partial_l \big(
\frac{\delta}{\delta J_l^{Q}} \frac{\delta}{\delta J_n^{P}} \big) \Big) \nonumber \\ 
&\times& \exp \Big( i\int d^d{\vec x}\, dt \, \lambda_0 
\frac{\delta}{\delta J_j^{\sigma^Q} } {\bf P}_{jn} \partial_l \big(
\frac{\delta}{\delta J_l^{P}} \frac{\delta}{\delta J_n^{Q}} \big) \Big) 
Z_0[{\bf J}^{P}, {\bf J}^{Q},
{\bf J}^{\sigma^P}, {\bf J}^{\sigma^Q}],
\end{eqnarray}
where $Z_0$ is the functional $Z$ evaluated at zero coupling: $\lambda_0 = 0$.
The functional derivatives have the property that, e.g., 
\begin{equation}
\frac{\delta J_k(\vec x, t)}{\delta J_l(\vec x', t')} = \delta_{kl}
\delta(\vec x- \vec x') \delta(t- t').
\end{equation}

Before continuing, special mention 
should be made regarding the treatment of the Jacobian
functional determinant ${\cal J}$ that appears in $Z$.
The determinant can be exponentiated by means of the identity $\det M = \exp(tr \ln M)$,
and it can be shown that the argument of the exponential is proportional to the
formally divergent object $\delta^d(0)$. This happens to be an ultraviolet or
short-distance divergence, and when renormalized by dimensional continuation, yields
zero, $\delta^d(0) = 0$ (Zinn-Justin 1996). 
Thus, in this regularization
scheme, the determinant 
renormalizes to yield a finite
and field independent constant (unity), and so 
contributes nothing to the normalized correlation 
functions calculated from $Z$. We henceforth drop
these determinants for the remainder of this discussion
(by setting ${\cal J} \rightarrow 1$). 
If one wants to investigate the short-distance or ultraviolet
limit of turbulent MHD however, then it is preferable to keep these
Jacobian determinants intact. {}For example, 
this is important when calculating the {\it effective action}
and {\it effective potential} associated 
with Eq. (\ref{full}).
The retention of these determinants can be done
with ghost fields, as described at length in
Hochberg et al. (1999b), where a number 
of details pertaining to the calculation 
of Jacobian determinants can be found. To our knowledge, the calculation of 
${\cal J}$ for MHD is an open problem. 
 
Up to this stage, we have assumed a completely
general noise distribution functional ${\cal P}$. 
{}For the remainder of this section, and throughout this paper,
only Gaussian noise is considered. {}For
Gaussian noise, $Z_0$ can be computed exactly and in closed form. 
In fact, we consider translationally
invariant Gaussian noise, that is
\begin{equation}
{\cal P}[{\vec \eta}] = {\cal N}
\exp \Big( -\frac{1}{2} \int d^d{\vec x} \, dt 
\int d^d{\vec x'}\, dt' \, \eta_i(\vec x, t)\, 
\Gamma_{ij}^{-1}(\vec x - \vec x',t - t')\,
\eta_j(\vec x', t') \Big),   
\end{equation}   
where ${\cal N}$ is a normalization constant and  
$\stackrel{\leftrightarrow}{\bf \Gamma}= \stackrel{\leftrightarrow}{\bf \Gamma}^T$
is the 
symmetric covariance matrix, which will be specified
below. 
By substituting this form of $\cal P$ into $Z_0$, 
and completing the square in the  
conjugate field variable, which is a $2d$-component vector field 
$\vec \sigma = (\vec \sigma^{P}, \vec \sigma^{Q})$,
exact integration can be done immediately
over the noise
field\footnote{Gaussian functional integration is treated in any number
of excellent texts devoted to (quantum) field theory. We recommend the
excellent coverage given in 
(Ramond 1981; Rivers 1987; Zinn-Justin 1996).}.

{}Following this procedure,
completing the square yields
\begin{equation}\label{completesquare1}
-\frac{1}{2} {\vec \eta} \cdot \stackrel{\leftrightarrow}{\bf \Gamma}^{-1}
\cdot {\vec \eta} - i {\vec \sigma} \cdot {\vec \eta} =
-\frac{1}{2}[{\vec \eta} + i\stackrel{\leftrightarrow}{\bf \Gamma}\cdot {\vec \sigma}]^T
\cdot \stackrel{\leftrightarrow}{\bf \Gamma}^{-1}  \cdot 
[{\vec \eta} + i\stackrel{\leftrightarrow}{\bf \Gamma} \cdot {\vec \sigma}]
-\frac{1}{2} {\vec \sigma} 
\cdot \stackrel{\leftrightarrow}{\bf \Gamma} \cdot {\vec \sigma},
\end{equation}
where $T$ denotes transpose. Next,
the functional integration over the noise leads to 
\begin{equation}\label{firstpass}
Z_0[{\bf J}, {\bf J}^{\sigma}] = \int [{\cal D}{\vec \sigma}]  [{\cal D}{\bf \Phi}] 
\exp \Big(-\frac{1}{2} {\vec \sigma} \cdot \stackrel{\leftrightarrow}{\bf \Gamma} \cdot
{\vec \sigma} + i {\vec \sigma} \cdot \stackrel{\leftrightarrow}{\bf G}^{-1}_0 \cdot
{\bf \Phi} + {\bf \Phi}\cdot{\bf J} + {\vec \sigma}\cdot{\bf J}^{\sigma} \Big),
\end{equation}
where we have used the fact that the noise distribution function ${\cal P}$ 
is normalized
to unity Eq. (\ref{normalized}). 
We have streamlined the notation even further by defining 
two-component fields for the Elsasser fields, their sources and their
conjugate fields respectively as 
${\bf \Phi} = 
({\bf P}, {\bf Q})$, ${\bf J} = 
({\bf J}^{P}, {\bf J}^{Q})$, and ${\bf J}^{\sigma} = 
({\bf J}^{\sigma^P}, {\bf J}^{\sigma^Q})$.
The (free) inverse response function in the space and time domain is defined to be
\begin{equation}\label{inverseresponse}
[\stackrel{\leftrightarrow}{\bf G}^{-1}_0]_{ij} =
\left( \matrix{\partial_{t} - \gamma_{+}\nabla^2 & -\gamma_{-}\nabla^2 \cr
               - \gamma_{-}\nabla^2  & \partial_{t} - \gamma_{+}\nabla^2 } 
\right)\delta_{ij}. 
\end{equation}

In Eq. (\ref{firstpass}) we have a quadratic form 
in the conjugate field, so we can
integrate over this field exactly by completing the square as
\begin{eqnarray}\label{completesquare2} 
-\frac{1}{2} {\vec \sigma} \cdot \stackrel{\leftrightarrow}{\bf \Gamma} \cdot
{\vec \sigma} + i {\vec \sigma} \cdot \stackrel{\leftrightarrow}{\bf G}^{-1}_0 \cdot
{\bf \Phi} &+& {\vec \sigma}\cdot{\bf J}^{\sigma} = \nonumber \\
&-& \frac{1}{2}\big({\vec \sigma} - i\stackrel{\leftrightarrow}{\bf \Gamma}^{-1} \cdot
[\stackrel{\leftrightarrow}{\bf G}_0^{-1}\cdot {\bf \Phi} - i {\bf J}^{\sigma} ] \big)^T 
\cdot
\stackrel{\leftrightarrow}{\bf \Gamma} \cdot 
\big({\vec \sigma} - i\stackrel{\leftrightarrow}{\bf \Gamma}^{-1}\cdot
[\stackrel{\leftrightarrow}{\bf G}_0^{-1}\cdot {\bf \Phi} 
- i {\bf J}^{\sigma} ] \big) \nonumber \\
&-& 
\frac{1}{2}(\stackrel{\leftrightarrow}{\bf G}_0^{-1}\cdot {\bf \Phi} - i {\bf J}^{\sigma})^T
\cdot \stackrel{\leftrightarrow}{\bf \Gamma}^{-1} \cdot
(\stackrel{\leftrightarrow}{\bf G}_0^{-1}\cdot {\bf \Phi} - i {\bf J}^{\sigma}).
\end{eqnarray}
Integrating over the conjugate fields yields 
(up to an overall and unimportant multiplicative constant)
\begin{equation}
Z_0[{\bf J},{\bf J}^{\sigma}]  \propto \int [{\cal D}{\bf \Phi}] 
\exp \Big( - \frac{1}{2}(\stackrel{\leftrightarrow}{\bf G}_0^{-1}\cdot
{\bf \Phi} - i {\bf J}^{\sigma})^T \cdot
\stackrel{\leftrightarrow}{\bf \Gamma}^{-1} \cdot
(\stackrel{\leftrightarrow}{\bf G}_0^{-1}\cdot {\bf \Phi} - i {\bf J}^{\sigma})
+ {\bf \Phi} \cdot {\bf J} \Big).
\end{equation}
{}Finally, by means of the identity
\begin{eqnarray}\label{completesquare3}
(\stackrel{\leftrightarrow}{\bf G}_0^{-1} &\cdot& {\bf \Phi} - i {\bf J}^{\sigma})^T
\cdot \stackrel{\leftrightarrow}{\bf \Gamma}^{-1} \cdot
(\stackrel{\leftrightarrow}{\bf G}_0^{-1}\cdot {\bf \Phi} - i {\bf J}^{\sigma}) =
\nonumber \\
& & ({\bf \Phi} - i\stackrel{\leftrightarrow}{\bf G}_0 \cdot {\bf J}^{\sigma})^T
\cdot (\stackrel{\leftrightarrow}{\bf G}_0^{-1} \cdot 
\stackrel{\leftrightarrow}{\bf \Gamma}^{-1} \cdot 
\stackrel{\leftrightarrow}{\bf G}_0^{-1})\cdot
({\bf \Phi} - i\stackrel{\leftrightarrow}{\bf G}_0 \cdot {\bf J}^{\sigma}),
\end{eqnarray}
we are led to make the constant shift change-of-variable 
${\bf \Phi'} = {\bf \Phi} -i\stackrel{\leftrightarrow}{\bf G}_0 \cdot {\bf J}^{\sigma}$,
integrate over the remaining field ${\bf \Phi'}$, and note that
$[{\cal D} {\bf \Phi}'] = [{\cal D} {\bf \Phi}]$.
Thus, up to an overall irrelevant constant prefactor, the free 
 $(\lambda_0 = 0)$ generating functional is
given by
\begin{equation}
Z_0[{\bf J}^{P}, {\bf J}^{Q},
{\bf J}^{\sigma^P}, {\bf J}^{\sigma^Q}] = \exp \Big(i{\bf J}^{\sigma} \cdot
\stackrel{\leftrightarrow}{\bf G}_0 \cdot {\bf J} \Big)
\exp \Big( \frac{1}{2}{\bf J}\cdot[\stackrel{\leftrightarrow}{\bf G}_0
\cdot \stackrel{\leftrightarrow}{\bf \Gamma} \cdot
\stackrel{\leftrightarrow}{\bf G}_0]\cdot {\bf J}\Big).
\end{equation}
This can be expressed in Fourier space as
\begin{eqnarray}
Z_0[{\bf J}^{P}, {\bf J}^{Q}&,&
{\bf J}^{\sigma^P}, {\bf J}^{\sigma^Q}]  =
\exp \Big( i\int \frac{d^d{\vec k}\, d\omega}{(2\pi)^{d+1}} \,
{\bf J}^{\sigma}(\vec k,\omega) \cdot \stackrel{\leftrightarrow}{\bf G}_0
(-\vec k,-\omega) \cdot {\bf J}(-\vec k,-\omega) \Big) \nonumber \\
&\times& 
\exp \Big( \frac{1}{2} \int \frac{d^d{\vec k}\, d\omega}{(2\pi)^{d+1}} \,
{\bf J}(\vec k,\omega) \cdot \stackrel{\leftrightarrow}{\bf G}_0
(\vec k,\omega) \cdot \stackrel{\leftrightarrow}{\bf \Gamma}(-\vec k,-\omega)
\cdot \stackrel{\leftrightarrow}{\bf G}_0 (-\vec k,-\omega) \cdot
{\bf J}(-\vec k,-\omega) \Big).
\end{eqnarray}

{}From the free generating functional, it 
is straightforward to read off the bare response and bare correlation
functions (or by taking the appropriate functional derivatives of $Z_0$ with
respect to the sources). In fact, the whole point in defining $Z$ the way we
did back in Eq. (\ref{Z}) was so we could compute the statistical average of products
of fields and conjugate fields. As we see, functionally differentiating $Z$ with
respect to any combination of sources brings down the corresponding factors of
fields under the functional integral. Then, setting all the sources to zero after 
differentiating
yields the correlation and response functions associated with this 
stochastic field theory. Let us now specify the explicit form for the
noise or covariance matrix as
\begin{equation}\label{covariance}
[\stackrel{\leftrightarrow}{\bf \Gamma}(\vec k,\omega)]_{mn} = D_0(k)\,
{\bf P}_{mn}(\vec k),
\end{equation}
which is the product of a $2 \times 2$-block noise matrix $D_0(k)$ times
the transverse projection operator. The presence of this projector
enforces the solenoidal character of the noise, and appears as well
in the original noise correlation functions above 
Eqs. (\ref{vnoise}) and (\ref{Bnoise}).
The noise matrix can be used to model an arbitrary spatially correlated
noise. To handle both spatial and temporal correlations in the noise, one would
write $D_0(k,\omega)$. In this paper, we consider spatially correlated noise
as specified 
by $2 \times 2$-block $D_0(k)$ given below in Eq. (\ref{correlation}).
The effects of random helicity (both kinetic and magnetic) can be studied by
an appropriate and simple extension of Eq. (\ref{covariance}); this is introduced
and analyzed in the 
main text. 

Thus, the bare response function in Fourier space is computed from
\begin{eqnarray}\label{correlationz}
\langle \sigma_m(\vec p_1,\omega_1)\Phi_n(\vec p_2,\omega_2)\rangle_0 &=& 
\frac{(2\pi)^{2(d+1)}}{Z_0[0]} \frac{\delta^2}{\delta J^{\sigma}_m(\vec p_1,\omega_1)
\delta J_n(\vec p_2,\omega_2)}|_{J^{\sigma}=J=0} Z_0[{\bf J}^{\sigma},{\bf J}]\nonumber \\
&=& i(2\pi)^{d+1}\delta^d(\vec p_1 + \vec p_2)\delta(\omega_1 + \omega_2)
[\stackrel{\leftrightarrow}{\bf G}_0(\vec p_2,\omega_2)]_{mn}.
\end{eqnarray}
The bare response function obtained by calculating the inverse
of Eq. (\ref{inverseresponse}) is
\begin{eqnarray}\label{bareresponse}
[\stackrel{\leftrightarrow}{\bf G}_0(\vec k,\omega)]_{mn} &=&
{\cal G}_0(\vec k,\omega) \delta_{mn},\nonumber \\
{\cal G}_0(\vec k,\omega) &=& \frac{1}{[-i\omega + k^2(\gamma_{+} + \gamma_{-})]
[-i\omega + k^2(\gamma_{+} - \gamma_{-})]}
\left( \matrix{ -i\omega + \gamma_{+}k^2 & - \gamma_{-}k^2 \cr
               - \gamma_{-}k^2 &   -i\omega + \gamma_{+}k^2 } \right),
\end{eqnarray}
where the convention is that momentum and frequency are taken to flow from 
the conjugate field
to the physical field (otherwise, one must flip the sign of $\omega$).
Next, the bare correlation function (in Fourier space) is computed from 
\begin{eqnarray}
\langle \Phi_m(\vec p_1,\omega_1)\Phi_n(\vec p_2,\omega_2)\rangle_0 &=& 
\frac{(2\pi)^{2(d+1)}}{Z_0[0]} \frac{\delta^2}{\delta J_m(\vec p_1,\omega_1)
\delta J_n(\vec p_2,\omega_2)}|_{J^{\sigma}=J=0} Z_0[{\bf J}^{\sigma},{\bf J}] ,\nonumber \\
&=& (2\pi)^{d+1}\delta^d(\vec p_1 + \vec p_2)\delta(\omega_1 + \omega_2)
[\stackrel{\leftrightarrow}{\bf C}_0(\vec p_2,\omega_2)]_{mn},
\end{eqnarray}

where,
\begin{eqnarray}\label{correlation}
[\stackrel{\leftrightarrow}{\bf C}_0(\vec k,\omega)]_{mn} &=&
{\cal C}_0(\vec k,\omega) {\bf P}_{mn}(\vec k),\nonumber \\
{\cal C}_0(\vec k,\omega) &=& 
{\cal G}_0(\vec k,\omega)D_0(k){\cal G}_0(-\vec k,-\omega), \nonumber
\\
D_0(k) &=& \left(\matrix {2k^{-y} A & 2k^{-y} B \cr
               2k^{-y} B & 2k^{-y} A }\right). 
\end{eqnarray}

To complete the specification of the full
generating functional $Z$, we must also express the non-linear interaction vertex in
Fourier space. This is obtained by Fourier-transforming the cubic terms
appearing in Eq. (\ref{fivefold}) or in Eq. (\ref{full}). Note there are {\it two} kinds of
cubic interaction, one corresponding to the $\sigma^{P}-{\bf P}-{\bf Q}$
vertex and the other corresponding to the $\sigma^{Q}-{\bf P}-{\bf Q}$
vertex (the distinguishing feature is the conjugate field). Of course, there
is only one (bare) cubic coupling parameter $\lambda_0$ (which we set to unity
at the end of the calculation). This means corrections only
to one of the two vertices
needs to be considered, even though {\it both} 
bare vertices are needed in building
up loop-expansions. Taking the convention that all momentum and 
frequency flow inward to
the vertex, the bare vertex corresponding to the cubic interaction
$\sigma_j^{P}(\vec k_1,\omega_1)-
P_n(\vec k_2,\omega_2) -Q_l(\vec k_3,\omega_3)$ as well
as 
the other cubic interaction
$\sigma_j^{Q}(\vec k_1,\omega_1)-Q_n(\vec k_2,\omega_2) -P_l(\vec k_3,\omega_3)$
is
\begin{equation}\label{barevertex}
i\, \lambda_0 {\bf P}_{l,jn}(\vec k_1) = 
i \,\lambda_0 \, (\vec k_1)_l {\bf P}_{jn}(\vec k_1) = i\, \lambda_0 \, (\vec k_1)_l
\Big(\delta_{jn} - \frac{(\vec k_1)_j(\vec k_1)_n}{(\vec k_1)^2} \Big). 
\end{equation}
The vertex is independent of the external frequencies.
Also, here it has been written so that it
depends only on the momentum carried by the conjugate field(s), since momentum
conservation implies $\vec k_1 + \vec k_2 + \vec k_3 = 0$.

In summary, in this Appendix the stochastic PDEs of MHD 
Eqs. (\ref{Elsasser3}) and (\ref{Elsasser4})
have been expressed in terms of a functional integral
representation.  {}From this
representation, the bare response function Eq. (\ref{bareresponse}), 
correlation functions Eq. (\ref{correlation}),
and the bare non-linear interaction vertex Eq. (\ref{barevertex})
have been identified and calculated.  As explained earlier,
this set of functions 
constitute the basic elements (see Fig. 1) upon which
a diagrammatic expansion (see Fig. 2) of the generating 
functional can be systematically developed.

\section{Response function renormalization}

This Appendix carries out the one-loop renormalization 
of the response function.
Transcribing the diagrammatic equation 
in Fig. (2a) for the one-loop corrected 
response function into the corresponding analytic expression by means
of the Feynman rules in Fig. 1 yields
\begin{equation}\label{response1}
[\stackrel{\leftrightarrow}{\bf G}(\vec k,\omega)]_{mn} =
[\stackrel{\leftrightarrow}{\bf G}_0(\vec k,\omega)]_{mn} - 
{\cal S} \lambda^2_0
[\stackrel{\leftrightarrow}{\bf G}_0(\vec k,\omega)]_{m \bar m}
[{\bf P}_{l,\bar n s}(\vec k) + {\bf P}_{s,\bar n l}(\vec k)]
I_{\bar m ls}(\vec k,\omega)
[\stackrel{\leftrightarrow}{\bf G}_0(\vec k,\omega)]_{\bar n n},
\end{equation}
where the symmetry factor for the one loop diagram ${\cal S} = 4$ and
the loop-integral is given by
\begin{equation}\label{resloop}
I_{\bar m ls}(\vec k,\omega) =
\int^{>} \frac{d^d\vec q}{(2\pi)^d} \int^{\infty}_{-\infty} \frac{d\Omega}{2\pi}\,
[{\bf P}_{\bar m ,lr}(\vec k - \vec q) + {\bf P}_{r,l \bar m}(\vec k - \vec q)]
{\cal G}_0(\vec k - \vec q, \omega - \Omega) {\cal C}_0(\vec q,\Omega) {\bf P}_{rs}
(\vec q).
\end{equation} 
The precise meaning of the momentum-shell integration will be given below. 
Using the factorization of the 
response function into the product of a block matrix
times a Kronecker delta as given in Eqs. (\ref{bareresponse}) and
(\ref{response1}), the following equation
is obtained for the one-loop correction
to the {\it inverse} response function,  
\begin{equation}\label{response2}
{\cal G}^{-1}(\vec k,\omega)\delta_{mn} = 
{\cal G}_0^{-1}(\vec k,\omega)\delta_{mn} + {\cal S}\lambda_0^2
[{\bf P}_{l, n s}(\vec k) + {\bf P}_{s, n l}(\vec k)]
I_{ m ls}(\vec k,\omega) + O(\lambda_0^4),
\end{equation}
where the renormalized $2 \times 2$ inverse block is parametrized as 
\begin{equation}\label{response3}
{\cal G}^{-1}(\vec k,\omega) = 
\left( \matrix{ -i\omega + \gamma_{+}^{<}k^2 &  \gamma_{-}^{<}k^2 \cr
 \gamma_{-}^{<}k^2 & -i\omega + \gamma_{+}^{<}k^2} \right).
\end{equation}
This structure follows from Fourier transforming Eq. (\ref{inverseresponse}).
The infrared (i.e., large-distance and long-time) 
renormalizability of the response function implies that the 
renormalized or corrected response function Eq. (\ref{response3}) must have the
identical mathematical structure as its bare 
counterpart ${\cal G}_0^{-1}$. It therefore has the same
frequency and momentum dependence as the bare response function and contains
the same number of parameters. Here 
$\gamma^{<}_{\pm} = \gamma^{<}_{\pm}(A, B, \gamma_{+},\gamma_{-},\lambda_0)$ 
are computable functions of the indicated
bare parameters which will be determined below.

Thus the renormalization of the response function yields the individual
renormalization of the two independent viscosities $\gamma_{+},\gamma_{-}$.
The renormalization of the two viscosities 
requires some amount of careful calculation.
We will go through the major points step by step.
To proceed we need to explicitly 
carry out the internal frequency integration that
appears in the loop
integral Eq. (\ref{resloop}). 
{}From Eq. (\ref{response3}), we see that the 
viscosities $\gamma_{\pm}$ are the coefficients of the
$k^2$ terms, so we can 
set the external frequency $\omega$ to zero from the outset. 
This frequency integral is straightforwardly calculated 
by means of the calculus of residues. We close the
contour in the lower half plane, where there are two simple poles, and obtain
\begin{eqnarray}\label{freq}
\int^{\infty}_{-\infty} \frac{d\Omega}{2\pi}\,
{\cal G}_0(&\vec k& - \vec q, - \Omega) {\cal C}_0(\vec q,\Omega) = \nonumber \\
& & \frac{D_0(q)}{8q^4}\,
\Big(1 + \frac{\vec q \cdot \vec k}{q^2} + O(k^2)\Big)
\left\{  \frac{1}{(\gamma_{+} - \gamma_{-})^2}
\left[ \matrix{1&-1\cr   
        -1&1} \right] +
\frac{1}{(\gamma_{+} + \gamma_{-})^2}
\left[ \matrix{1& 1\cr   
         1&1} \right] \right\}, \nonumber \\
&=&
\frac{1}{4}q^{-y-4}\Big(1 + \frac{\vec q \cdot \vec k}{q^2} + O(k^2)\Big)
\left\{  \frac{A-B}{(\gamma_{+} - \gamma_{-})^2}
\left[ \matrix{1&-1\cr   
        -1&1} \right] +
\frac{A+B}{(\gamma_{+} + \gamma_{-})^2}
\left[ \matrix{1& 1\cr   
         1&1} \right] \right\}. 
\end{eqnarray}
The result displayed in the first line holds for all 
spatially correlated Gaussian noise $D_0(q)$. We should point out this
was evaluated assuming a positive viscosity, $\nu > 0$, 
and a positive resistivity, $\nu_B > 0$. A change in sign will change the
location of the poles and correspondingly modify the frequency integration.  
The second line results from substituting 
the particular form for the noise spectrum used
in this paper, Eq. (\ref{correlation}).
If one wants to treat
noise with temporal correlations, i.e., noise 
spectra of the form $D_0(q,\Omega)$, then the frequency integral
must be evaluated in a case-by-case fashion, and will generally involve both 
poles and branch cuts. 

Substituting this result back into Eq. (\ref{resloop}),
we turn next to the remaining and final integration over the internal loop momentum.
{}From Eq. (\ref{response3}) the renormalization of the parameters $\gamma_{\pm}$ requires
that we expand the one-loop contribution up to second order in the external momentum.
The projection operator in Eq. (\ref{response2}) is linear in $\vec k$, so
the loop integral only need be expanded to first order in external momentum.
It is important to note that this integral depends on the
external momentum not only via the integrand but also through its
domain of integration.
This is because, when integrating over the momentum shell, we must
ensure that
all combinations of internal and external momenta appearing in 
Eq. (\ref{resloop})
remain within the band or shell of momenta to be integrated over. 
In the case of the present integral, this means we must integrate over the
{\it intersection} of 
the domains $\Lambda/s \leq |\vec q| \leq \Lambda$ and
$\Lambda/s \leq |\vec k - \vec q| \leq \Lambda$, where $s > 1$ is a measure
of the fraction of modes to be eliminated in the coarse-graining.  
To first order in $\vec k$, the
second inequality can be written as $\Lambda/s + k\cos \theta < q < 
\Lambda + k\cos \theta$, where $\theta$ is the angle between $\vec k$ and
$\vec q$. There are two cases to consider: when $\cos \theta > 0$, the
intersection of the two intervals can be expressed as the difference
$[\Lambda/s,\Lambda] - [\Lambda/s,\Lambda/s +  k\cos \theta]$, and
when $\cos \theta < 0$, the intersection can be written as
$[\Lambda/s,\Lambda] - [\Lambda + k\cos \theta , \Lambda]$.
This means the complete momentum-shell integration 
valid up to $O(k^2)$ can be written as
\begin{equation}\label{momshell}
\int^{>} \frac{d^d\vec q}{(2\pi)^d} = \int d\Omega_d \,
\Big(\int_{\Lambda/s}^{\Lambda} - \int_{\Lambda/s}^{\Lambda/s + k \cos \theta}
- \int_{\Lambda + k \cos \theta}^{\Lambda} \Big)
\frac{dq\, q^{d-1}}{(2\pi)^d} + O(k^2).
\end{equation}
{}Furthermore, note it is convenient 
to carry out the integrations over the momentum modulus before
working out the angular integrations over the element of solid angle $d\Omega_d$. 

{}From Eqs. (\ref{freq}) and (\ref{resloop}), the remaining integration
to be carried out after the frequency integration is performed is
\begin{eqnarray}\label{bigI}
I &=&
\int^{>} \frac{d^d\vec q}{(2\pi)^d}
[{\bf P}_{m ,lr}(\vec k - \vec q) + {\bf P}_{r,l m}(\vec k - \vec q)]
{\bf P}_{rs}(\vec q) \,q^{-y-4} \Big(1 + \frac{\vec k \cdot \vec q}{q^2} + O(k^2)\Big),\\
 &=& I_1 + I_2 + I_3,
\end{eqnarray}
where we break the full integral $I$  down into somewhat smaller and
more manageable pieces as
\begin{eqnarray}\label{piece1}
I_1 &=& \int^{>} 
\frac{d^d \vec q}{(2\pi)^d}\, q^{-y-4}\Big(-q_m {\bf P}_{ls}(\vec q) \Big)
+ O(k^2),\\
\label{piece2}
I_2 &=& \int^{>} \frac{d^d \vec q}{(2\pi)^d}\, q^{-y-4} 
\Big( -q_m {\bf P}_{ls}(\vec q) \Big) \frac{\vec k \cdot \vec q}{q^2}
+ O(k^2), \\
\label{piece3}
I_3 &=& \int^{>} \frac{d^d \vec q}{(2\pi)^d}\, q^{-y-4}\Big( k_m {\bf P}_{ls}(\vec q)
-(k_s - {\vec k}\cdot{\vec q}\,\frac{q_s}{q^2})\frac{q_m q_l}{q^2} + k_r
{\bf P}_{lm}(\vec q){\bf P}_{rs}(\vec q) \Big) + O(k^2).
\end{eqnarray}
The following projection operator product expansions proved 
to be useful in arriving at these expressions:
\begin{eqnarray}\label{projectexpand}
{\bf P}_{a,bc}(\vec k - \vec q){\bf P}_{dc}(\vec q) &=& 
(k_a - q_a)\Big( \delta_{bd} - \frac{q_b q_d}{q^2} \Big) -
\frac{q_a q_b}{q^2} \Big( k_d - \frac{\vec q \cdot \vec k\, q_d}{q^2}\Big) 
+ O(k^2), \nonumber \\
{\bf P}_{a,bc}(\vec k - \vec q){\bf P}_{ad}(\vec q) &=& k_a{\bf P}_{bc}(
\vec q){\bf P}_{ad}(\vec q) + O(k^2).
\end{eqnarray}
The integrand of $I_1$ is independent of $k$ and the contribution from
the interval $[\Lambda/s, \Lambda]$ vanishes upon angular averaging. Only the latter
two contributions in Eq. (\ref{momshell}) will contribute to the angular
averaging and we find that
\begin{equation}\label{I1}
I_1 = [\big(\frac{\Lambda}{s}\big)^{d-y-4} - \Lambda^{d-y-4}]
\left\{ \frac{1}{2}\frac{S_d}{d(2\pi)^d} k_m \delta_{ls}
- \frac{1}{2}\frac{S_d}{d(d + 2)(2\pi)^d}(k_m \delta_{ls} + k_l \delta_{ms}
+ k_s \delta_{ml} ) \right\},
\end{equation}
where $S_d$ is defined in Eq. (\ref{sphere}).
The integrand in $I_2$ is already linear in $k$, so it suffices to work out the
contribution from the interval $[\Lambda/s, \Lambda]$. This yields 
\begin{equation}\label{I2}
I_2 = \frac{2}{(d-y-4)} I_1.
\end{equation}
Lastly, the integrand in $I_3$ is also linear in $k$. Integrating over
$[\Lambda/s,\Lambda]$ and averaging over the angles yields
\begin{eqnarray}\label{I3}
I_3 = \frac{[\Lambda^{d-y-4} - (\frac{\Lambda}{s})^{d-y-4}]}{(d-y-4)}
\Big( \frac{(d-1)S_d}{d(2\pi)^d} k_m \delta_{ls}
&+& \frac{(d-3)S_d}{d(2\pi)^d} k_s \delta_{lm}  \nonumber \\
&+&  \frac{2 S_d}{d(d + 2)(2\pi)^d}(k_m \delta_{ls} + k_l \delta_{ms}
+ k_s \delta_{ml} ) \Big).
\end{eqnarray}
Various identities needed for the angular integrations are collected in
Eq. (\ref{sphere}).
This completes the calculation of the one-loop integral. Combining all the
results from Eqs. (\ref{bigI}), (\ref{piece1}),
(\ref{piece2}), (\ref{piece3}),
(\ref{I1}) ,(\ref{I2}), and (\ref{I3}), we obtain 
the one-loop matrix equation
\begin{eqnarray}\label{matrix}
{\cal G}^{-1}(\vec k, 0)\delta_{mn} = 
{\cal G}_0^{-1}(\vec k, 0)\delta_{mn} & + & {\cal S}\frac{\lambda_0^2}{4}
\frac{(d^2-y-4)S_d}{d(d + 2)(d-y-4)(2\pi)^d}
[\Lambda^{d-y-4} - \big( \frac{\Lambda}{s} \big)^{d-y-4}] \nonumber \\
& & \left\{  \frac{A-B}{(\gamma_{+} - \gamma_{-})^2}
\left[ \matrix{1&-1\cr   
        -1&1} \right] +
\frac{A+B}{(\gamma_{+} + \gamma_{-})^2}
\left[ \matrix{1& 1\cr   
         1&1} \right] \right\}\times k^2 \,{\bf P}_{mn}(\vec k).
\end{eqnarray}

We can project out individual equations for each of the two viscosities by
multiplying Eq. (\ref{matrix}) through by ${\bf P}_{ml}(\vec k)$ and using the
idempotency of this projection operator, 
${\bf P}_{mn}(\vec k){\bf P}_{ml}(\vec k) 
= {\bf P}_{nl}(\vec k)$.
Thus we find that each viscosity renormalizes as
\begin{eqnarray}\label{elements}
\gamma^{<}_{+} &=& \gamma_{+} + {\cal S}\frac{\lambda_0^2}{4}
[\Lambda^{d-y-4} - (\frac{\Lambda}{s})^{d-y-4}]
\frac{(d^2-y-4)S_d}{d(d+2)(d-y-4)(2\pi)^d} \left\{ \frac{(A-B)}
{(\gamma_{+} - \gamma_{-})^2}
+ \frac{(A+B)}{(\gamma_{+} + \gamma_{-})^2} \right\}, \\
\label{elements2}
\gamma^{<}_{-} &=& \gamma_{-} + {\cal S}\frac{\lambda_0^2}{4}
[\Lambda^{d-y-4} - (\frac{\Lambda}{s})^{d-y-4}]
\frac{(d^2-y-4)S_d}{d(d+2)(d-y-4)(2\pi)^d} \left\{ \frac{-(A-B)}
{(\gamma_{+} - \gamma_{-})^2}
+ \frac{(A+B)}{(\gamma_{+} + \gamma_{-})^2} \right\}.
\end{eqnarray}
Individual equations for the sum $(\gamma_{+} + \gamma_{-})$ 
and difference $(\gamma_{+} - \gamma_{-})$ can be easily obtained by taking the
sum and difference, respectively,  of
Eqs. (\ref{elements}) and (\ref{elements2}). This
yields the individual renormalizations of the fluid viscosity $\nu$ and magnetic
resistivity $\nu_B$. 

\section{Noise spectrum renormalization}

{}From the diagrammatic representation of the one-loop correction
to the noise spectral function in Fig. (2b),
(the noise spectral function is obtained by amputating the external
legs from the correlation function) 
which is a $2 \times 2$ block matrix,
we obtain the following expression for the one-loop noise spectrum in 
terms of the bare response,
correlation and vertex functions, 
\begin{equation}\label{noisespec1}
D(k){\bf P}_{rs}(\vec k) = D_0(k){\bf P}_{rs}(\vec k) +
{\cal S} \lambda_0^2
[{\bf P}_{l,rn}(\vec k) + {\bf P}_{n,rl}(\vec k)] I_{lm,nj}(\vec k,\omega)
[{\bf P}_{m,sj}(\vec k) + {\bf P}_{j,sm}(\vec k)],
\end{equation} 
where the symmetry factor in this case is ${\cal S} = 2$ and 
the associated one-loop integral is given by
\begin{eqnarray}\label{noise2}
I_{lm,nj}(\vec k,\omega) &=& \int^{>} \frac{d^d \vec q}{(2\pi)^d}\,
\int^{\infty}_{-\infty} \frac{d\Omega}{2\pi} \,
[\stackrel{\leftrightarrow}{\bf C}_0(\vec k - \vec q,\omega - \Omega)]_{lm}
[\stackrel{\leftrightarrow}{\bf C}_0(\vec q, \Omega)]_{nj}\nonumber \\
&=& \int^{>} \frac{d^d \vec q}{(2\pi)^d}\, 
\int^{\infty}_{-\infty} 
\frac{d\Omega}{2\pi} \, {\bf P}_{lm}(\vec k - \vec q){\bf P}_{nj}(\vec q)
{\cal C}_0(\vec k - \vec q,\omega - \Omega) 
{\cal C}_0(\vec q,\Omega),\nonumber \\ 
{\cal C}_0(\vec k - \vec q,\omega - \Omega) 
{\cal C}_0(\vec q,\Omega) &=&
{\cal G}_0(\vec k - \vec q,\omega - \Omega)D_0(\vec k - \vec q)
{\cal G}_0(\vec q - \vec k, \Omega - \omega)\nonumber \\
&\times& {\cal G}_0(\vec q,\Omega)D_0(q){\cal G}_0(-\vec q,-\Omega).
\end{eqnarray} 
The infrared renormalizability of the noise spectral function implies that the 
renormalized spectral function must have the
identical mathematical structure as its bare counterpart $D_0(k)$ in 
Eq. (\ref{correlation}).
It must therefore have the same
frequency and momentum dependence as the bare noise spectral function and contain
the same number of parameters,
\begin{equation}\label{noise1}
D(k) = \left(\matrix {2k^{-y} A^{<} & 2k^{-y} B^{<} \cr
               2k^{-y} B^{<} & 2k^{-y} A^{<} }\right).
\end{equation}
Here $A^{<} = A^{<}(A,B,\gamma_{+}, \gamma_{-},\lambda_0)$ and 
$B^{<} = B^{<}(A,B,\gamma_{+}, \gamma_{-},\lambda_0)$ are calculable in terms of the bare
parameters and will be determined below.
 
We are interested in evaluating this loop integral in the
hydrodynamic limit (i.e., for small $k$ and small $\omega$). 
However, there is already an
overall  
quadratic factor
of $k^2$ multiplying the loop integral arising 
from the product of individual projection operator factors
appearing in Eq. (\ref{noisespec1}). 
As can be seen from Taylor expanding about $k = 0$, 
the loop integral itself can only contribute
a constant plus positive {\it integer} powers of external momentum $\vec k$ in the
hydrodynamic limit. Thus, the complete one-loop 
correction in the long-wavelength regime is a power series in 
$k$, and it starts off at quadratic order. {}For the type
of noise of interest to us, we now show that this one loop correction
is {\it irrelevant} in the hydrodynamic regime. 
In fact, by setting $\vec k$ and $\omega$ to zero 
inside the loop-integral Eq. (\ref{noise2}), it gives
\begin{eqnarray}\label{noise3}
I_{lm,nj}(0,0) &=& \int^{>} \frac{d^d \vec q}{(2\pi)^d}\, 
\int^{\infty}_{-\infty} 
\frac{d\Omega}{2\pi} \, {\bf P}_{lm}(- \vec q){\bf P}_{nj}(\vec q)
{\cal C}_0(- \vec q,- \Omega) 
{\cal C}_0(\vec q,\Omega)\nonumber \\
&=& \int d\Omega_d \, {\bf P}_{lm}(-\vec q){\bf P}_{nj}(\vec q)
\int_{\Lambda/s}^{\Lambda}\frac{q^{d-1}dq}{(2\pi)^d}\, 
\int_{-\infty}^{\infty} \frac{d\Omega}{2\pi}\,\, 
{\cal C}_0(- \vec q,- \Omega) 
{\cal C}_0(\vec q,\Omega)\nonumber \\
&=& 
\int d\Omega_d \, {\bf P}_{lm}(-\vec q){\bf P}_{nj}(\vec q)\,
\int_{\Lambda/s}^{\Lambda} \frac{q^{d-1}dq}{(2\pi)^d} {\cal M}(q).
\end{eqnarray}
Here
\begin{equation}
 {\cal M}(q) =  
\left(\matrix {M_1(q) & M_2(q) \cr
               M_2(q) & M_1(q) }\right),
\end{equation}
is the $2 \times 2$ 
block-matrix function of the {\it modulus}  $q = |\vec q|$,
which results
from integrating the product of the bare
correlation functions in Eq. (\ref{noise2}) 
over the internal loop frequency. This can be
straightforwardly calculated by means of residues (there are four simple
poles in both the upper and lower half-planes, respectively), 
but we need not do so
for the purposes of the present discussion. 
Note that the integration over the sphere factors out from the modulus and
frequency integrations. This is because the product of projection operators
depends only on angles, while the product of the correlation block-matrices
depends only on frequency and momentum modulus. 
Integrating
the product of the projection operators that appear in Eq. (\ref{noise3}) 
over the unit sphere using the identities in
Eq. (\ref{sphere}) yields the result
\begin{equation}
\int d\Omega_d \, {\bf P}_{lm}(-\vec q){\bf P}_{nj}(\vec q) =
\frac{S_d}{d(d + 2)} \Big( (d^2-3)\delta_{lm}\delta_{nj} 
+ [ \delta_{ln}\delta_{mj} + \delta_{lj}\delta_{nm}] \Big).
\end{equation}
After using the properties of the projection operators
to further simplify the final expression, it leads to 
\begin{eqnarray}
[{\bf P}_{l,rn}(\vec k) + {\bf P}_{n,rl}(\vec k)] I_{lm,nj}(0,0)
[{\bf P}_{m,sj}(\vec k) &+& {\bf P}_{j,sm}(\vec k)] = \nonumber \\
& & 2k^2 \times {\bf P}_{rs}(\vec k) \Big(\frac{S_d(d^2-2)}{d(d+2)}\Big)
\int_{\Lambda/s}^{\Lambda} \frac{q^{d-1}dq}{(2\pi)^d} {\cal M}(q). 
\end{eqnarray}
The one-loop correction is of order $O(k^2)$
and is proportional to the same projector as the bare noise spectrum.
Inserting the above line
into Eq. (\ref{noisespec1}) gives
\begin{equation}\label{Dfinal} 
D(k) = D_0(k) +
{\cal S} \lambda_0^2 \,
2k^2 \times  \Big(\frac{S_d(d^2-2)}{d(d+2)(2\pi)^d}\Big)
\int_{\Lambda/s}^{\Lambda}  q^{d-1} \,dq {\cal M}(q) + O(k^3). 
\end{equation}
Referring to the form of the 
noise spectrum matrix in Eq. (\ref{correlation}), 
we see that the one-loop correction
to the noise spectrum is {\it irrelevant} in the long-wavelength limit
provided $y > -2$ (The case $y = -2$ is akin to 
Model A of Forster, Nelson and 
Stephen 1977). The Taylor expansion of $I_{lm,nj}(\vec k,0)$ 
about $\vec k = 0$ (taking into account contributions from the domain
of integration as well, see discussion surrounding Eq. (\ref{momshell}) ) 
will just
introduce positive powers of the external momentum.   
Thus we conclude that at one-loop as $k \rightarrow 0$ and for $y > -2$
\begin{eqnarray}
D(k) &=& D_0(k)  \nonumber \\
\label{Anoise}
\Rightarrow A^{<} &=& A,  \\
\label{Bnoise}
{\rm and}\,\, B^{<} &=& B.
\end{eqnarray}

Note that for noise spectra having $y < -2$, the one-loop correction will
induce relevant operators in the hydrodynamic limit. {}For example, if $y = -3$,
the one-loop correction goes as $O(k^2)$ and in fact {\it dominates} over the 
the
initial bare spectrum in the infrared. 
To be able to renormalize consistently 
in these other situations requires adding the appropriate
dominant relevant operators to the noise spectrum.   
{}For the case $y = -2$, one needs only complete the calculation sketched
in Eq. (\ref{noise3}) and insert this into Eq. (\ref{Dfinal}). In this case, one would
find that both $A$ and $B$ receive nonzero corrections in the hydrodynamic limit. 

\section{Vertex renormalization}

The diagrammatic expansion for the one-loop correction to the vertex
function, or coupling constant, is depicted
in Fig. (2c). As can be seen from this figure, there
are three distinct one-loop diagrams 
contributing to the correction.
Momentum and frequency conservation implies that the vertex function
can depend at most on two independent external momenta and two
independent external frequencies. With the momentum and frequency
flow assignment taken as shown in Fig. (2c),
where momentum and frequency enter 
via the conjugate field and exit via the two physical fields,
the equation for the one-loop vertex correction is 
\begin{equation}\label{vertexloop}
\lambda^{<} {\bf P}_{l,nm}(\vec k) = \lambda_0 {\bf P}_{l,nm}(\vec k)
- {\cal S} \lambda_0^3 [{\bf P}_{j,nt}(\vec k) + {\bf P}_{t,nj}(\vec k)]
I_{lmjt}(\vec k,\vec k_2,\vec k - \vec k_2; \omega,\omega_2,\omega - \omega_2),
\end{equation}
where the diagrammatic symmetry factor is ${\cal S} = 4$. Here
$\lambda^{<} = \lambda^{<}(A,B,\gamma_{+},\gamma_{-},\lambda_0)$ 
is the renormalized
coupling parameter, which in principle 
can depend on all the bare parameters appearing
in the equations of motion.
The loop-integral representing the sum
of the three types of (amputated) triangle graphs can be written as
\begin{eqnarray}\label{vertex1}
I_{lmjt}&(&\vec k,\vec k_2 , \vec k - \vec k_2 ; \omega,\omega_2,\omega - \omega_2) = 
\nonumber \\
&{}& {\rm tr} \int^{>} \frac{d^d \vec q}{(2\pi)^d}\, \int^{\infty}_{-\infty} 
\frac{d\Omega}{2\pi} \, \left\{
{\cal C}_0(\vec q,\Omega){\bf P}_{jk}(\vec q)
[{\bf P}_{k,rl}(\vec k_2 - \vec q) + {\bf P}_{l,rk}(\vec k_2 - \vec q)]
{\cal G}_0(\vec q - \vec k_2, \Omega- \omega_2)\delta_{rs} \right. \nonumber \\
&\times& [{\bf P}_{s,pm}(\vec k- \vec q) + {\bf P}_{m,ps}(\vec k- \vec q)]
{\cal G}_0(\vec q - \vec k, \Omega- \omega)\delta_{tp}\nonumber \\
&+& {\cal G}_0(-\vec q,-\Omega)\delta_{jk}
[{\bf P}_{l,kr}(\vec q) + {\bf P}_{r,kl}(\vec q)]
{\cal C}_0(\vec q - \vec k_2, \Omega - \omega_2){\bf P}_{rs}(\vec q - \vec k_2)\nonumber \\
&\times& [{\bf P}_{s,pm}(\vec k- \vec q) + {\bf P}_{m,ps}(\vec k- \vec q)]
{\cal G}_0(\vec q - \vec k, \Omega - \omega)\delta_{tp}\nonumber \\
&+& {\cal G}_0(-\vec q,-\Omega)\delta_{jk}
[{\bf P}_{l,kr}(\vec q) + {\bf P}_{r,kl}(\vec q)]
{\cal G}_0(\vec k_2 - \vec q, \omega_2- \Omega)\delta_{rs}
[{\bf P}_{p,sm}(\vec q- \vec k_2) + {\bf P}_{m,sp}(\vec q - \vec k_2)] \nonumber \\
&\times& \left. {\cal C}_0(\vec k - \vec q, \omega - \Omega)
{\bf P}_{tp}(\vec k - \vec q) \right\},
\end{eqnarray}
which follows from translating the diagrams into their associated
mathematical expressions using the basic elements derived above for the
bare response Eq. (\ref{bareresponse}), 
bare correlation Eq. (\ref{correlation}) and bare vertex functions 
Eq. (\ref{barevertex}). 
The trace (tr)
is taken over the
product of the $2 \times 2$ block matrices. 
We are interested in evaluating this integral in the
hydrodynamic limit (i.e., for small $k$ and small $\omega$). As there is already a factor
linear in $\vec k$ multiplying the loop integral, 
and the bare vertex is itself linear in momentum, 
we can immediately take the
limit $\vec k = \vec k_2 \rightarrow 0$ and
$\omega = \omega_2 \rightarrow 0$ {\it inside} 
the loop integral right from the outset and focus attention on $I_{lmjt}(0,0,0;0,0,0)$. 
This
results in a tremendous simplification. The properties of the projection operators
allow one to further reduce products such as
\begin{equation}
[{\bf P}_{l,jr}(\vec q) + {\bf P}_{r,jl}(\vec q)]
[{\bf P}_{p,rm}(\vec q) + {\bf P}_{m,rp}(\vec q)]
{\bf P}_{tp}(\vec q) = q_l q_m {\bf P}_{jt}(\vec q) = q^2 {\hat n}_l {\hat n}_m 
{\bf P}_{jt}(\vec q),
\end{equation}
and likewise for the other products that one encounters in Eq. (\ref{vertex1}).

The angular
integrations immediately can be computed by using
\begin{equation}
\int d\Omega_d \, {\hat n}_l {\hat n}_m 
{\bf P}_{jt}(\vec q) = 
\frac{S_d}{d(d + 2)}\Big( (d+1) \delta_{jt}\delta_{lm} - [\delta_{lj}\delta_{mt}
+ \delta_{lt}\delta_{jm}] \Big),
\end{equation}
with the result that
\begin{eqnarray}\label{vertex2}
\lambda^{<} {\bf P}_{l,nm}(\vec k) &=& \lambda_0 {\bf P}_{l,nm}(\vec k)
+ \lambda_0^3 \big(\frac{-2 S_d}{d(d + 2)}\big)
[{\bf P}_{l,nm}(\vec k) + {\bf P}_{m,nl}(\vec k)] \times \nonumber \\ 
&{\rm tr}&
\int^{>} \frac{q^{d+1} dq}{(2\pi)^d}\, \int^{\infty}_{-\infty} 
\frac{d\Omega}{2\pi} \, \left\{ 
{\cal C}_0(\vec q,\Omega){\cal G}_0(\vec q,\Omega){\cal G}_0(\vec q,\Omega)  
- {\cal G}_0(-\vec q,-\Omega){\cal C}_0(\vec q,\Omega)
{\cal G}_0(\vec q,\Omega) \right. \nonumber \\
&+& \left. {\cal G}_0(-\vec q,-\Omega){\cal G}_0(-\vec q,-\Omega){\cal C}_0(-\vec q,-\Omega)
\right\}.
\end{eqnarray}

The frequency integral can be performed by the method of residues. The integrand has
both simple and double poles in the complex frequency plane; the 
integration contours can be closed in either
the upper or lower half-plane. 
The first and third factors appearing under the frequency
integral have simple poles in the upper and lower half-plane, respectively. 
These
separate contributions can be neatly combined to yield
\begin{eqnarray}\label{combine}
\int^{\infty}_{-\infty} 
\frac{d\Omega}{2\pi} \, ( 
{\cal C}_0(\vec q,\Omega){\cal G}_0(\vec q,\Omega){\cal G}_0(\vec q,\Omega)  
&+& {\cal G}_0(-\vec q,-\Omega){\cal G}_0(-\vec q,-\Omega){\cal C}_0(-\vec q,-\Omega)
) \nonumber \\ 
&=& \frac{D_0(q)}{8q^6} \left\{
\frac{1}
{(\gamma_{+} - \gamma_{-})^3}
\left[ \matrix{1 & -1\cr 
        -1 & 1} \right] 
+ \frac{1}
{(\gamma_{+} + \gamma_{-})^3}
\left[ \matrix{1&1\cr 
         1&1} \right] 
\right\}.
\end{eqnarray}
On the other hand,
the middle factor has double poles in both the upper and lower
half-planes and its contribution to the frequency integral turns out
to yield
\begin{eqnarray}\label{middle}
\int^{\infty}_{-\infty} 
\frac{d\Omega}{2\pi} \,\, {\cal G}_0(-\vec q,-\Omega){\cal C}_0(\vec q,\Omega)
{\cal G}_0(\vec q,\Omega)  
= \frac{D_0(q)}{8q^6} \left\{
\frac{1}
{(\gamma_{+} - \gamma_{-})^3}
\left[ \matrix{1 & -1\cr 
        -1 & 1} \right] 
+ \frac{1}
{(\gamma_{+} + \gamma_{-})^3}
\left[ \matrix{1&1\cr 
         1&1} \right] 
\right\}.
\end{eqnarray}
Subtracting Eq. (\ref{middle}) from Eq. (\ref{combine}) 
as per Eq. (\ref{vertex2}) yields a null result for the
frequency integral that holds 
for all spatially correlated Gaussian noise $D_0(q)$. Thus the vertex
coupling parameter does not renormalize in the long-wavelength limit and we
have that
\begin{equation}\label{vertex4}
\lambda^{<} = \lambda_0.
\end{equation}

This non-renormalization of the nonlinear coupling parameter holds also for
the Navier-Stokes equation driven by arbitrary spatially correlated Gaussian
noise and is ascribed to the 
Galilean invariance of the stochastic NS 
equation (Forster et al. 1977). The importance
of Galilean invariance for the Burgers equation (and the KPZ) equation is
further addressed in Medina et al. (1989).
We suspect that the non-renormalization of the nonlinearity holds to all
orders in the loop expansion for stochastic MHD. 
To establish this one could investigate the
renormalization of the vertex coupling via suitable Ward identities, as was 
done for the case of the KPZ equation (Frey and Tauber 1994). 
We leave that problem for a separate
discussion elsewhere.  
Temporally correlated noise
breaks the Galilean invariance and can (
and usually does) lead to non-trivial renormalization
of $\lambda_0$.

\section{Integrating over the d-sphere}

The various angular integrations and angular averages
that are needed in the renormalization group calculations are collected here. 
The area
element of the unit sphere in $d$-dimensions is 
\begin{equation}
d\Omega_d = d\phi \, \sin \theta_1 d\theta_1 \, \sin^2 \theta_2 d\theta_2 \,
\cdots \sin^{(d-2)} \theta_{d-2} d\theta_{d-2},
\end{equation}
where $0 \leq \phi < 2\pi$ and $0 \leq \theta_j < \pi$, for $j = 1,2, \cdots (d - 2)$.
Let $S_d$ represent the area of the unit $d$-sphere, $\hat n_j$ denote a unit
vector in the $j$-th direction, and 
$\Gamma$ denote the Gamma function. Then
\begin{eqnarray}\label{sphere}
\int d\Omega_d &=& \frac{2\pi^{d/2}}{\Gamma(\frac{d}{2})} = S_d, \nonumber \\
\int d\Omega_d \, {\hat n}_i {\hat n}_j &=& \frac{S_d}{d} \delta_{ij}, \nonumber \\
\int d\Omega_d \, {\hat n}_i {\hat n}_j {\hat n}_n {\hat n}_m &=&
\frac{S_d}{d(d + 2)} \big( \delta_{ij} \delta_{mn} + 
\delta_{im} \delta_{jn} + \delta_{in} \delta_{jm} \big).
\end{eqnarray} 
The angular average of the product of an 
odd number of unit vectors over the unit sphere vanishes
identically.

\medskip
\noindent
Amit, D.J. 1978 
{\it Field Theory, the Renormalization Group, and
Critical Phenomena}. McGraw-Hill.

\medskip
\noindent
Barab\'asi A.-L., and Stanley, H.E. 1995 
{\it Fractal Concepts in
Surface Growth}. 
Cambridge.

\medskip
\noindent
Biskamp, D. 1993
{\it Nonlinear Magnetohydrodynamics}.
Cambridge.

\medskip
\noindent
Brandenburg, A., Enqvist, K. and Olesen, P. 1996 
Large-scale
magnetic fields from hydromagnetic turbulence in the very early universe.
Phys. Rev. D{\bf 54}, 1291-1300.

\medskip
\noindent
Burlaga, L.F. 1991 Intermittent
turbulence in the solar-wind.  J. Geophys. Res. {\bf 96} (A4),
5847-5851.

\medskip
\noindent
Camargo, S.J.  and Tasso, H. 1992 
Renormalization
group in magnetohydrodynamic turbulence. 
Phys. Fluids B{\bf 4} (5), 1199-1212.

\medskip
\noindent
Cardy, J. 1996 
{\it Scaling and Renormalization in 
Statistical Physics}.
Cambridge.

\medskip
\noindent
Dannevik, W.P., Yakhot, V.  and
Orszag, S.A. 1987
Analytical theories of turbulence and the $\epsilon$
expansion. 
Phys. Fluids {\bf 30} (7), 2021-2029.

\medskip
\noindent
De Dominicis, C.  and  Peliti, L. 1978
Field-theory renormalization and critical dynamics above 
$T_c$: helium, antiferromagnets and liquid-gas systems.
Phys. Rev. {\bf B18}, 353-376.

\medskip
\noindent
DeDominicis, C. and Martin, P.C. 1979 
Energy spectra of certain randomly-stirred fluids.
Phys. Rev. A
{\bf 19}, 419-422.

\medskip
\noindent
DeYoung, D.S. 1980
Turbulent Generation of Magnetic Fields in
Extended Extragalactic Radio Sources. 
Ap. J. {\bf 241}, 81-97.

\medskip
\noindent
DeYoung, D.S. 1992
Galaxy-Driven Turbulence and the Growth of Intracluster Magnetic
Fields. 
Ap. J. {\bf 386}, 464-472.

\medskip
\noindent
Elsasser, W.M. 1950 
The hydromagnetic equations.
Phys. Rev. {\bf 79}, 183.

\medskip
\noindent
Enqvist, K. 1998
Primordial Magnetic Fields.
Int. J. Mod. Phys. D{\bf 7}, 331-350.

\medskip
\noindent
Eyink, G.L. 1994
The renormalization group method in
statistical hydrodynamics. 
Phys. Fluids {\bf 6} (9), 3063-3078.

\medskip
\noindent
Forster, D.,  Nelson, D.R.  and Stephen, M.J. 1977
Large-distance and long-time properties of a randomly stirred fluid.
Phys. Rev. {\bf A16}, 732-749.

\medskip
\noindent
Fournier, J.-D., Sulem, P.-L.
and
Pouquet, A. 1982
Infrared properties of forced magnetohydrodynamic
turbulence. 
J. Phys. A: Math. Gen. {\bf 15}, 1393-1420.

\medskip
\noindent
Frey, E. and Tauber, U.C. 1994
Two-loop renormalization-group analysis of the Burgers-
Kardar-Parisi-Zhang equation.
Phys. Rev. E{\bf 50}, 1024-1044.

\medskip
\noindent
Frisch, U., Pouquet, A., Leorat, J. and Mazure, A. 1975
Possibility of an inverse cascade of magnetic helicity
in magnetohydrodynamic turbulence.
J. Fluid Mech. {\bf 68}, 769-778.

\medskip
\noindent
Frisch, U.  and Sulem, P.L. 1984
Numerical simulation of the inverse cascade in
two-dimensional turbulence. 
Phys. Fluids {\bf 27}, 
1921-1923.

\medskip
\noindent
Frisch, U. 1995 
{\em Turbulence}. 
Cambridge.

\medskip
\noindent
Gross, D.J. 1981 
in {\it Methods in Field Theory}, Les Houches 
Summer School 1975 Session XXVIII, edited by
R. Balian and J. Zinn-Justin. 
North-Holland/World Scientific

\medskip
\noindent
Hochberg, D., P\'erez-Mercader, J., Molina-Par\'\i s,
C. and Visser, M. 1999a
Renormalization group improving the effective action: a
review. 
Int. J. Mod. Phys. A{\bf 14}, 1485-1521.

\medskip
\noindent
Hochberg, D., Molina-Par\'\i s, C., P\'erez-Mercader,
J. and Visser, M. 1999b
Effective action for stochastic partial differential equations.
Phys. Rev. {\bf E}60, 6343-6360.

\medskip
\noindent
Kraichnan, R.H. 1967 
Inertial ranges in two-dimensional turbulence.
Phys. Fluids {\bf 10}, 1417-1423.

\medskip
\noindent
Kraichnan, R.H. 1973 
Helical turbulence
and absolute equilibrium. 
J. Fluid Mech. {\bf 59}, 745-752.

\medskip
\noindent
Lilly, D. 1969 
Numerical simulation of
two-dimensional turbulence.
Phys. Fluids Suppl. {\bf 12} II, 240-249.

\medskip
\noindent
Longcope, D.W. and Sudan, R.N. 1991 
Renormalization
group analysis of reduced magnetohydrodynamics with application to subgrid
modeling. Phys. Fluids B{\bf 3} (8), 1945-1962.

\medskip
\noindent
Ma, S.-K. 1976 
{\it Modern Theory of Critical Phenomena}.
Addison-Wesley

\medskip
\noindent
Martin, P.C., Siggia, E.D. and Rose, H.A. 1973
Statistical dynamics of classical systems.
Phys. Rev. {\bf A8}, 423-437.

\medskip
\noindent
McComb, W.D. 1995 
{\it The Physics of Fluid Turbulence}.
Oxford.

\medskip
\noindent
Medina, E., Hwa, T., Kardar, M. and Zhang, Y-C. 1989
Burgers equation with correlated noise: Renormalization-group
analysis and applications to directed polymers and interface growth.
Phys. Rev. {\bf A39}, 3053-3075.

\medskip
\noindent
Papaloizou, J.C.B  and Lin, D.N.C. 1995
Theory of accretion disks I: angular momentum
transport processes. 
Annu. Rev. Astron. Astrophys. {\bf 33}, 505-540. 

\medskip
\noindent
Pouquet, A., Frisch, U.  and Leorat J. 1976
Strong MHD helical turbulence and the nonlinear
dynamo effect.
J. Fluid Mech. {\bf 77}, 321-354.

\medskip
\noindent
Pouquet, A. 1978
On two-dimensional magnetohydrodynamic turbulence.
J. Fluid Mech. {\bf 88}, 1-16.

\medskip
\noindent
Pouquet, A. and Patterson, G.S. 1978 
Numerical simulation of helical magnetohydrodynamic
turbulence.
J. Fluid Mech. {\bf 85},
305-323.

\medskip
\noindent
Ramond, P. 1981 {\it Field Theory: A Modern Primer} Benjamin-Cummings,
Reading, Massachusetts. 

\medskip
\noindent
Rivers, R.J. 1987 
{\it Path integral methods 
in quantum field theory}.
Cambridge.

\medskip
\noindent
Sommeria, J. 1986 
Experimental study of the two-dimensional inverse
energy cascade in a square box.
J. Fluid Mech. {\bf 170}, 139-168.

\medskip
\noindent
Van Ballegooijen, A.A. 1985
Electric
currents in the solar corona and the existence of magnetostatic
equilibrium. 
Ap. J. 
{\bf 298}, 421-430.

\medskip
\noindent
Van Ballegooijen, A.A. 1986 
Cascade of magnetic energy as a mechanism of coronal heating.
Ap. J. {\bf 311}, 1001-1014.

\medskip
\noindent
Wilson K.G. and Kogut, J. 1974
The renormalization group and the $\epsilon$ 
expansion.
Physics Reports 12{\bf C}, 75-200.

\medskip
\noindent
Yakhot, V.  and Orszag, S.A. 1986
Renormalization-group analysis of turbulence.
Phys. Rev. Lett. {\bf 57}, 1722-1724.

\medskip
\noindent
Zinn--Justin, J. 1996
{\it Quantum field theory and critical phenomena}.
Oxford.



\newpage

\begin{figure}
\begin{center}
 \epsfig{file=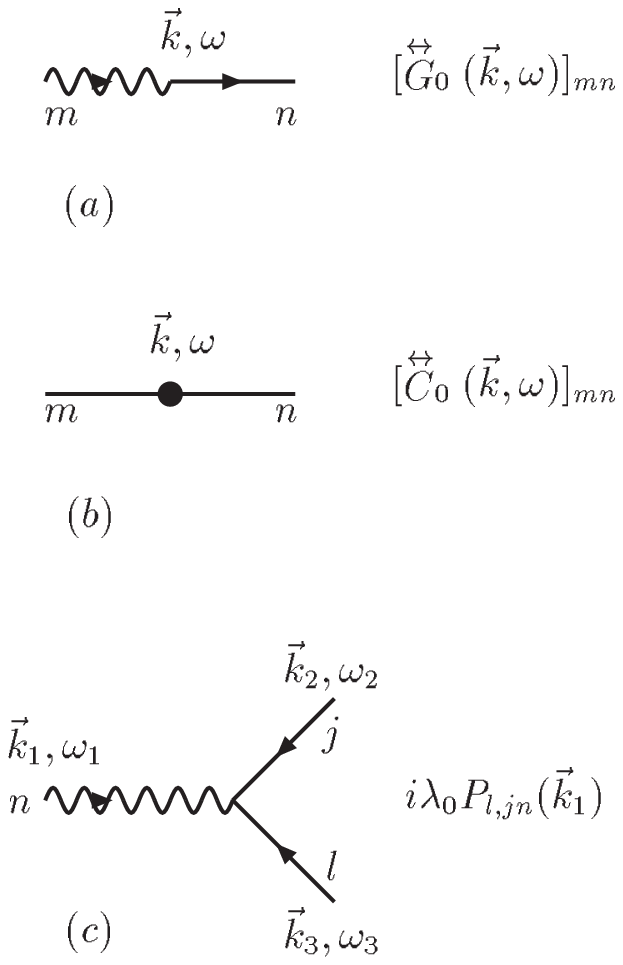, width=7in,bbllx=65pt,bblly=400pt,%
         bburx=530pt, bbury=790pt}
\end{center}
\begin{center}
FIG 1: Diagrammatic (Feynman-type) rules for the perturbation theory
expansion of MHD.
\end{center}
\end{figure}

\begin{figure}[b]
\epsfysize=31cm
{\centerline{\epsfbox{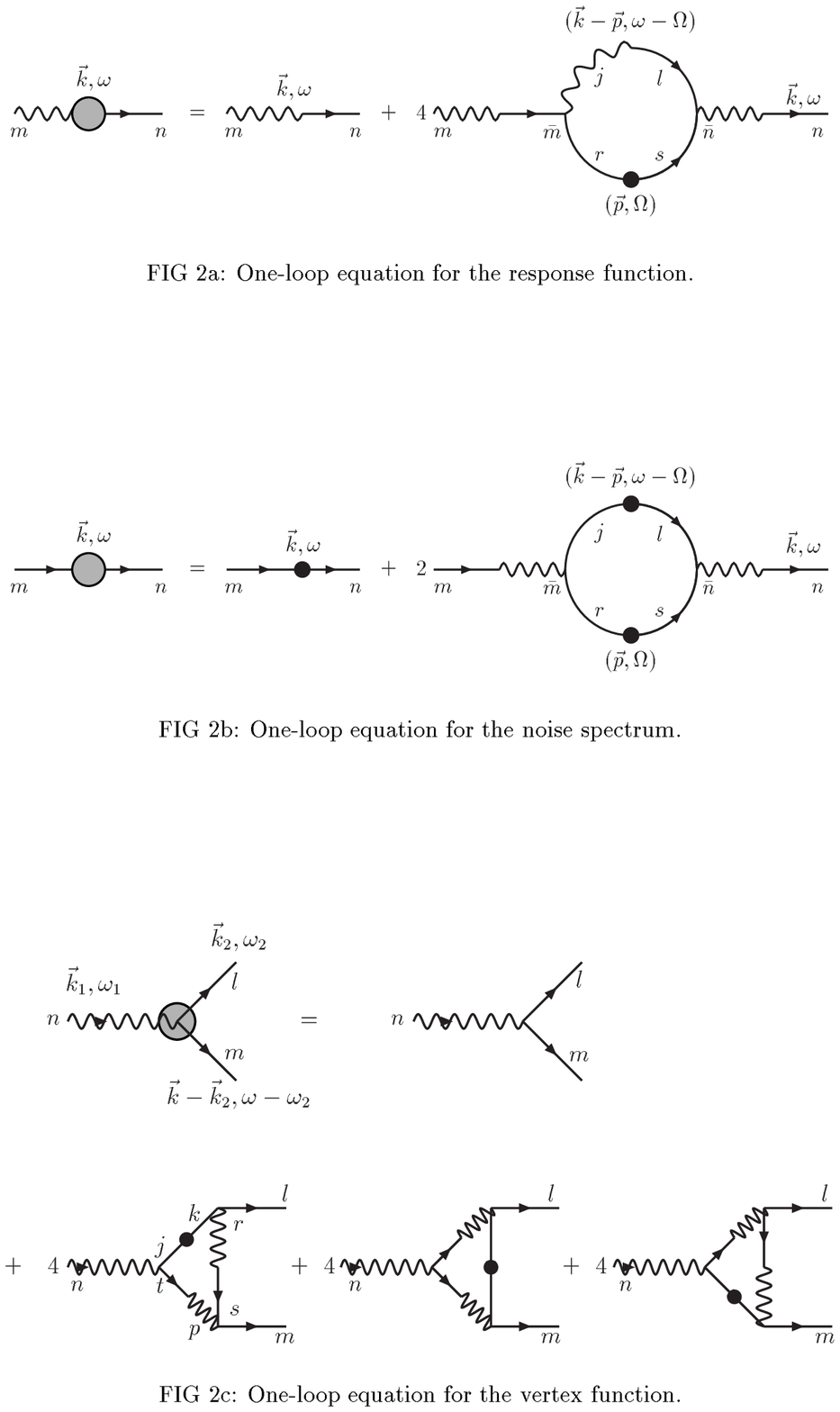}}}

\vspace{1cm}

\end{figure}

\end{document}